\documentclass[10pt,journal,compsoc]{IEEEtran}

\ifCLASSOPTIONcompsoc
  \usepackage[nocompress]{cite}
\else
  \usepackage{cite}
\fi
\ifCLASSINFOpdf
\else
\fi
\usepackage{graphicx}
\usepackage{algorithm}
\usepackage[noend]{algpseudocode}
\usepackage{subfig}
\usepackage[fleqn]{amsmath}
\usepackage{relsize}
\usepackage{mathtools}

\usepackage{ragged2e}
\usepackage{booktabs} 
\usepackage[percent]{overpic}
\usepackage{relsize}
\usepackage{multirow}
\usepackage{amssymb}
\usepackage{bm}
\usepackage{enumitem}
\usepackage{dblfloatfix}
\usepackage{hhline}
\usepackage{amsmath}
\usepackage[left=0.65in,right=0.65in,top=0.65in,bottom=0.65in,footskip=.25in]{geometry}
\usepackage[normalem]{ulem}
\usepackage{tikz}

\newcommand*\circled[1]{\tikz[baseline=(char.base)]{
		\node[shape=circle,fill,inner sep=2pt] (char) {\textcolor{white}{#1}};}}

\begin{document}

%
\title{An Analytical Model for Performance and Lifetime Estimation of Hybrid DRAM-NVM Main Memories}
%
%
%
%

\newcommand{\textoverline}[1]{$\overline{\mbox{#1}}$}
\newcommand{\specialcell}[2][c]{%
	\begin{tabular}[#1]{@{}l@{}}#2\end{tabular}}

\author{Reza Salkhordeh,
        Onur Mutlu, and~Hossein~Asadi\vspace{-.6cm}
}

\IEEEtitleabstractindextext{%
	{\justify
\begin{abstract}
Emerging \emph{Non-Volatile Memories} (NVMs) have promising advantages
(e.g., lower idle power, higher density, and non-volatility) over the
existing predominant {main} memory technology, DRAM. Yet, NVMs also have
disadvantages (e.g., longer latencies, higher active power, and limited
endurance). System architects are therefore examining hybrid DRAM-NVM
main memories to enable the advantages of NVMs while avoiding the
disadvantages as much as possible. Unfortunately, the hybrid memory
design space is very large and complex due to the
existence of very different types of NVMs and their {rapidly-changing}
characteristics. Therefore, optimization of performance and lifetime of hybrid
memory based computing platforms  and their experimental evaluation
using traditional simulation methods can be very time-consuming and sometimes even
impractical.
As such, it is necessary to develop a fast and flexible analytical
model to estimate the performance and lifetime of hybrid memories on various
workloads.

This paper presents an analytical model for hybrid memories based on
Markov decision processes.
The proposed model estimates the hit ratio and lifetime for various 
configurations of DRAM-NVM hybrid main memories.
Our model also provides accurate estimation of the effect of data migration
policies on the hybrid memory hit ratio (i.e., percentage of accesses supplied by either DRAM or NVM), one of the most
important factors in hybrid memory performance and lifetime.
Such an analytical model can aid designers to tune hybrid memory configurations
to improve performance and/or lifetime. 
We present several optimizations that make our model more efficient while
maintaining its accuracy. 
Our experimental evaluations conducted using the PARSEC benchmark suite
show that the proposed model \emph{(a)} accurately predicts the 
hybrid memory hit ratio compared to the state-of-the-art hybrid memory 
simulators with an average ({maximum})
error of 4.61\% (13.6\%) {on} a commodity server
(equipped with 192GB main memory and quad-core Xeon processor), \emph{(b)} 
{accurately}
{estimates} the NVM lifetime {with} an average ({maximum}) error of 2.93\%  (8.8\%), 
and \emph{(c)}
is on average (up to) 4x (10x) faster than conventional state-of-the-art simulation platforms for hybrid memories.
\end{abstract}}

}

\maketitle

\IEEEdisplaynontitleabstractindextext

%
\IEEEpeerreviewmaketitle


\ifCLASSOPTIONcompsoc
\IEEEraisesectionheading{\section{Introduction}\label{sec:introduction}}
\else
\section{Introduction}
\label{sec:introduction}
\fi
\vspace{-.17cm}
{Large-scale} data-intensive applications increasingly require large and efficient 
main memory {due} their large data {footprints}.
Traditionally, \emph{Dynamic Random Access Memory} (DRAM) has been used as the
predominant {main} memory technology in computer systems {due to} its low 
cost
per bit (\$/GB), low access latency, and symmetric performance on read and write accesses.
DRAM technology, however, suffers from major shortcomings, such as high idle 
power, low scalability, and reduced reliability {due to its fundamental 
dependence on} charge storage \cite{7927156,6582088,5430747,Lee:2010:PCM:1785414.1785441}.

To address the shortcomings of DRAM, emerging \emph{Non-Volatile Memories} 
(NVMs) offer promising characteristics, such as low idle power, high density, 
and non-volatility 
\cite{Lee:2009:APC:1555754.1555758,7155523,6557176,Qureshi:2009:SHP:1555754.1555760,cheshmikhani,5430747,Lee:2010:PCM:1785414.1785441}.
Example emerging NVM technologies include \emph{Phase Change Memory} (PCM), \emph{Spin-Transfer Torque Magnetic Random-Access Memory} (STT-MRAM), Metal Oxide Resistive RAM, and memristors.
NVMs, however, have several drawbacks, such as limited endurance, longer latencies, and high active power for write requests, which prevents them from \emph{completely} replacing DRAM as main memory {\cite{6582088,Qureshi:2009:SHP:1555754.1555760}}.

\begin{figure}[t]
	\centering
	\vspace{-.05cm}
	\includegraphics[scale=0.45]{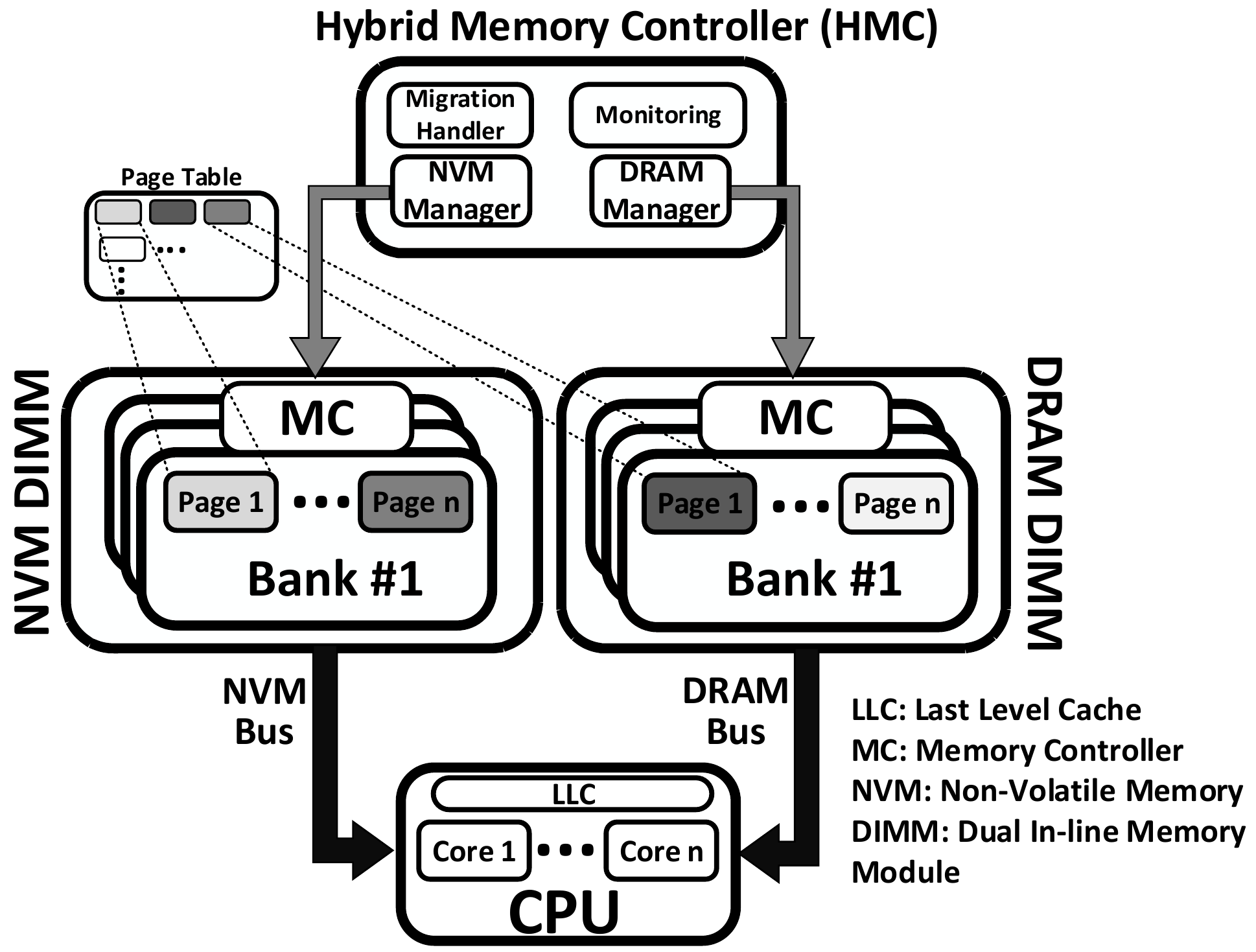}
	\caption{Hardware structure of {an example} HMA}
	\vspace{-.65cm}
	\label{fig:hma}
\end{figure}

\begin{figure*}[!b]
	\centering
	\vspace*{-.4cm}
	\subfloat[CLOCK-DWF with Various ``\emph{Expiration}'' Threshold Values (1, 2, 4, 8, $\infty$)] {\includegraphics[width=.483\textwidth]{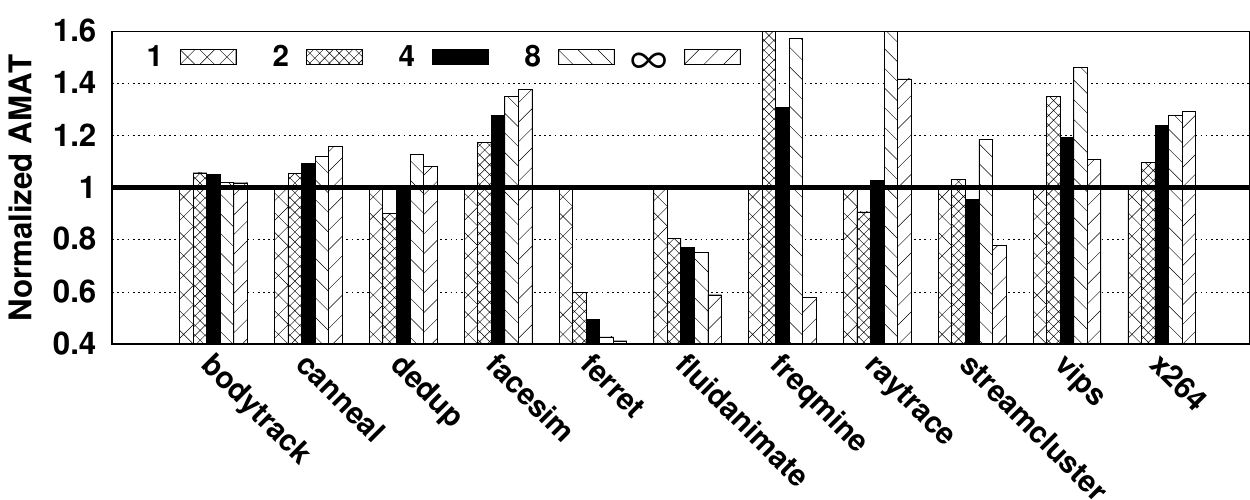}%
		\label{fig:clockdwf-param}}
	\hfill
	\subfloat[Two-LRU with Various ``\emph{Migration}'' Threshold Values (1, 4, 16, 32, $\infty$)] {\includegraphics[width=.483\textwidth]{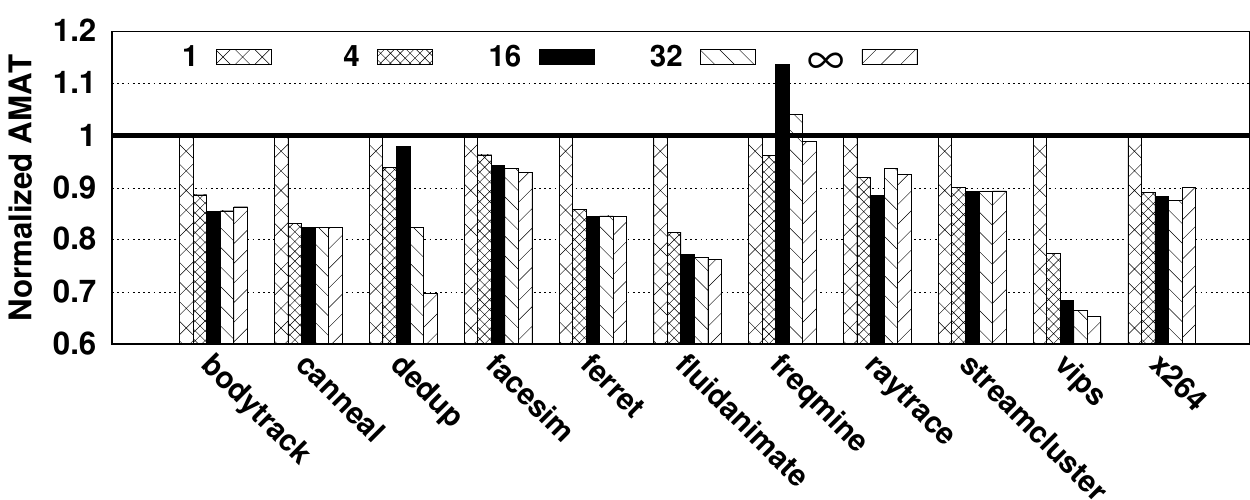}%
		\label{fig:twolru-param}}
	\vspace{-0.2cm}
	\caption{Effect of HMA parameters on AMAT}
	\label{fig:param-sens}
	\vspace{-0.6cm}
\end{figure*}

In order to take advantage of the promising characteristics of NVMs while minimizing the effect of their limitations, previous studies suggested employing \emph{Hybrid Memory Architectures} (HMAs) composed of DRAM and NVMs in a single level or multiple levels of the memory hierarchy \cite{Ramos:2011:PPH:1995896.1995911,6172197,8048927,Yu:2017:BBD:3123939.3124555,Su:2015:HEM:2818950.2818974,6378661,Dhiman:2009:PHP:1629911.1630086,Qureshi:2009:SHP:1555754.1555760,clockdwf,date16,Lee:2015:MMP:2695664.2695675,chen2016dram,7753305,7856636,Agarwal:2017:TAP:3037697.3037706,6522338,Lin:2016:MTP:2872362.2872401,7056027,6657042,6968756,7011373,7284066,7855888,7011375,Qureshi:2012:FLT:2457472.2457502,7011374}.
Fig. \ref{fig:hma} {shows the} general hardware structure commonly used by HMAs.
In this example HMA, each memory device has a
separate controller and is connected to the memory bus using a traditional DRAM 
channel.\footnote{\emph{{The hybrid} memory controller} can be implemented 
either as a hardware memory controller or a module in the Operating System.
Migrating data pages from DRAM to NVM and vice versa is done by the same memory bus via \emph{Direct Memory Access} (DMA).}
In HMAs, management policies can significantly affect system performance.
This necessitates significant effort to carefully optimize such policies.
In addition to selecting the most suitable policies, our analysis in this paper reveals that the internal configurations and parameters of the selected policies also significantly affect system performance,
which is evaluated {using the} \emph{Average Memory Access Time} (AMAT) {metric} in this 
work.
This metric, in turn, depends {mainly} on the latency of memory devices and the hit ratio of the hybrid memory.
The hybrid memory hit ratio is defined as the ratio of accesses to the virtual 
memory {(excluding accesses that hit in processor caches)} that are supplied by either DRAM or NVM.

To show the importance of the internal configurations of hybrid {memory management systems}, 
we evaluate the performance of two different {software-based} management mechanisms used in {HMAs} 
(CLOCK-DWF \cite{clockdwf} and TwoLRU \cite{date16}) by assigning different 
values to their major configuration parameters.
As we show in our analysis in Fig. \ref{fig:param-sens}, performance\footnote{Throughout the paper, we use the terms \emph{performance} and \emph{AMAT}  interchangeably.} of HMAs heavily depends on the selected values for parameters used in each HMA management mechanism.
The performance difference is greater than 30\% (up to 150\%) in many cases, 
which {indicates} the importance of selecting the most suitable values for 
{HMA} parameters to reduce the AMAT.

With such a significant effect HMA parameters have on performance, memory 
architects put a lot of effort into
finding optimized parameters for HMA management mechanisms across a wide variety of workloads.
Such studies usually result in {sub}-optimal configurations for HMAs since they 
explore and experiment {in} a \emph{limited design space}.
{With} various configurations of HMAs and different  types of NVMs, the 
design space is very large, 
which makes design and optimization of HMAs using traditional simulation-based 
methods very time-consuming and {sometimes impractical}.
In recent years, this {problem has become} even more challenging {due to} 
rapid changes in NVM characteristics.
Any change {in NVM} characteristics requires re-running time-consuming 
simulations to find optimized HMA parameters.
{Thus}, using traditional simulation-based methods to find suitable 
parameters for emerging HMAs for a specific NVM and/or workload has become 
either {very} cumbersome or extremely time-consuming {due to} the wide 
variety of design choices and fast evolution of NVMs.
This shortcoming exists in HMAs {described} in \cite{clockdwf,date16}.


To alleviate the shortcomings of simulation techniques, it is very beneficial for system designers to have a fast but accurate {\emph{analytical}} performance model for HMAs.
Analytical modeling of an HMA enables designers to \emph{(a)} {perform} accurate {performance} analysis of any arbitrary HMA architecture, \emph{(b)} explore various tradeoffs, and \emph{(c)} quickly predict the {HMA} hit ratio and NVM lifetime for a given HMA configuration.

To our knowledge, {\emph{no}} previous study has attempted to analytically model HMAs.
However, several analytical models have been proposed in previous studies for 
\emph{other} levels of {the} memory hierarchy, such as CPU cache management 
policies \cite{Guo:2006:AMC:1140103.1140304,7095785,5749733,6375779}.
Gui et. al. \cite{Guo:2006:AMC:1140103.1140304} propose an analytical model to predict the hit ratio for \emph{Least Recently Used} (LRU)-based cache management policies based on circular sequence profiling \cite{Chandra:2005:PIC:1042442.1043432}.
Another study attempts to model LRU-based policies with less profiling overhead \cite{7095785}.
{Due to the fundamentally-different performance and endurance 
characteristics of NVMs used in main memory and the need to model the effects 
of the migration policy between NVM and DRAM, to our knowledge, \emph{none} of 
the previously suggested analytical models are applicable to HMAs.}

In this paper, we present the first analytical model for HMAs.
Our model can accurately a) predict the {HMA} hit ratio,
b) predict the lifetime of the NVM used in the HMA, and
c) provide an estimation of the effect of various parameters of {the HMA}  
architecture, such as migration policy, on the {HMA} hit ratio.
The proposed model is applicable to a wide range of HMAs (with a structure similar to the one presented in Fig. \ref{fig:hma}).\footnote{{Appendix A presents} the assumptions on HMAs that can be
	evaluated {with} the proposed analytical model.}
The proposed model can also accurately estimate the effect of migration 
probability on the hybrid memory hit ratio.
This can be used to estimate the effect of internal parameters of the 
examined HMA on the hybrid memory hit ratio.
To accurately estimate the hit ratio of a hybrid memory, the proposed model 
uses three major inputs: a) an access profile of the workload, b) 
probability of evicting each data page when {the} HMA needs to free up space {in 
main} memory, and c) the probability of promotion from NVM to DRAM when a data 
page is accessed in NVM.
We extract the first input from the workload by using a trace analyzer
while the other two inputs are extracted from the target HMA.

To evaluate the performance and accuracy of the proposed model, we have applied 
it to TwoLRU \cite{date16} and CLOCK-DWF \cite{clockdwf} architectures, which 
are the most recent software-based HMAs.\footnote{HMAs with memory controller 
modifications are not {evaluated in our} experiments since their full 
implementation in simulators is rather complex and error-prone, which can lead 
to inaccurate results for evaluating the proposed model.
In other words, HMAs that require hardware modifications {do} not have a ground-truth baseline
to which we can compare our model to.}
Experimental results over the PARSEC and the SPEC CPU2006 benchmark suites show 
that the proposed model can accurately predict the hybrid memory hit ratio and 
AMAT of {these two} HMAs {with} an average error of 4.61\% (13.6\% at 
most), and  2.99\% (11.3\% at most), respectively.
The proposed model can also estimate the NVM lifetime {with} an average 
error of 2.93\% (8.8\% at most).
{Our model is on average 10x (20x at most) faster than performing {simulations} 
when estimating {the} hit ratio of 1000 different HMA configurations.}

The main contributions of this paper are as follows:
\begin{itemize}[leftmargin=*]
	\item We introduce the first analytical model for hybrid memories.
	{Our model} can accurately predict the hit ratio of general hybrid 
	memories.
	\item The proposed model provides an easy-to-use framework for designers to 
	explore various trade-offs between hit ratio and major configuration 
	{parameters} of an arbitrary HMA.
	Such analysis is {critical} in exploring hybrid memory {designs} and 
	optimizing the configuration of {a} hybrid memory system.
	\item The proposed model can accurately estimate the NVM lifetime without 
	any additional computation and/or memory overhead {over} hit ratio 
	estimation.
	\item We develop an open-source framework, based on our model, which can be 
	used by architects and designers to propose {and evaluate new} HMA 
	designs, {taking advantage of emerging NVMs}.\footnote{http://dsn.ce.sharif.edu/software/}
\end{itemize}

\vspace{-.3cm}
\section{Previous Studies}
\label{sec:previouswork}
\vspace{-.1cm}
Numerous analytical models have been proposed in previous studies for {the} 
memory hierarchy and LRU-based cache management policies 
\cite{Dan:1990:AAL:98460.98525,Guo:2006:AMC:1140103.1140304,7095785,807544,Vera:2004:FAF:973097.973099,mobs1,Gulur:2014:AAM:2591971.2591995,4079516,5452069,Vera:2002:LSW:874076.876456,CaBetacaval:2003:ECM:782814.782836,7970187}.
The main goal of such models is to predict the virtual memory or \emph{Last Level Cache} (LLC) hit ratio by profiling the running workload. 
Most of the previous analytical studies use hit ratio as a main estimator for 
performance  as it can accurately {represent} the performance of {the} 
memory hierarchy.
Utilization is another predictor for memory performance which is used in an analytical performance model of main memory \cite{mobs1}.
Predicting latency using analytical {modeling} has also been examined in previous studies using {queuing} models \cite{Gulur:2014:AAM:2591971.2591995}.

In \cite{Kotera:2008:MCA:1509084.1509086}, an analysis over memory workloads 
shows that the stack distance of memory accesses follows {a Zipf 
distribution} law, which {is also} observed in other application domains 
such as web caching and disk I/O 
\cite{Che:2006:HWC:2312147.2313846,Yang:2016:WSZ:2940403.2908557}.
A {low-overhead} hardware mechanism for calculating the stack distance 
{is} proposed in \cite{Sen:2013:ROM:2465529.2465756}.
The accuracy of modelling LRU-based caching techniques is discussed in \cite{Fricker:2012:VAA:2414276.2414286}.
Extension and generalization of such techniques is investigated in \cite{Garetto:2016:UAP:2935322.2896380}, where temporal locality is included in the suggested model.
The behavior of LRU and \emph{First-In First-Out} (FIFO) policies can also be 
predicted using {the} low-complexity method presented in 
\cite{Dan:1990:AAL:98460.98525}.
A model for estimating CPU cache miss ratio for age-based eviction policies 
{is presented} in \cite{7446067}.
Estimating cache miss ratio based on reuse distance is suggested in \cite{4079516,multireuse,7970187}.
In \cite{7920821}, reuse distance is modified to support hybrid cache line sizes.
An analytical model for cache eviction policies {is} proposed in 
\cite{Guo:2006:AMC:1140103.1140304}, which uses workload profiling and Markov 
processes in order to predict the cache hit ratio.
The profiler in \cite{Guo:2006:AMC:1140103.1140304} is based on \emph{circular 
sequence profiling} \cite{Chandra:2005:PIC:1042442.1043432} that is obtained 
{using} a single run {of} the workload and can be used for predicting 
{the} hit ratio {of} many eviction policies.
This profiler collects the total number of accesses and {the} number of unique data pages between each two consecutive accesses to a data page,
which is also employed in \cite{5452069}.
To reduce the profiling overhead, a profiling technique is proposed in \cite{7095785}, which enables {the} prediction of cache hit ratio under various eviction policies and cache associativities.
{Queuing} models can also be employed to predict the behavior of DRAM main memories \cite{Gulur:2014:AAM:2591971.2591995}.
{Such queuing models} can {be used for} many on-chip schedulers by 
re-arranging accesses in the trace file while maintaining the high accuracy of 
the technique.

Since HMAs consist of \emph{more than one} memory module and have 
\emph{multiple governing policies} for managing data pages, {the models 
discussed in this section} \emph{cannot} accurately predict the {hybrid 
memory} behavior.
In addition, the effect of {page} migrations on the memory behavior cannot be modelled 
by the traditional analytical models proposed for {main memory or caches}.

There are numerous HMAs proposed in previous studies,
which require hardware modifications and/or controller re-programming \cite{Ramos:2011:PPH:1995896.1995911,Su:2015:HEM:2818950.2818974,6378661,Dhiman:2009:PHP:1629911.1630086,Qureshi:2009:SHP:1555754.1555760,7516053,6172197,Agarwal:2015:PPS:2694344.2694381}.
A simple {write-only DRAM} cache for NVM main memory is proposed in 
\cite{chen2016dram} to increase NVM lifetime.
\cite{Qureshi:2009:SHP:1555754.1555760} {proposes a} lazy write technique to reduce the number of writes in NVM by using DRAM as an intermediate cache.
\cite{6378661} {proposes a} row-buffer-miss-aware HMA, which improves 
performance by moving data blocks that frequently cause row buffer misses to 
DRAM.
HpMC \cite{Su:2015:HEM:2818950.2818974} {proposes} a hybrid inclusive/exclusive 
HMA in the memory {controller that} attempts to switch DRAM between a cache 
for NVM and a separate module visible to applications to reduce the energy 
consumption.
A modification of {the} CLOCK data structure \cite{tanenbaum2009modern} with two CLOCK handles is proposed in \cite{Lee:2015:MMP:2695664.2695675} to reduce the number of writes to NVM.
Using two CLOCK {handles} along with {another} CLOCK {for frequency of accesses} is proposed in 
\cite{7336177} to improve performance, lifetime, and energy consumption of 
hybrid memories.
CLOCK-DWF \cite{clockdwf} uses two CLOCK data structures to manage DRAM and NVM 
memory modules and {decide page} demotions from DRAM to NVM.
To reduce the migration cost in terms of performance and endurance, UH-MEM 
\cite{8048927} and TwoLRU \cite{date16} aim to limit the 
migrations by estimating their benefit {and} cost.

%
%


%
\vspace{-.25cm}
\section{Characterization of Trace Files}
\label{sec:inputs}
\begin{figure}[h]
	\vspace{-0.4cm}
			
			\centering
			\includegraphics[scale=0.45]{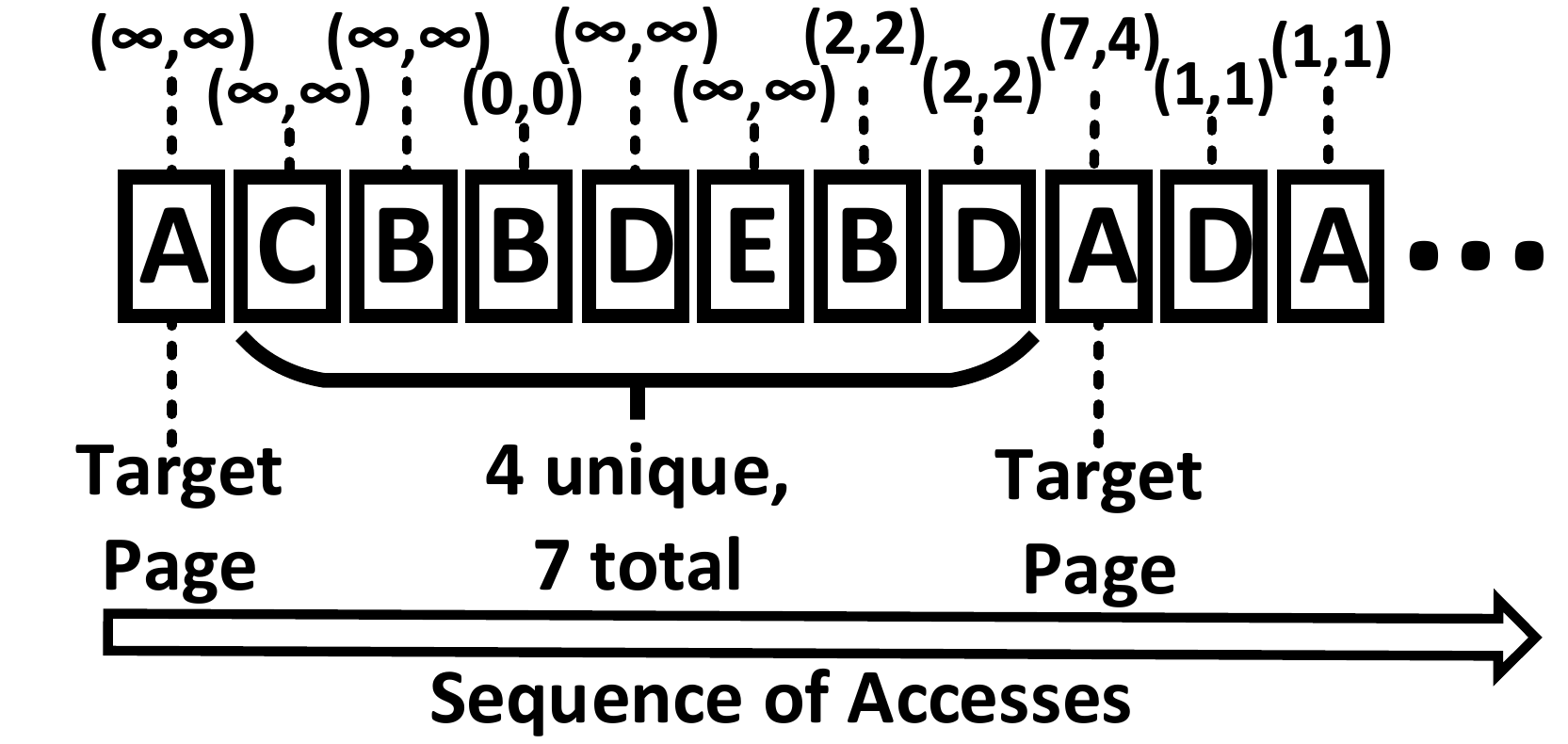}	
			\vspace{-0.1cm}
			\caption{Calculating sequence profiles}
			\label{fig:profiling}
			
\vspace{-0.5cm}
\end{figure}

\vspace{-.1cm}
{Before presenting the proposed analytical model, we first describe the 
methods for extracting the required inputs of the proposed model from {a memory access} trace 
file.
The proposed model requires the frequency of \emph{sequences} for estimating 
the behavior of HMAs.}
{A \emph{sequence} is defined as a set of accesses between two 
consecutive accesses to a data page.}
{As an example, consider an} {arbitrary sequence of accesses to the 
data pages in Fig. \ref{fig:profiling}.
The corresponding \emph{sequence} for the second access to page ``A" is} \textless C, B, B, D, E, B, D\textgreater\space.
{It contains the accesses between two consecutive accesses to page ``A".}
{In this access sequence, page ``A" is called the} \emph{target data 
page}.
{This notation is used throughout the paper.}

{The proposed analytical model also requires} DRAM hit probability in case 
of a hit {in the HMA} ($P_{hitDRAM|hit}$).
The required computations for calculating $P_{hitDRAM|hit}$ has O(n) time complexity.
In the remainder of this section, {we first present} the {algorithm for 
extracting sequences of requests} and {then} propose {a low-overhead 
mechanism for estimating} $P_{hitDRAM|hit}$.

\vspace{-.1cm}
\subsection{{Extracting Sequences}}
\label{sec:extract}
\vspace{-.1cm}
{The goal of sequence profiling is to find the number of total and unique 
accesses between two consecutive accesses to {each} data page.
For each sequence of accesses between two consecutive accesses to a data page, 
the sequence profiler counts 1) the number of accesses in the sequence (denoted as $r$), and 2) the number of unique data pages accessed 
in the sequence (denoted as $u$).
The sequence profiler computes {the} (r,u) pair for consecutive accesses to the 
same data page.
It does not store sequences or the corresponding (r,u) pairs for each data 
page.
Only the frequency of observing each pair is stored by the profiler.}
{The $u$ parameter is similar to the \emph{reuse distance} {metric}, which is employed in caching algorithms \cite{ahmadian:eci}.
As an example, there are a total of seven accesses between two 
accesses to page ``A" 
belonging to four different data pages in Fig. \ref{fig:profiling}.}
Therefore, the second access to page ``A" belongs to {the} (7, 4) pair.
The second access to page B belongs to (0, 0) since the second access is 
exactly after the first access.

\begin{figure}[!h]
	\vspace{-0.6cm}
	\centering
	\includegraphics[scale=0.66]{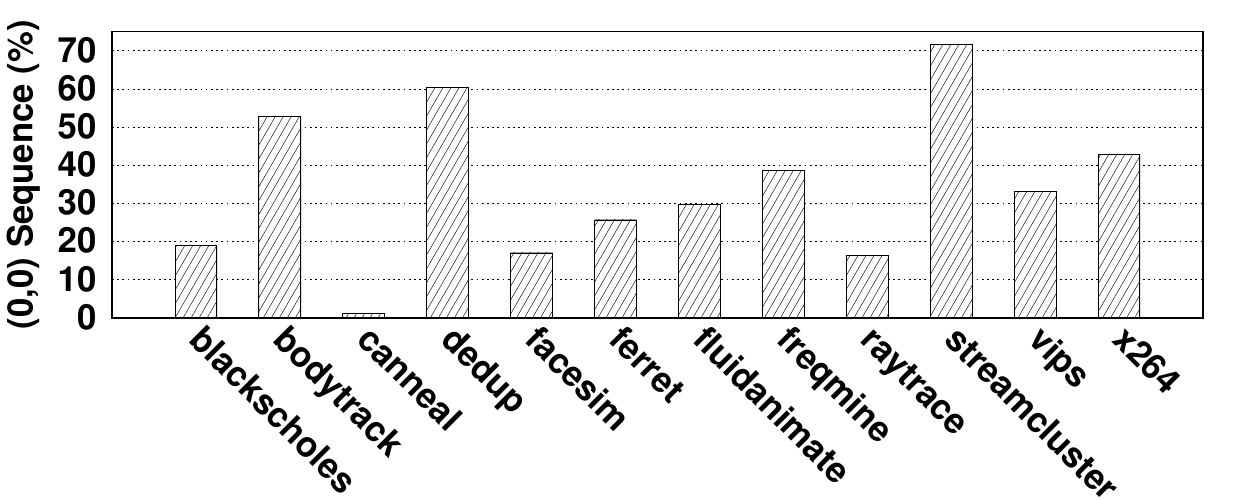}
	\caption{Percentage of consecutive accesses to a data page}
	\label{fig:zerounique}
	\vspace{-0.2cm}
\end{figure}

This profiling has $O(r\bar{u})$ complexity, where $\bar{u}$ is the average number of unique data pages in sequences.
{Algorithm \ref{alg:seqprofiler} shows how (r,u) pairs are extracted from 
the trace file.
The \emph{profiler} function is called for each request in the trace file and the $pairs$ variable holds the number of occurrences of each (r,u) pair.
For each request, if the address was accessed before, this request will be considered a consecutive {request} and the sequence for the previous access is calculated.
This data page is also added to the request sequences for all other data pages.}
Our analysis shows that unlike {the} \emph{Last Level Cache} (LLC), a significant 
percentage of accesses to virtual memory {belongs} to {the} (0,0) 
pair for 
traces obtained from {the} PARSEC benchmark
suite, {as shown in Fig. \ref{fig:zerounique}.}
This is {due to} the {coarse granularity} of data pages in virtual memory 
(4KB/8KB), {as opposed} to the {fine granularity} of {cache blocks in} 
LLC {(32/64B)}.

\newlength{\oldtextfloatsep}\setlength{\oldtextfloatsep}{\textfloatsep}
\setlength{\textfloatsep}{0pt}
\begin{algorithm}[!h]
	\caption{Sequence Profiler Algorithm}\label{alg:seqprofiler}
	\scriptsize
	\begin{algorithmic}[1]
		\State map $\gets$ Map()
		\State pairs $\gets$ Map()
		\Procedure{profiler}{Request}
		\If{Request.address in map}
		\State sequence $\gets$ map[Request.address]
		\State pairs[len(sequence), len(unique(sequence))] += 1
		\EndIf
		\State map[Request.address] $\gets$ new Array()
		\For{seq in sequences}
		\State seq.append(Request.address)
		\EndFor
		\EndProcedure
	\end{algorithmic}
\end{algorithm}

\setlength{\textfloatsep}{\oldtextfloatsep}


\vspace{-.3cm}
\subsection{DRAM Hit Estimation {Model}}
\vspace{-.1cm}
In addition to {the frequency of pairs}, the proposed analytical model 
requires hit probability in DRAM in case of a hit access in hybrid memory 
$(P_{hitDRAM|hit})$.
This probability enables the proposed model to stay memory-less and still 
accurately predict the hit ratio {of the HMA}.
Calculating {an} accurate value of $P_{hitDRAM|hit}$ requires running {different workloads} {on the} HMA and extracting the hit ratio of DRAM.
This {goes against} the goal of using {an} analytical model, which {requires} predicting the hit ratio \emph{without} running the workload on the target architecture.
{The profiler calculates} an estimation of $P_{hitDRAM|hit}$.
Previous analytical models that can predict the position of a data page in a queue after a certain number of accesses can be used to determine $P_{hitDRAM|hit}$.
Such models, however, increase the {computational} complexity from an $n^{th}$ degree polynomial to {an} ${(n+m)}^{th}$ degree polynomial, where $n$ and $m$ are maximum number of accesses in a sequence and maximum number of accesses to any data page, respectively.
{To address such complexity}, we propose a formula for predicting $P_{hitDRAM|hit}$ with {low complexity and less overhead}.
To predict $P_{hitDRAM|hit}$, {we consider} three scenarios: a) all accesses to NVM result in a migration, called FREE-MIG, b) no migration is allowed from NVM to DRAM, called NO-MIG, and c) there is a probability for {an access} to NVM to result in a migration from NVM to DRAM, called MIG-PROB. 
{We describe these} three scenarios next.

\begin{figure*}[t]
	\centering
	\subfloat[Allowing all migrations (FREE-MIG)]{\includegraphics[width=.33\textwidth]{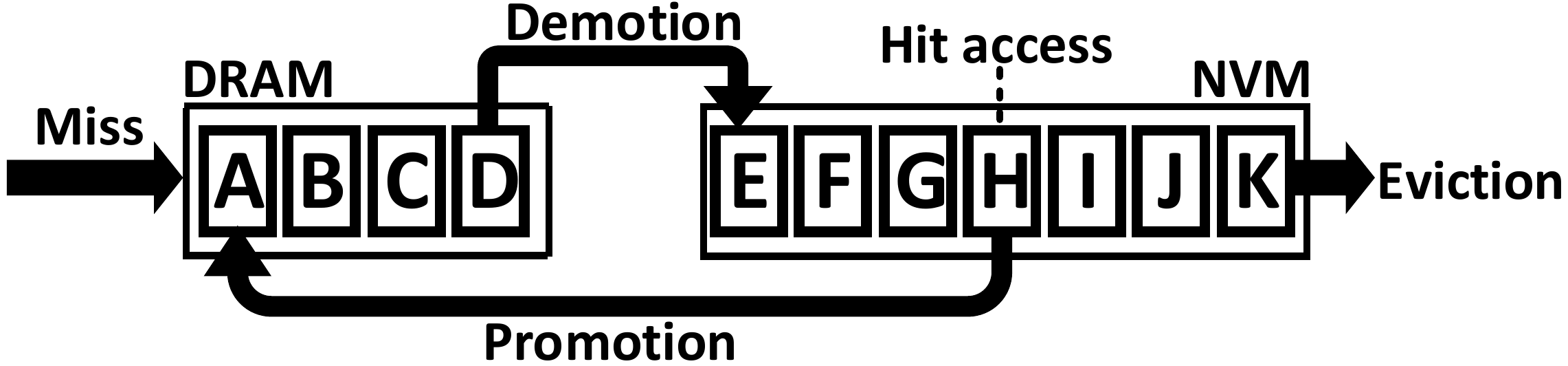}%
		\label{fig:allmig}}
	\hfill
	\subfloat[Preventing all migrations (NO-MIG)]{\includegraphics[width=.33\textwidth]{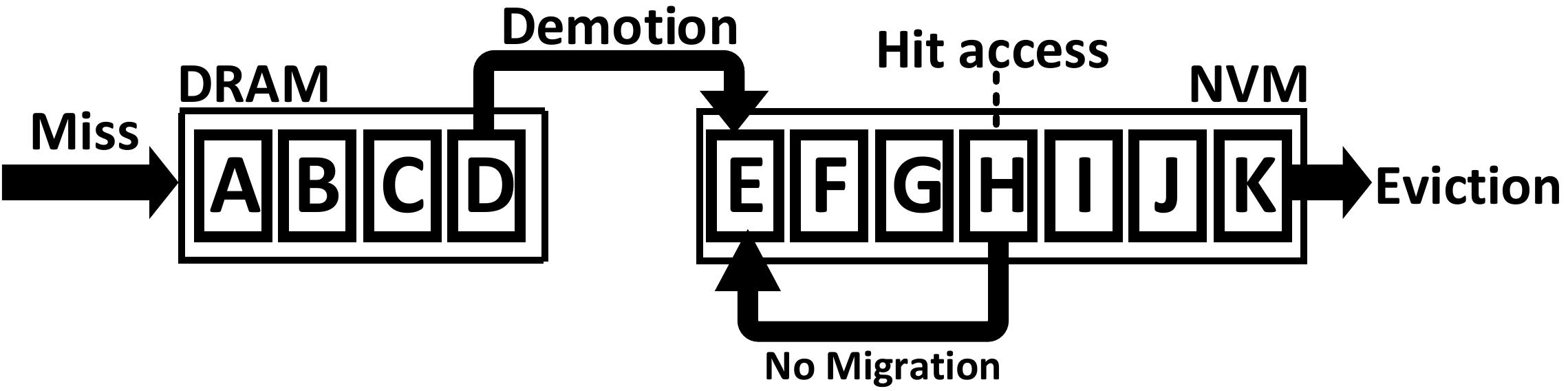}%
		\label{fig:nomig}}
	\hfill
	\subfloat[Partially preventing migrations (MIG-PROB)]{\includegraphics[width=.33\textwidth]{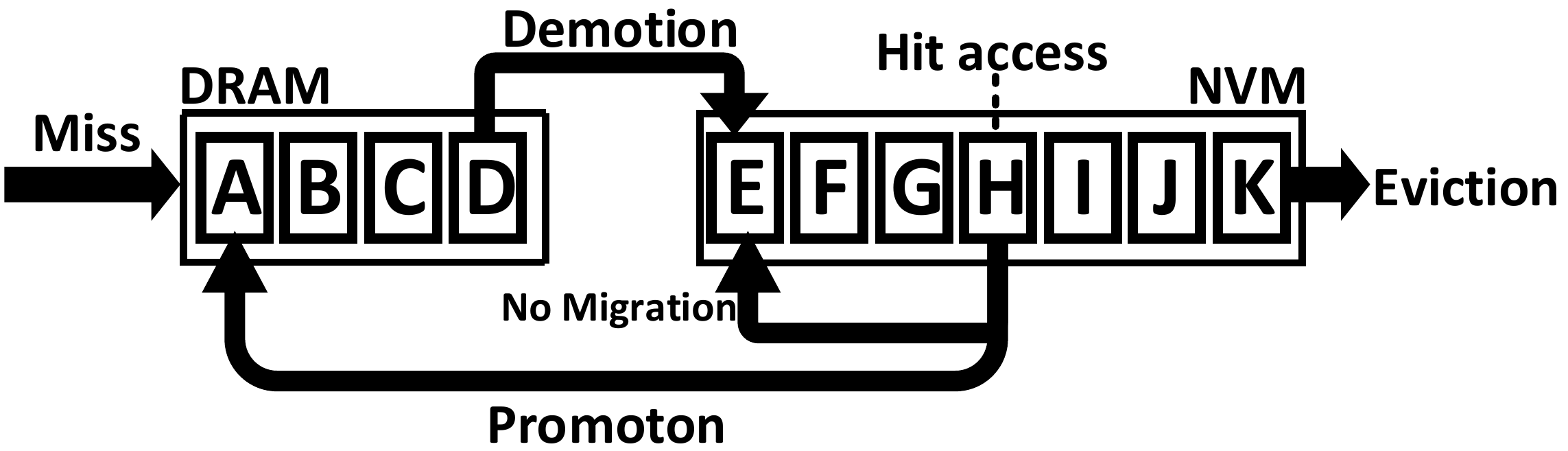}%
		\label{fig:withmig}}
	\vspace{-0.1cm}
	\caption{Various policies {for} NVM to DRAM {page} migration}
	\label{fig:migpolicies}
	\vspace{-0.5cm}
\end{figure*}

\vspace{-.2cm}
\subsubsection{FREE-MIG}
\vspace{-.1cm}
The number of unique accesses in a sequence determines the position of the 
target data page in DRAM/NVM when {the page} is hit again.
A sequence  of accesses starts from an access to the target data page and ends at the
second access to the same data page.
In the \emph{FREE-MIG} scenario (shown in Fig. \ref{fig:allmig}), {we use} a simple LRU queue.
{When} a sequence starts, the target data page is at the head of the LRU queue.
The position of the target data page changes \emph{only} when a new data page (either present in the memory and after the target data page in the queue or a miss access) is accessed.
Thus, the number of unique accesses can determine the position of the target data page when the sequence ends.
If the number of unique data pages is less than the DRAM size, the target data page will hit in DRAM.

To estimate the DRAM hit probability after extracting sequences, the profiler groups sequences by their number of unique accesses ($u$) and creates a probability array ($probArr(i)$) for all values between 0 and {the} maximum value of $u$.
The index of this array is the number of unique accesses and its value stores {the} percentage of sequences with the {corresponding} number of unique accesses.
Note that Prob[i] represents only the number of unique 
accesses, 
regardless of whether the page resides in the HMA.
{A value of 0.3} for $probArr(1)$ means that by randomly selecting an access from {a} trace file and computing its sequence profile, {it has 30\% probability to have only one unique access} (similar to the third access to page D in Fig. \ref{fig:profiling}).
To calculate $P_{hitDRAM}$, we first assume that DRAM and NVM are managed by a single LRU queue, where pages in DRAM occupy the first $DRAMSIZE$ positions of the queue (scenario \emph{a}).
Therefore, each hit in NVM results in migrating the data page to DRAM and demoting the last data page in {the} DRAM queue to NVM, similar to Fig. \ref{fig:allmig}.
In this scenario, $P_{hitDRAM}$ {is} calculated based on $probArr$ using 
{Equation \ref{eq:phitdramsinglelru}, Equation 
\ref{eq:phitdramsinglelru2}, and Equation \ref{eq:phitdramsinglelru3}.}
{$P_{Dbasic}$ and $P_{Nbasic}$ denote the probability of {a} hit in DRAM and 
NVM, respectively, when memory is managed by a simple LRU} {replacement 
mechanism.}
{Table \ref{tbl:notation} also summarizes the key notational elements used in equations throughout the paper.}

\begin{figure}[h]
	\footnotesize
	\vspace{-.4cm}
	\begin{align}
	\label{eq:phitdramsinglelru}
	P_{Dbasic} &= \sum_{i := 0}^{DRAMSIZE} ProbArr[i] \\
	\label{eq:phitdramsinglelru2}
	P_{Nbasic} &= \sum_{i := DRAMSIZE+1}^{TOTALSIZE} ProbArr[i] \\
	\label{eq:phitdramsinglelru3}
	P_{Missbasic} &= 1-(P_{Dbasic}+P_{Nbasic})
	\end{align}
	\vspace{-.6cm}
\end{figure}

	\begin{table}[!h]
	\hspace{-3cm}
	\centering
	\caption{Description of Notations}
	\label{tbl:notation}
	\hskip-0.9cm\begin{tabular}{|l|l|}
		\hline
		\textbf{Notation} & \textbf{Description} \\ \hline\hline
		$unique$            & Probability that a request is the first access \\ &to a data page. \\ \hline
		$eviction$            & Eviction probability of the data page. \\ \hline
		$ProbArr[index]$         & Array containing percentage of requests with \\ &$index$ number of unique accesses. \\ \hline
		$P_{Dbasic}$         & Probability that an access hits in DRAM \\ &when normal LRU is employed.\\ \hline
		$P_{Nbasic}$         & Probability that an access hits in NVM\\& when normal LRU is employed.\\ \hline
		$P_{Missbasic}$         & Probability that an access misses from HMA \\&when  normal LRU is employed.\\ \hline
		$P_{Dnomig}$         & Probability that an access hits in DRAM \\ &when no migration is allowed.\\ \hline
		$P_{Nnomig}$         & Probability that an access hits in NVM \\&when no migration is allowed.\\ \hline		
		$P_{D}$         & Probability that an access hits in DRAM\\& when migrations are allowed.\\ \hline
		$P (unique)$	& Probability that the next access is a unique\\& access.\\ \hline
		$P(hitDRAM|hit)$	& Probability that an access that hits in HMA \\&hits in 	DRAM.\\ \hline			
		$P_{DRAMevictionsource}$	& Probability that a data page is moved to the \\ & head of the NVM queue due to eviction from \\& DRAM.\\
		\hline
		$P_{NVMhitsource}$	& Probability that a data page is moved to the\\ & head of the NVM queue due to a hit in NVM.\\
		\hline
		$Rlat_{x}$ & Read latency for device x.\\
		\hline
		$Wlat_{x}$ & Write latency for device x.\\
		\hline
		$R_{x}$ & Number of read accesses to device x.\\
		\hline
		$W_{x}$ & Number of write accesses to device x.\\
		\hline
		$Miss$ & Number of miss accesses.\\
		\hline
		$Mig_{toNVM}$ & Number of migrations from DRAM to NVM.\\
		\hline
		$DRAMSIZE$ & DRAM size.\\
		\hline
		$TOTALSIZE$ & Total size of DRAM and NVM.\\
		\hline
		$P_{mig}$ & Probability that an access to a data page in \\ & NVM results in a migration.\\
		\hline
		$Pagefactor$ & Number of writes required for migrating a\\ & page.\\
		\hline
	\end{tabular}
	\vspace{-.5cm}
\end{table}

\vspace{-.2cm}
\subsubsection{NO-MIG}
\vspace{-.1cm}
In the second scenario (shown in Fig. \ref{fig:nomig}), no migration is allowed from NVM to DRAM.
{We use a} demotion arc in the \emph{NO-MIG} scenario since a new data 
page will be added to DRAM only if the access misses in 
the HMA (i.e., both DRAM and NVM) and thus, one of the data pages currently residing in 
DRAM should be evicted from DRAM to NVM to free up space for {the newly-accessed}
data page.
By not allowing migrations, the NVM hit ratio will increase since data pages {that}
hit in the NVM will stay there and such recently 
accessed data pages {likely} have {a} high probability of {being} {re-referenced} in 
the near future. 
Fig. \ref{fig:migdist} compares $probArr$ values for scenarios \emph{(a)} and \emph{(b)} for {the} PARSEC benchmark suite.
The y-axis shows the values in $probArr$ array for all array indexes.
{$P(DRAM_{mig})$ and $P(DRAM_{noMig})$ denote DRAM hit probability in \emph{FREE-MIG} and \emph{NO-MIG}, respectively.}
We observe that the increase in {the} NVM hit ratio causes all DRAM data pages to 
be accessed less frequently ({leading to} almost 20\% reduction in the number of 
DRAM accesses {compared to \emph{FREE-MIG}}).
{As Fig. \ref{fig:migdist} shows,} when no migration is allowed, the \emph{Most Recently Used} (MRU) position in DRAM {($P(DRAM_{noMig}[0])$)} still receives more than three times {the} accesses {(on 
average)} than the MRU position in NVM {($P(NVM_{noMig}[0])$)} since newly-accessed data 
pages are moved to DRAM and such data pages are accessed frequently.

{An} accessed data {page} in NVM {has} {a hit probability of} $P_{Dbasic} + P_{Nbasic}$  {for its} upcoming accesses in {\emph{FREE-MIG}} (as defined in Equation \ref{eq:phitdramsinglelru}).
Since the accessed data pages in NVM will remain in NVM in {\emph{NO-MIG}}, 
{the} NVM hit ratio will be higher than {that} in {\emph{FREE-MIG}}.
A normal eviction from DRAM in {\emph{NO-MIG}} will have the same hit probability as {in} \emph{FREE-MIG}.
The NVM hit ratio can be calculated using Equation \ref{eq:phitdramnomigbase}, where
$P_{DRAMevictionsource}$ and $P_{NVMhitsource}$ are probabilities of each source of moving a data page to the head of {the} NVM LRU queue.
The eviction probability from DRAM ($P_{DRAMeviction}$) is the same as {the} HMA miss probability since each miss will result in moving a data page to DRAM and consequently will result in evicting a data page to NVM.
{NVM} hit ratio is calculated via $P(hit|at\_queue\_head)$ where $at\_queue\_head$ is equal to $P_{NVMhit}+P_{miss}$ in this scenario.
Therefore, the NVM hit ratio in {\emph{NO-MIG}} can be calculated using Equation \ref{eq:phitdramnomig}.
{The description of notations used in Equation \ref{eq:phitdramnomig} is reported in Table \ref{tbl:notation}.}
Our analysis shows that {the} migration policy has negligible effect on the total memory hit ratio, which enables us to easily calculate {the} DRAM hit ratio based on the miss probability of {\emph{FREE-MIG}} and NVM hit ratio of {\emph{NO-MIG}} (Equation \ref{eq:phitdramnomig}).

\begin{figure}[!h]
	\footnotesize
	\vspace{-.5cm}
	\begin{align}
	\label{eq:phitdramnomigbase}
	P_{Nnomig} &= P_{DRAMevictionsource} * P_{Nbasic} + P_{NVMhitsource} *\nonumber \\
	& (P_{Dbasic} + P_{Nbasic})
	\end{align}
	\vspace{-0.2cm}
	\begin{align}
	\label{eq:phitdramnomig}
	P_{Nnomig}&= P_{Missbasic}/(P_{Missbasic} + P_{Nbasic}) * P_{Nbasic} \nonumber\\
	&+ P_{Nbasic}/(P_{Missbasic} + P_{Nbasic}) *  (P_{Dbasic} + P_{Nbasic})\nonumber\\
	P_{Dnomig} &= 1 - P_{Nnomig} - P_{Missbasic}
	\end{align}
	\vspace{-.6cm}
\end{figure}

\begin{figure*}[t]
	\centering
	\subfloat[blackscholes]{\begin{overpic}[width=0.242\textwidth,tics=10]{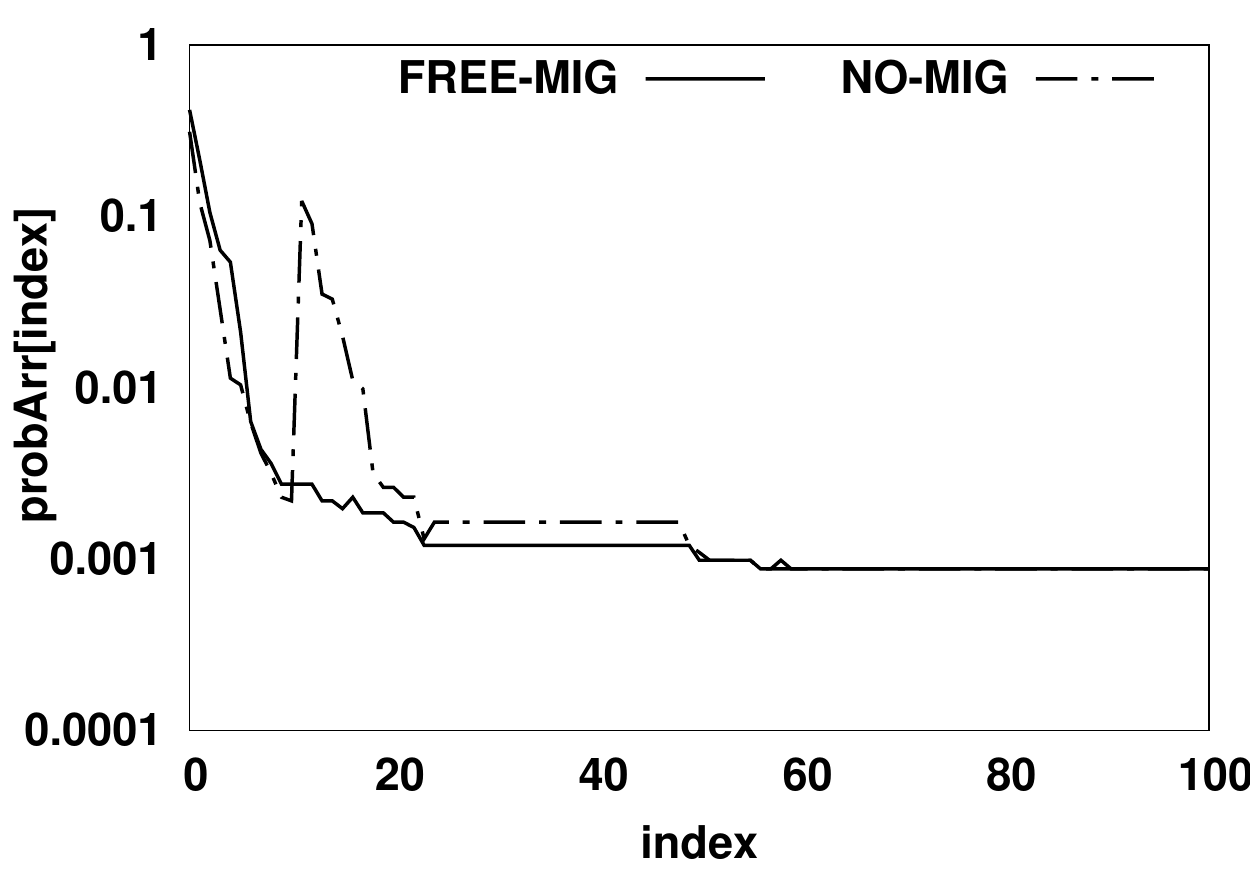}
			\put (27,55) {\scalebox{.45}{ $ \mathbf{P\boldsymbol{(}D\!R\!A\!M_{mig}\boldsymbol{)} \!\boldsymbol{=} \!1.83\! \times \!P\boldsymbol{(}D\!R\!A\!M_{noMig}\boldsymbol{)}}  $ }}
			\put (27,45) {\scalebox{.42}{ $\mathbf{P\boldsymbol{(}D\!R\!A\!M_{noMig}\boldsymbol{[}0\boldsymbol{]}\boldsymbol{)} \!\boldsymbol{=} \!2.06\!\! \times \!\!P\boldsymbol{(}N\!V\!M_{noMig}\boldsymbol{[}0\boldsymbol{]}}\boldsymbol{)}$}}
		\end{overpic}
		\vspace{-.4cm}
		\label{fig:accessdistblackscholes}}
	\hfill
	\subfloat[bodytrack]{\begin{overpic}[width=0.242\textwidth,tics=10]{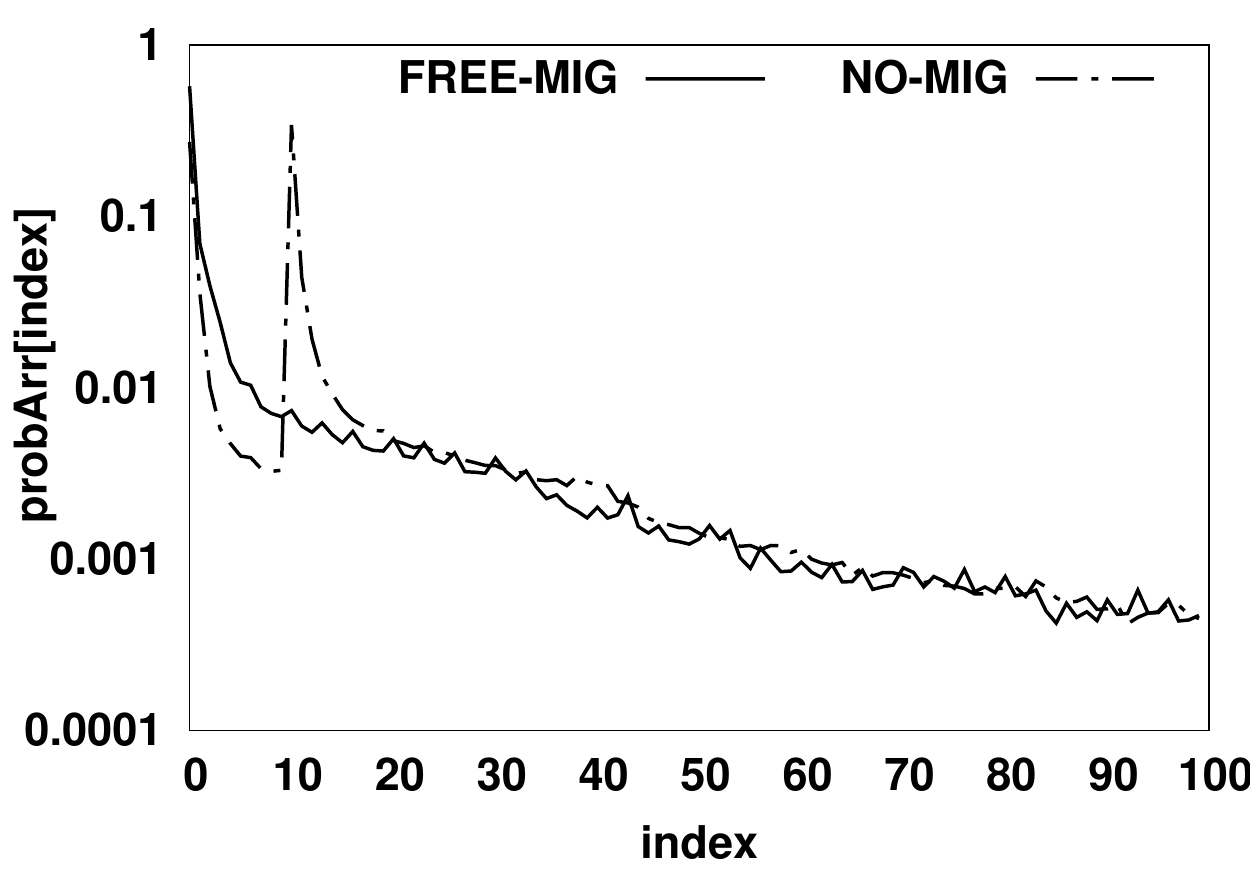}
			\put (27,55) {\scalebox{.45}{ $ \mathbf{P\boldsymbol{(}D\!R\!A\!M_{mig}\boldsymbol{)} \!\boldsymbol{=} \!2.21\! \times \!P\boldsymbol{(}D\!R\!A\!M_{noMig}\boldsymbol{)}}  $ }}
			\put (27,45) {\scalebox{.42}{ $\mathbf{P\boldsymbol{(}D\!R\!A\!M_{noMig}\boldsymbol{[}0\boldsymbol{]}\boldsymbol{)} \!\boldsymbol{=} \!1.68\!\! \times \!\!P\boldsymbol{(}N\!V\!M_{noMig}\boldsymbol{[}0\boldsymbol{]}}\boldsymbol{)}$}}
		\end{overpic}
		\vspace{-.4cm}
		\label{fig:accessdistbodytrack}}
	\hfill
	\subfloat[canneal]{\begin{overpic}[width=0.242\textwidth,tics=10]{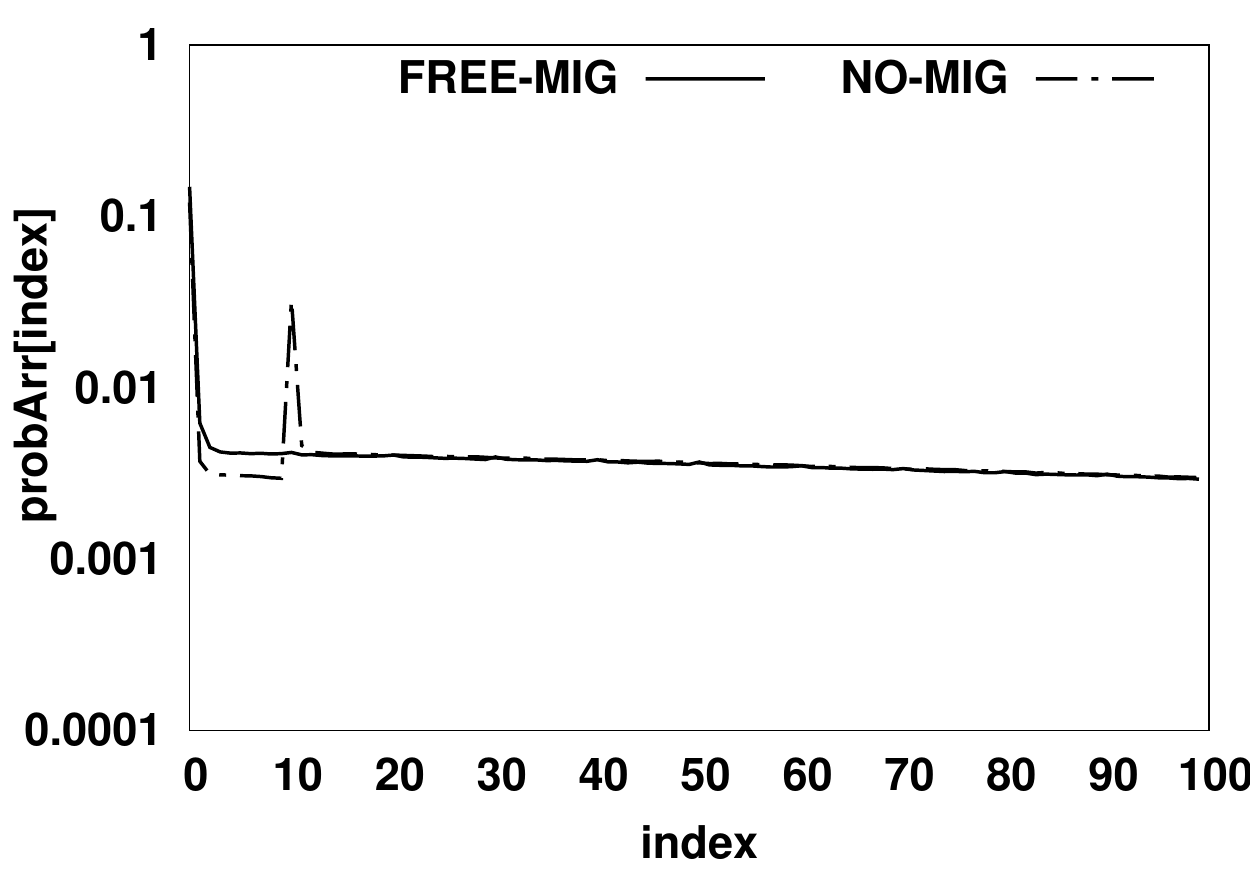}
			\put (27,55) {\scalebox{.45}{ $\mathbf{P\boldsymbol{(}D\!R\!A\!M_{mig}\boldsymbol{)} \!\boldsymbol{=} \!1.26\! \times \!P\boldsymbol{(}D\!R\!A\!M_{noMig}\boldsymbol{)}} $}}
			\put (27,45) {\scalebox{.42}{ $\mathbf{P\boldsymbol{(}D\!R\!A\!M_{noMig}\boldsymbol{[}0\boldsymbol{]}\boldsymbol{)} \!\boldsymbol{=} \!4.57\!\! \times \!\!P\boldsymbol{(}N\!V\!M_{noMig}\boldsymbol{[}0\boldsymbol{]}}\boldsymbol{)}$}}
		\end{overpic}
		\vspace{-.4cm}
		\label{fig:accessdistcanneal}}
	\hfill
	\subfloat[dedup]{\begin{overpic}[width=0.242\textwidth,tics=10]{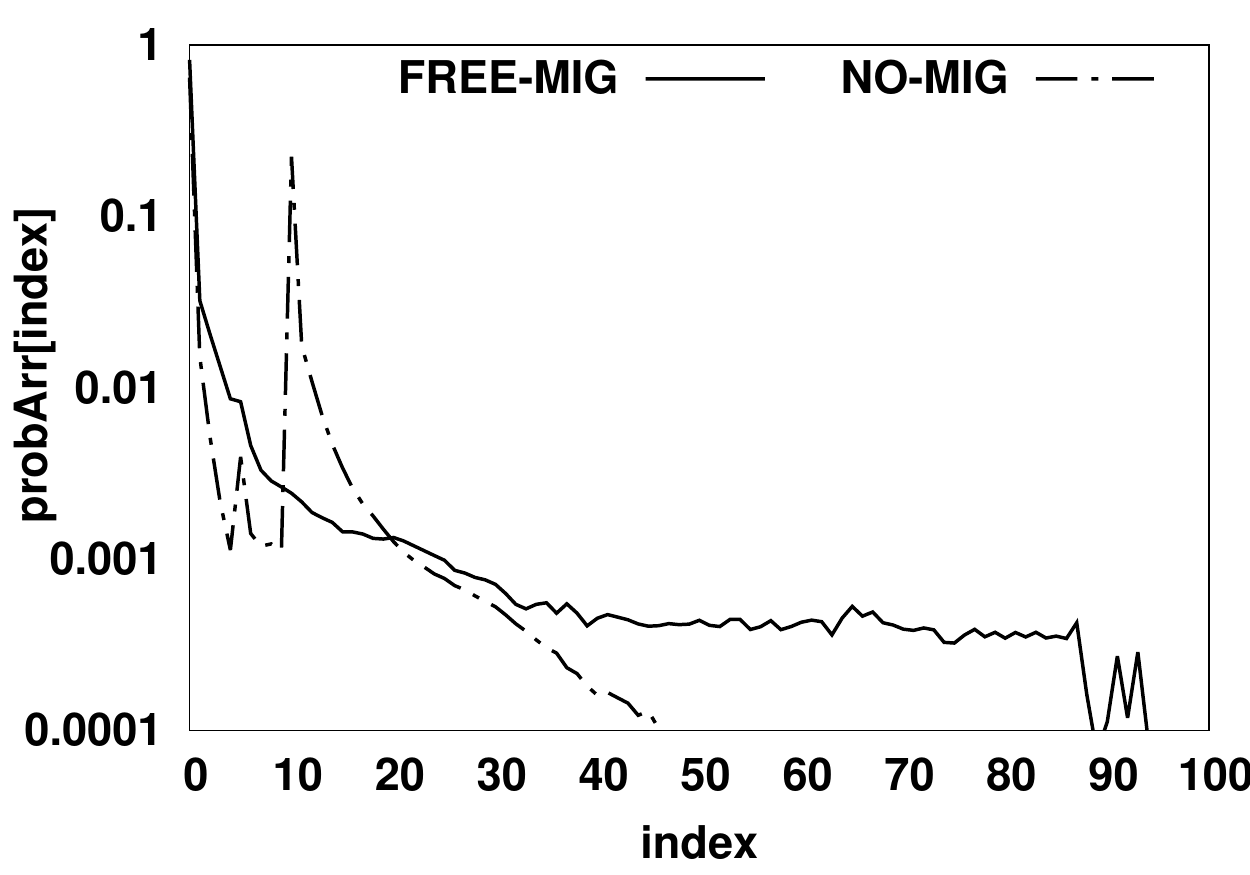}
			\put (27,55) {\scalebox{.45}{ $ \mathbf{P\boldsymbol{(}D\!R\!A\!M_{mig}\boldsymbol{)} \!\boldsymbol{=} \!1.34\! \times \!P\boldsymbol{(}D\!R\!A\!M_{noMig}\boldsymbol{)}}  $ }}
			\put (27,45) {\scalebox{.42}{ $\mathbf{P\boldsymbol{(}D\!R\!A\!M_{noMig}\boldsymbol{[}0\boldsymbol{]}\boldsymbol{)} \!\boldsymbol{=} \!3.66\!\! \times \!\!P\boldsymbol{(}N\!V\!M_{noMig}\boldsymbol{[}0\boldsymbol{]}}\boldsymbol{)}$}}
		\end{overpic}
		\vspace{-.4cm}
		\label{fig:accessdistdedup}}
	\hfill
	\vspace{-.4cm}
	\subfloat[facesim]{\begin{overpic}[width=0.242\textwidth,tics=10]{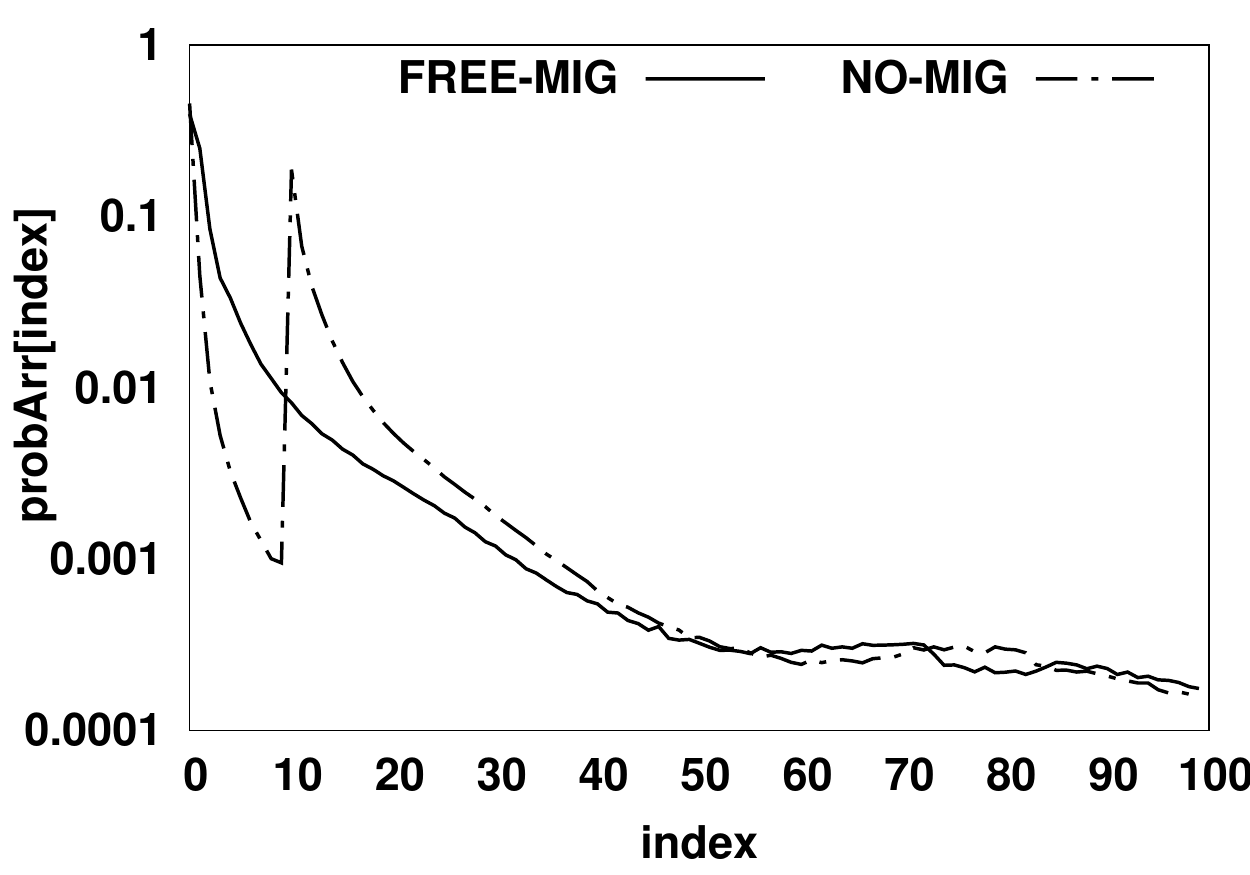}
			\put (27,55) {\scalebox{.45}{ $\mathbf{P\boldsymbol{(}D\!R\!A\!M_{mig}\boldsymbol{)} \!\boldsymbol{=} \!1.67\! \times \!P\boldsymbol{(}D\!R\!A\!M_{noMig}\boldsymbol{)}} $}}
			\put (27,45) {\scalebox{.42}{ $\mathbf{P\boldsymbol{(}D\!R\!A\!M_{noMig}\boldsymbol{[}0\boldsymbol{]}\boldsymbol{)} \!\boldsymbol{=} \!2.1\!\! \times \!\!P\boldsymbol{(}N\!V\!M_{noMig}\boldsymbol{[}0\boldsymbol{]}}\boldsymbol{)}$}}
		\end{overpic}
		\vspace{-.4cm}
		\label{fig:accessdistfacesim}}
	\hfill
	\subfloat[ferret]{\begin{overpic}[width=0.242\textwidth,tics=10]{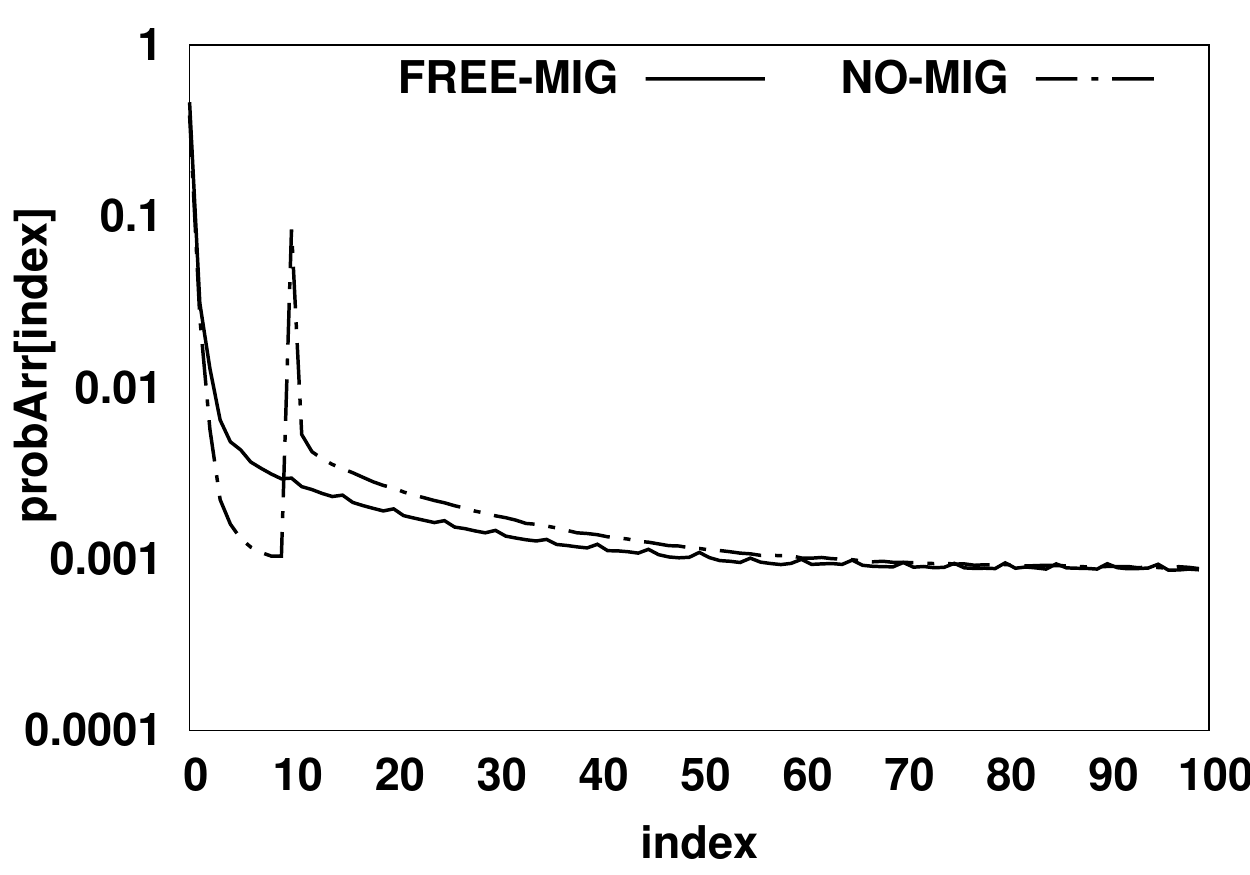}
			\put (27,55) {\scalebox{.45}{ $\mathbf{P\boldsymbol{(}D\!R\!A\!M_{mig}\boldsymbol{)} \!\boldsymbol{=} \!1.24\! \times \!P\boldsymbol{(}D\!R\!A\!M_{noMig}\boldsymbol{)}} $}}
			\put (27,45) {\scalebox{.42}{ $\mathbf{P\boldsymbol{(}D\!R\!A\!M_{noMig}\boldsymbol{[}0\boldsymbol{]}\boldsymbol{)} \!\boldsymbol{=} \!5.51\!\! \times \!\!P\boldsymbol{(}N\!V\!M_{noMig}\boldsymbol{[}0\boldsymbol{]}}\boldsymbol{)}$}}
		\end{overpic}
		\vspace{-.4cm}
		\label{fig:accessdistferret}}
	\hfill
	\subfloat[fluidanimate]{\begin{overpic}[width=0.242\textwidth,tics=10]{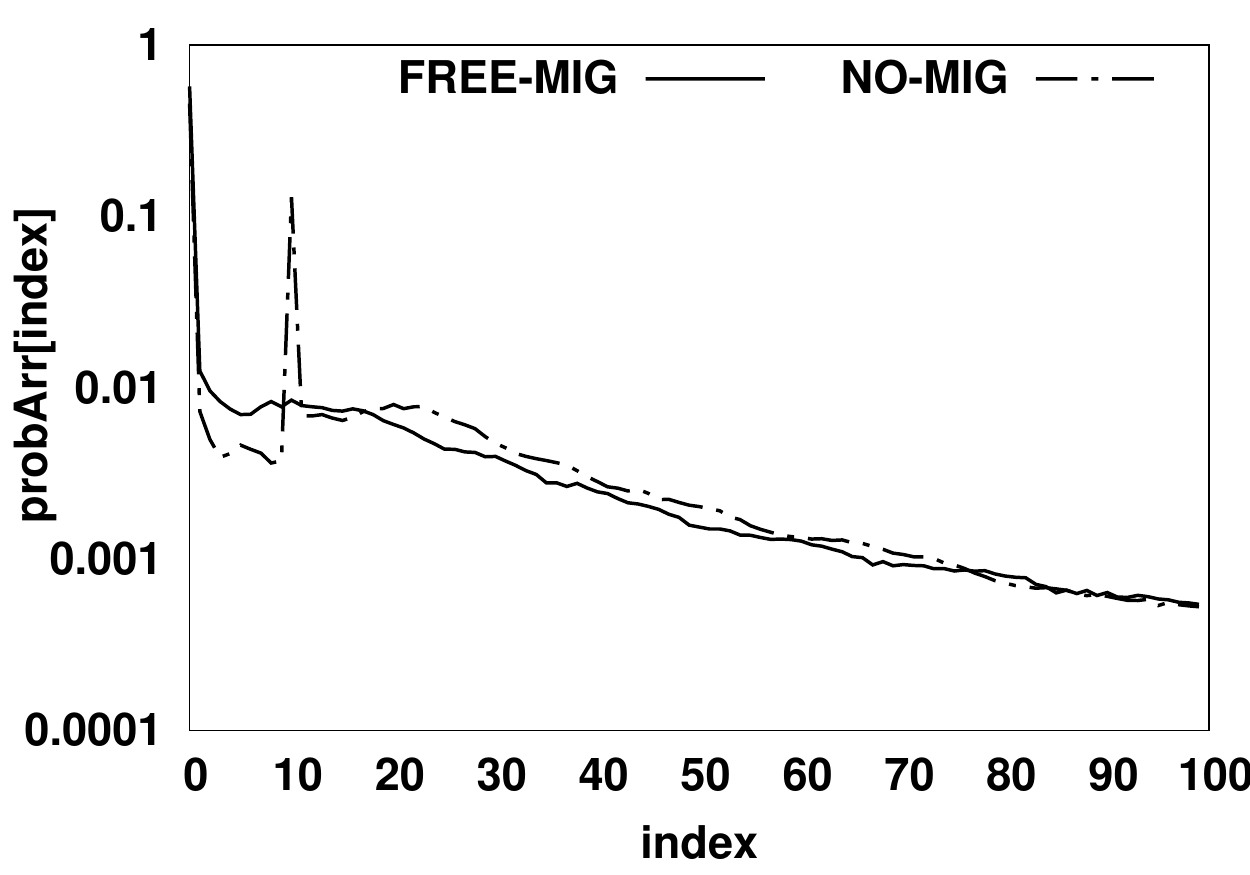}
			\put (27,55) {\scalebox{.45}{ $\mathbf{P\boldsymbol{(}D\!R\!A\!M_{mig}\boldsymbol{)} \!\boldsymbol{=} \!1.3\! \times \!P\boldsymbol{(}D\!R\!A\!M_{noMig}\boldsymbol{)}} $}}
			\put (27,45) {\scalebox{.42}{$\mathbf{P\boldsymbol{(}D\!R\!A\!M_{noMig}\boldsymbol{[}0\boldsymbol{]}\boldsymbol{)} \!\boldsymbol{=} \!4.43\!\! \times \!\!P\boldsymbol{(}N\!V\!M_{noMig}\boldsymbol{[}0\boldsymbol{]}}\boldsymbol{)}$}}
		\end{overpic}
		\vspace{-0.4cm}
		\label{fig:accessdistfluidanimate}}
	\hfill
	\subfloat[freqmine]{\begin{overpic}[width=0.242\textwidth,tics=10]{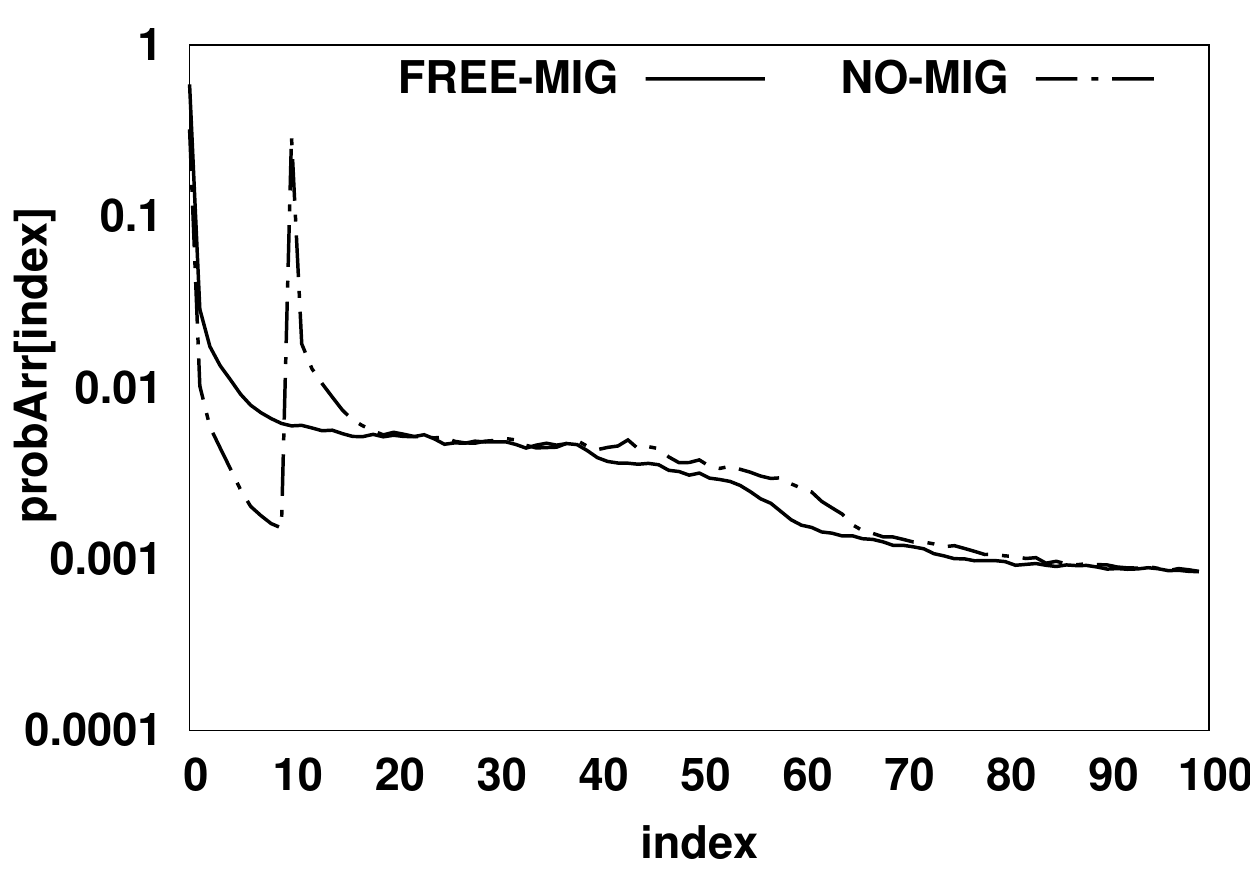}
			\put (27,55) {\scalebox{.45}{ $\mathbf{P\boldsymbol{(}D\!R\!A\!M_{mig}\boldsymbol{)} \!\boldsymbol{=} \!1.95\! \times \!P\boldsymbol{(}D\!R\!A\!M_{noMig}\boldsymbol{)}} $}}
			\put (27,45) {\scalebox{.42}{$\mathbf{P\boldsymbol{(}D\!R\!A\!M_{noMig}\boldsymbol{[}0\boldsymbol{]}\boldsymbol{)} \!\boldsymbol{=} \!2\!\! \times \!\!P\boldsymbol{(}N\!V\!M_{noMig}\boldsymbol{[}0\boldsymbol{]}}\boldsymbol{)}$}}
		\end{overpic}
		\vspace{-.4cm}
		\label{fig:accessdistfreqmine}}
	\hfill
	\vspace{-.4cm}
	\subfloat[raytrace]{\begin{overpic}[width=0.242\textwidth,tics=10]{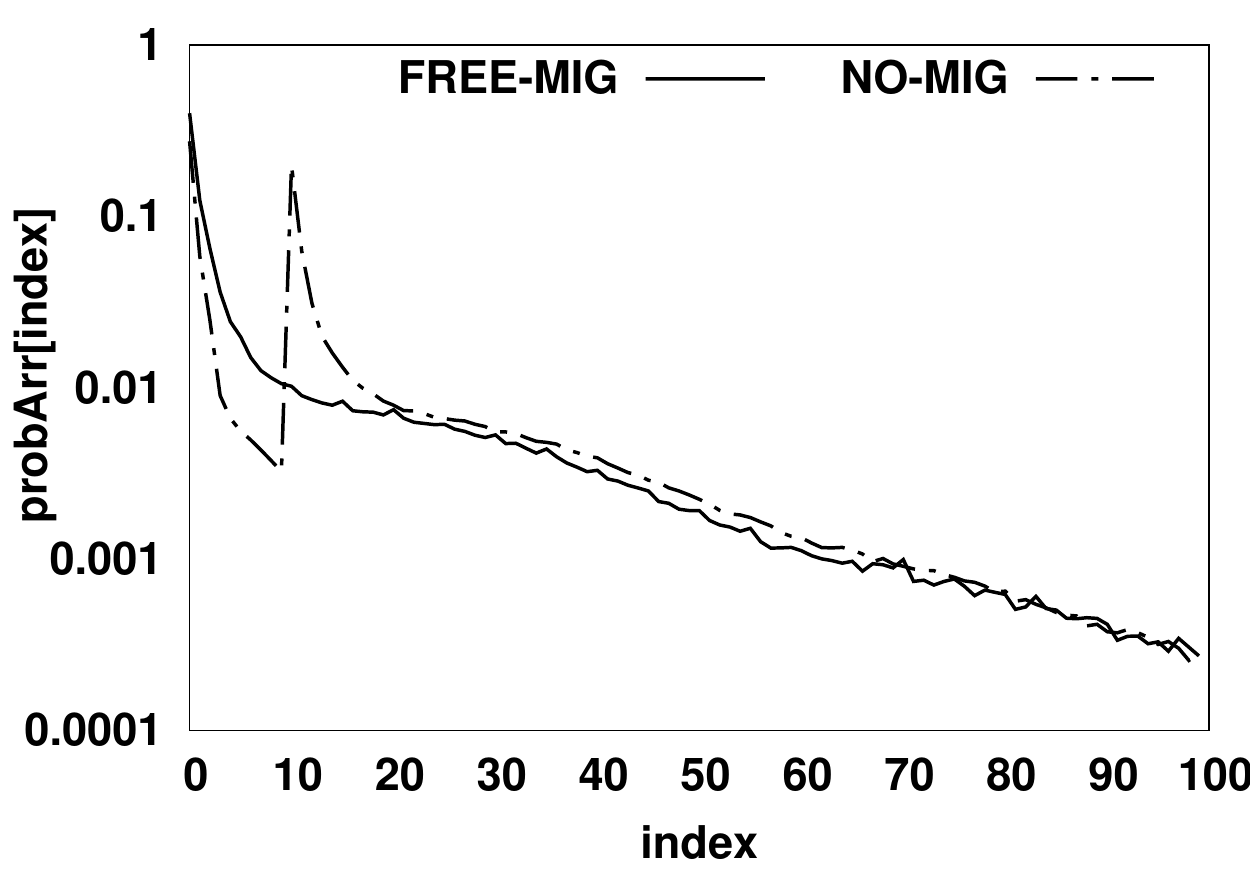}
			\put (27,55) {\scalebox{.45}{ $\mathbf{P\boldsymbol{(}D\!R\!A\!M_{mig}\boldsymbol{)} \!\boldsymbol{=} \!1.82\! \times \!P\boldsymbol{(}D\!R\!A\!M_{noMig}\boldsymbol{)}} $}}
			\put (27,45) {\scalebox{.42}{ $\mathbf{P\boldsymbol{(}D\!R\!A\!M_{noMig}\boldsymbol{[}0\boldsymbol{]}\boldsymbol{)} \!\boldsymbol{=} \!2.01\!\! \times \!\!P\boldsymbol{(}N\!V\!M_{noMig}\boldsymbol{[}0\boldsymbol{]}}\boldsymbol{)}$}}
		\end{overpic}
		\vspace{-.4cm}
		\label{fig:accessdistraytrace}}
	\hfill
	\subfloat[streamcluster]{\begin{overpic}[width=0.242\textwidth,tics=10]{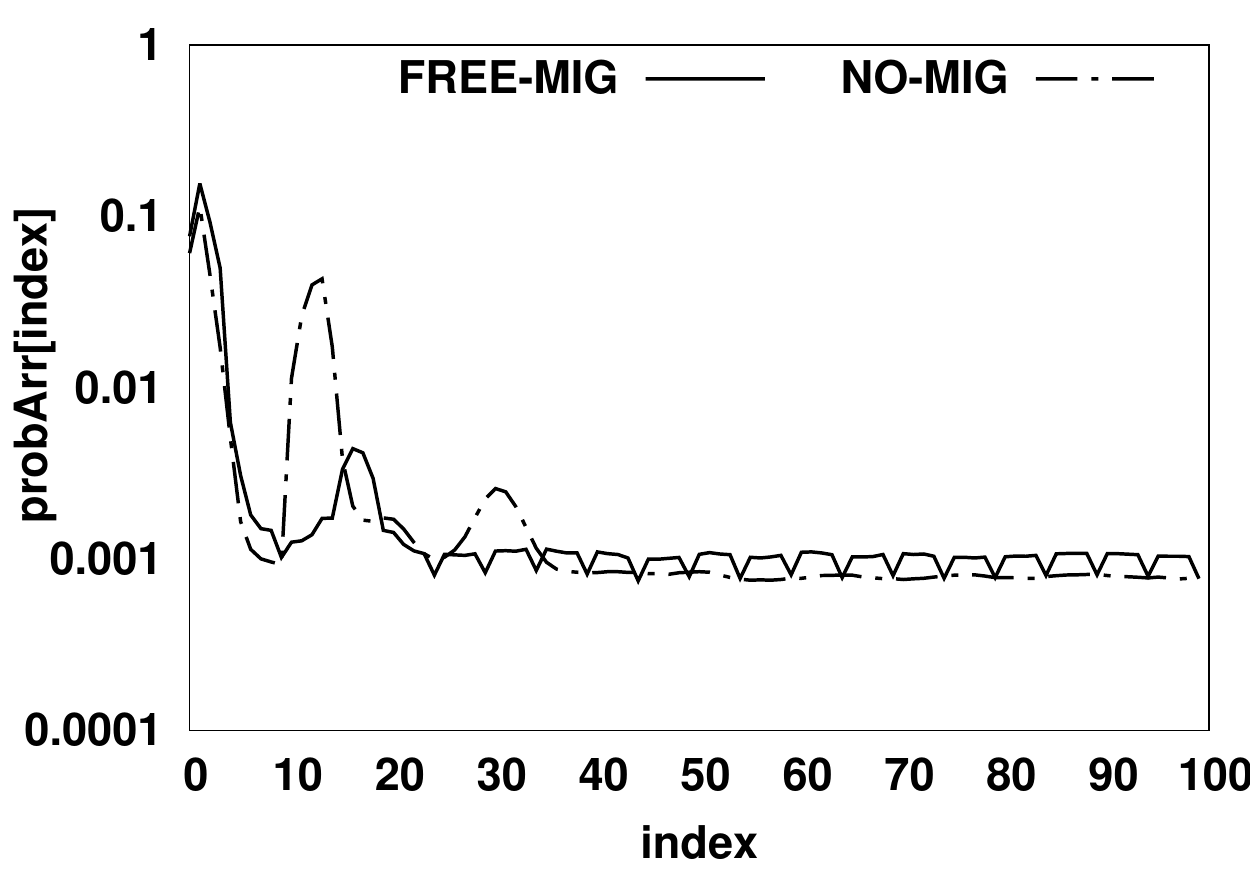}
			\put (27,55) {\scalebox{.45}{ $ \mathbf{P\boldsymbol{(}D\!R\!A\!M_{mig}\boldsymbol{)} \!\boldsymbol{=} \!1.56\! \times \!P\boldsymbol{(}D\!R\!A\!M_{noMig}\boldsymbol{)}}  $ }}
			\put (27,45) {\scalebox{.42}{ $\mathbf{P\boldsymbol{(}D\!R\!A\!M_{noMig}\boldsymbol{[}0\boldsymbol{]}\boldsymbol{)} \!\boldsymbol{=} \!6.74\!\! \times \!\!P\boldsymbol{(}N\!V\!M_{noMig}\boldsymbol{[}0\boldsymbol{]}}\boldsymbol{)}$}}
		\end{overpic}
		\vspace{-.4cm}
		\label{fig:accessdiststreamcluster}}
	\hfill
	\subfloat[vips]{\begin{overpic}[width=0.242\textwidth,tics=10]{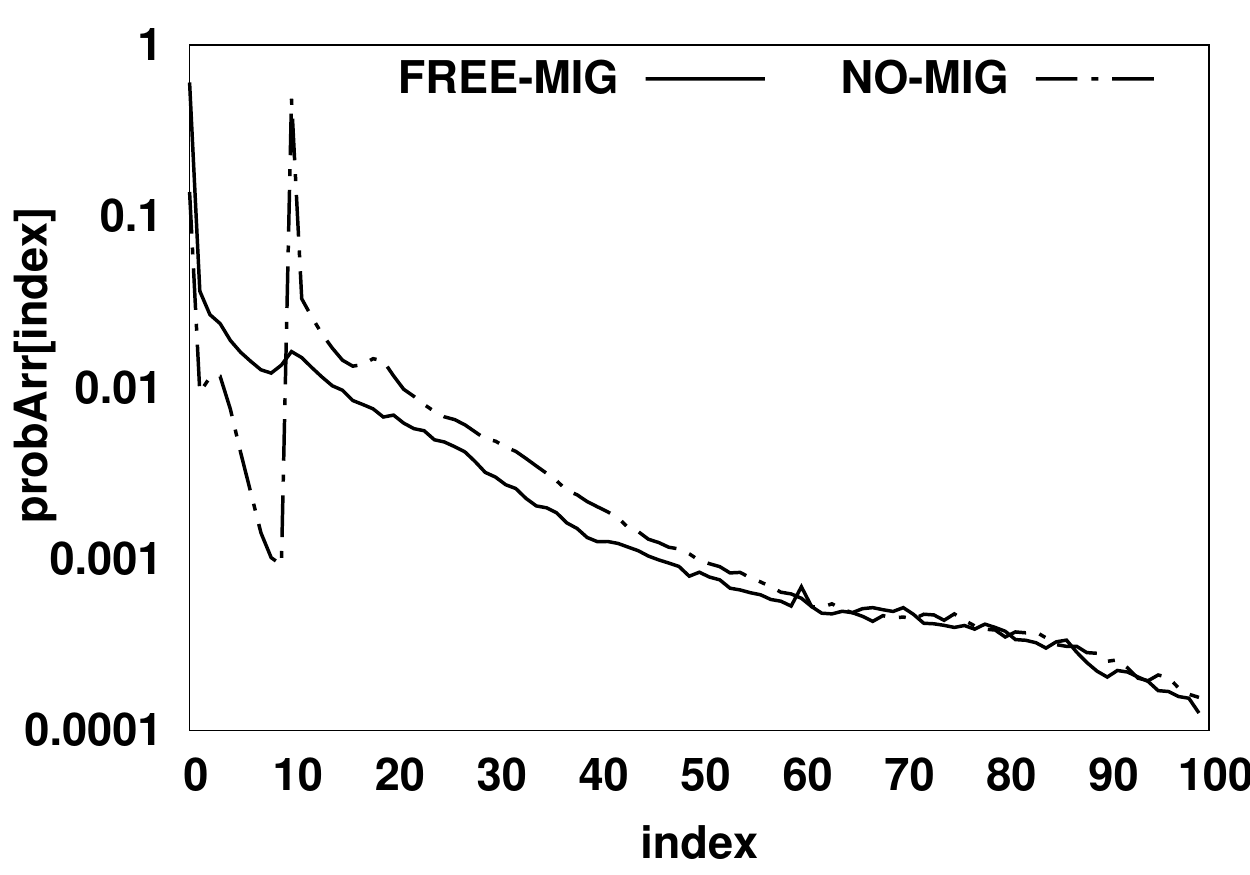}
			\put (27,55) {\scalebox{.45}{ $ \mathbf{P\boldsymbol{(}D\!R\!A\!M_{mig}\boldsymbol{)} \!\boldsymbol{=} \!4.13\! \times \!P\boldsymbol{(}D\!R\!A\!M_{noMig}\boldsymbol{)}}  $ }}
			\put (27,45) {\scalebox{.42}{ $\mathbf{P\boldsymbol{(}D\!R\!A\!M_{noMig}\boldsymbol{[}0\boldsymbol{]}\boldsymbol{)} \!\boldsymbol{=} \!1.24\!\! \times \!\!P\boldsymbol{(}N\!V\!M_{noMig}\boldsymbol{[}0\boldsymbol{]}}\boldsymbol{)}$}}
		\end{overpic}
		\vspace{-.4cm}
		\label{fig:accessdistvips}}
	\hfill
	\subfloat[x264]{\begin{overpic}[width=0.242\textwidth,tics=10]{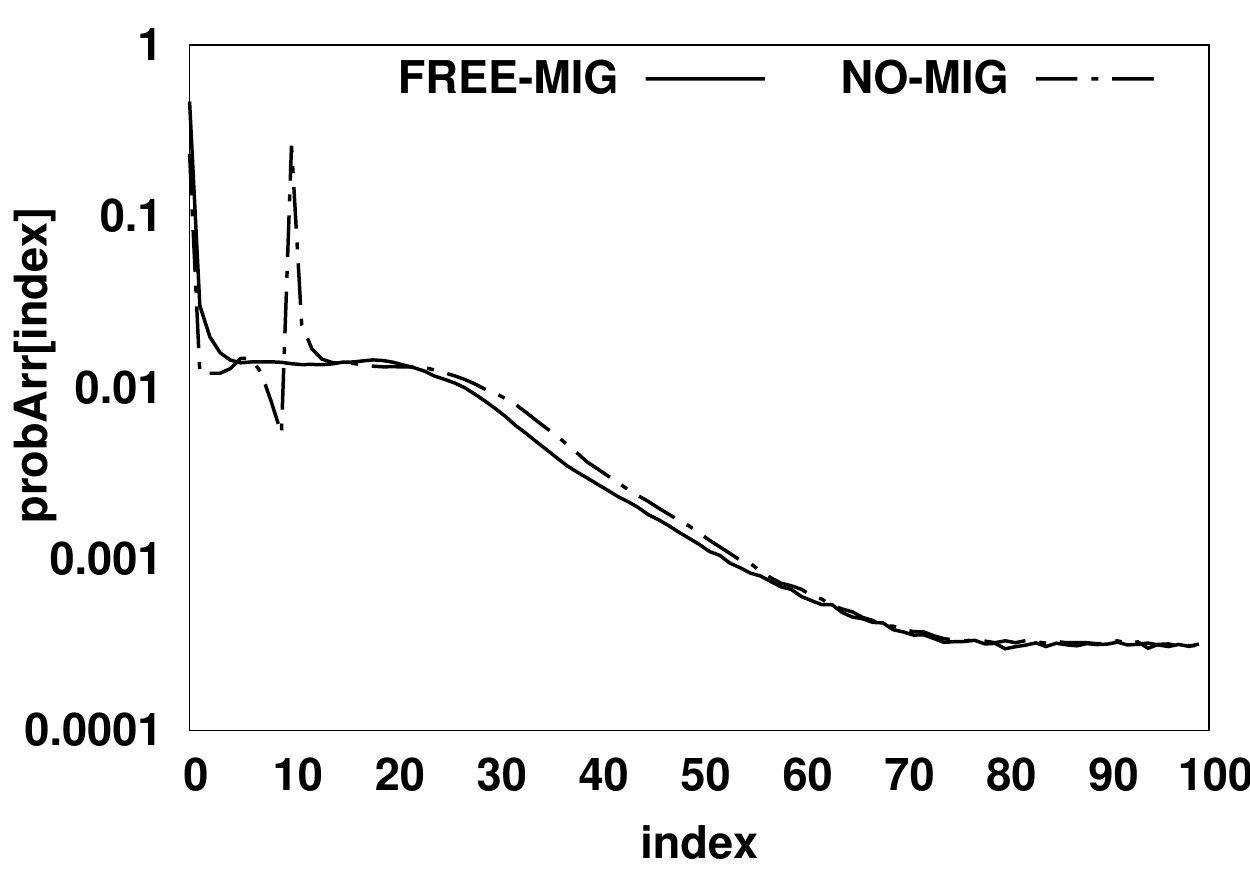}
			\put (27,55) {\scalebox{.45}{ $\mathbf{P\boldsymbol{(}D\!R\!A\!M_{mig}\boldsymbol{)} \!\boldsymbol{=} \!1.83\! \times \!P\boldsymbol{(}D\!R\!A\!M_{noMig}\boldsymbol{)}} $}}
			\put (27,45) {\scalebox{.42}{ $\mathbf{P\boldsymbol{(}D\!R\!A\!M_{noMig}\boldsymbol{[}0\boldsymbol{]}\boldsymbol{)} \!\boldsymbol{=} \!1.8\!\! \times \!\!P\boldsymbol{(}N\!V\!M_{noMig}\boldsymbol{[}0\boldsymbol{]}}\boldsymbol{)}$}}
		\end{overpic}
		\vspace{-.4cm}
		\label{fig:accessdistx264}}
	\vspace{-0.1cm}
	\caption{Access distribution of various migration policies {(in each 
			subfigure,} {probArr[index] is reported for index values ranging 
			from 0 to 100)}}
	\vspace{-0.53cm}
	\label{fig:migdist}
\end{figure*}
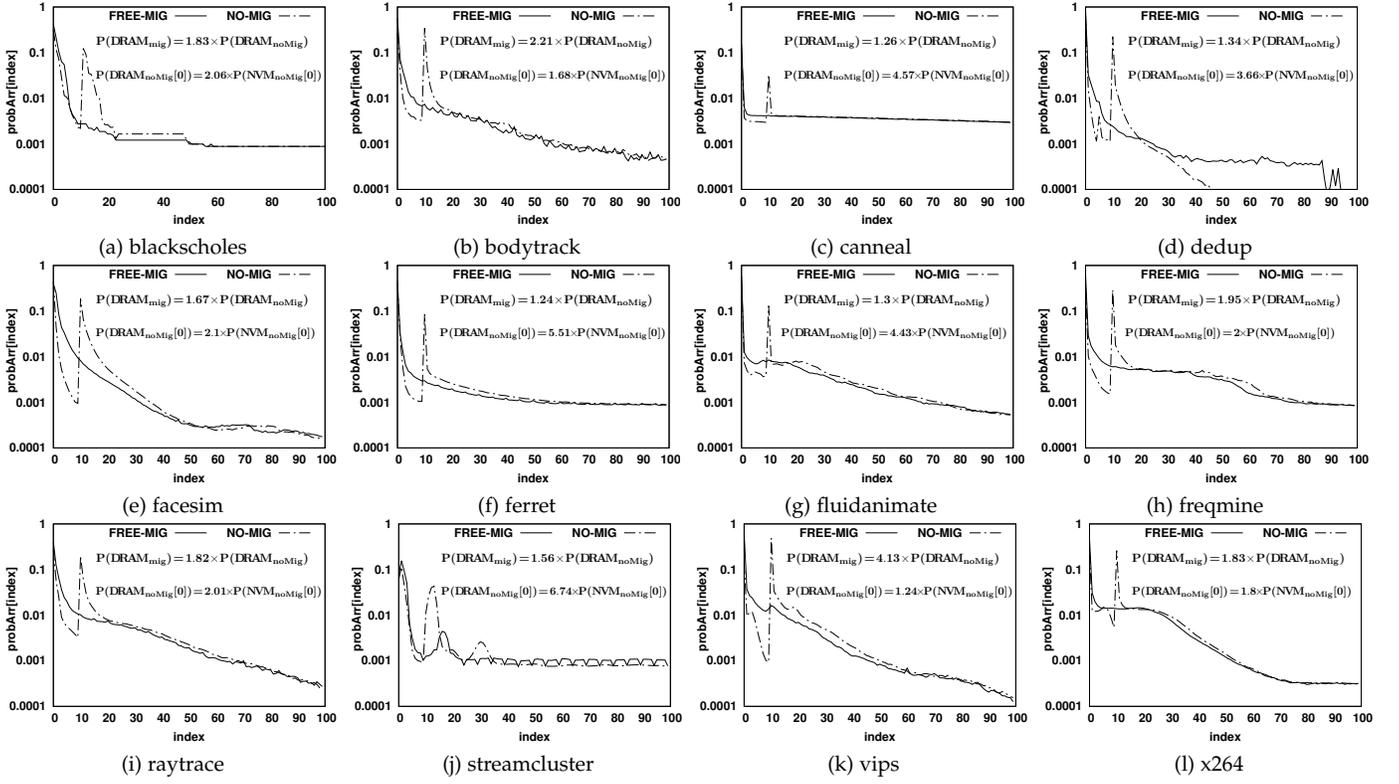

\subsubsection{MIG-PROB}
\vspace{-.1cm}
The third scenario (shown in Fig. \ref{fig:withmig}) is the general case where 
{the} migration probability {($P_{mig}$)} decides whether {or not} to migrate a page into DRAM when it is accessed in NVM.
If migration is allowed, we use the equation of {\emph{FREE-MIG}}.
{Otherwise, we use} the equation of {\emph{NO-MIG}}.
Therefore, the hit probability of DRAM can be calculated using Equation \ref{eq:phitdram}.

\begin{figure}[h]
	\footnotesize
	\vspace{-.3cm}
	\begin{align}
	\label{eq:phitdram}
	P_{D} &= P_{Dnomig} * (1 - P_{mig}) + P_{Dbasic} * P_{mig}
	\end{align}
	\vspace{-.6cm}
\end{figure}

\begin{figure}[!h]
	\centering
	\vspace{-0.2cm}
	\includegraphics[scale=0.7]{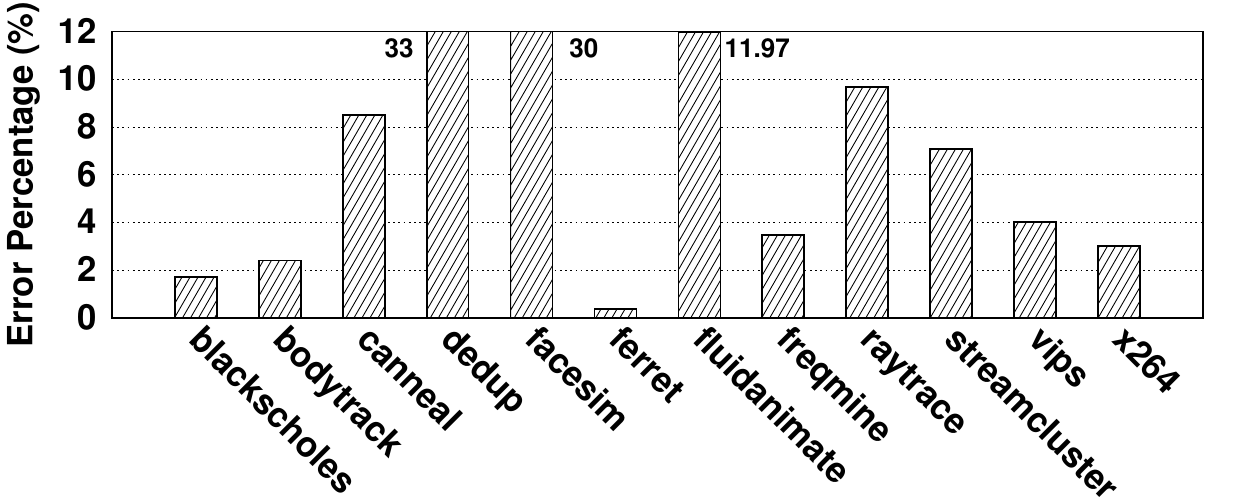}
	\vspace{-0.3cm}
	\caption{Percentage error in predicting $P(hitDRAM|hit)$ using {our} DRAM hit rate estimation model}
	\label{fig:pdramerror}
	\vspace{-0.6cm}
\end{figure}

In order to evaluate the accuracy of {our} DRAM hit rate estimation model, Fig. 
\ref{fig:pdramerror} presents the percentage error in our mechanism's 
prediction of the DRAM hit rate.
The error is calculated by comparing the value of $P(hitDRAM|hit)$ in 
our model and the simulation method.
The error reported in Fig. \ref{fig:pdramerror} will not \emph{directly} 
translate into error in our analytical model for hybrid memories, since even if 
the place of a data page is inaccurately predicted,
there is a probability that the proposed model correctly decides {whether} the 
request will be hit or miss.
This is dependent on the number of unique pages in sequence profiles.
If an access has a few unique pages in its sequence, there is a high 
probability that the data page remains in the HMA under most of the 
eviction policies.



%
\section{Workflow}
\label{sec:workflow}
\vspace{-.15cm}
In this section, we present the overall workflow of our {HMA hit ratio estimation} model
using a simple example.
The process of estimating {the} HMA hit ratio consists of two stages:
\textbf{1)} collecting inputs and \textbf{2)} applying the proposed analytical 
model
to the inputs.
The first stage requires {\textbf{1)}} analyzing a trace file {containing accesses to the memory} and 
{\textbf{2)}} examining the {HMA} to extract {the} necessary information.
In the second stage, {we configure} the proposed analytical model, which is 
based on Markov {decision} processes, {using} the extracted information 
and then estimate the HMA hit ratio {by} solving the {analytical model}.

\begin{figure}[b]
	\centering
	\vspace{-.1cm}
	\includegraphics[scale=0.3]{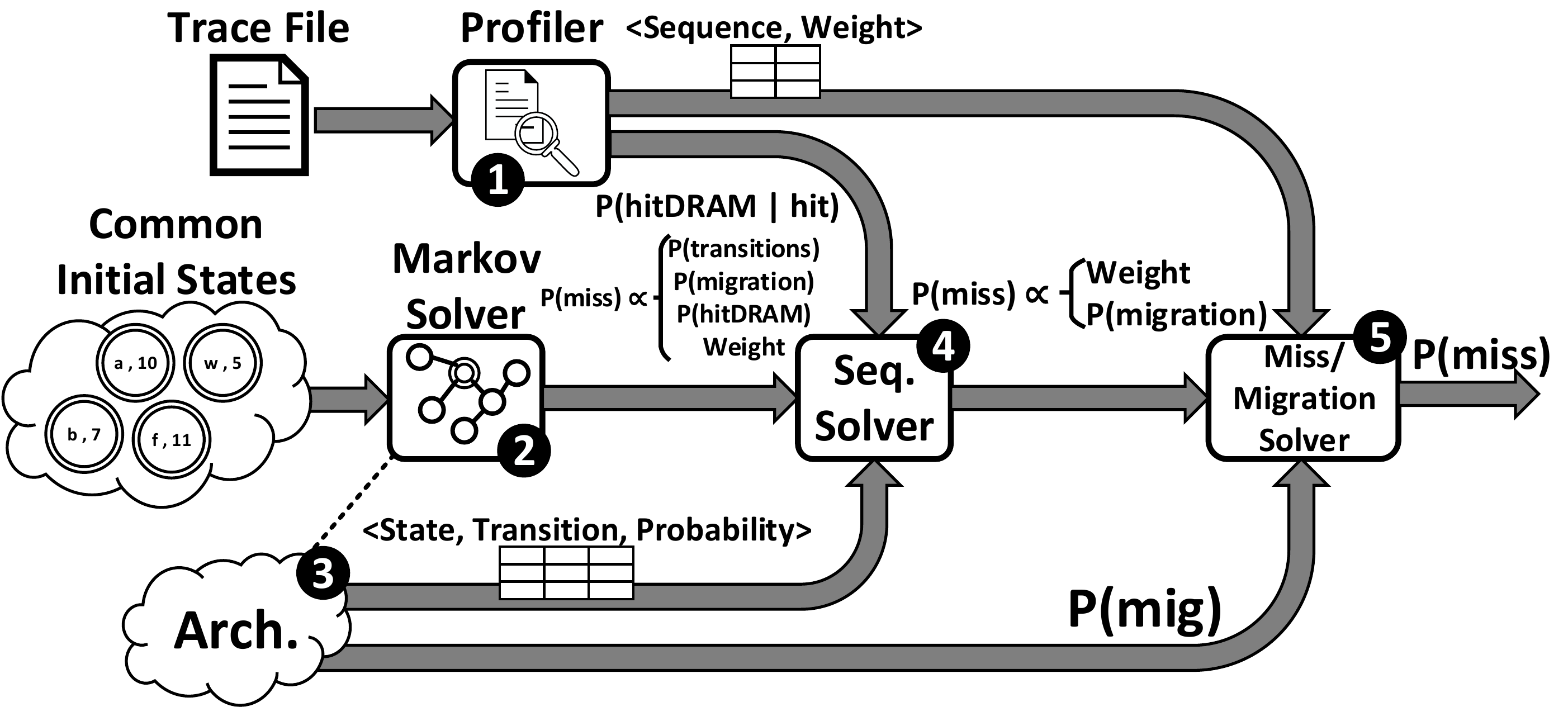}
	\vspace{-0.1cm}	
	\caption{Overall workflow of the proposed model}
	\label{fig:process}
	\vspace{-0.1cm}	
\end{figure}


Fig. \ref{fig:process} demonstrates the overall workflow of the proposed model.
\begin{footnotesize}\circled{1}\end{footnotesize} \emph{Profiler} collects the 
required information from the trace file, which consists of \emph{(a)} {the 
frequency of (r,u)} and \emph{(b)} an estimation of the hit probability in 
DRAM in case of {an} HMA hit.
Section \ref{sec:inputs} {already presented} the detailed algorithm {for the} 
\emph{profiler}.
{If we assume Fig. \ref{fig:profiling} shows an example trace file,} the profiler 
extracts the sequences and their corresponding weights, as reported in 
Table \ref{tbl:extractedweights}.
In the {example} trace file shown in Fig. \ref{fig:profiling}, {the} (1,1) pair is 
found twice out of a total of {the} 11 requests.
Therefore, the weight of (1,1) pair is $\frac{2}{11}$.
{The} special sequence \textless$\infty$,$\infty$\textgreater\space denotes the first access to each data page, which will be a miss regardless of the examined HMA.
Once such information is extracted, {the} trace {file is} no longer 
required for further steps.

In the proposed model, {the} HMA miss ratio is computed using the formula 
shown in Equation \ref{eq:missformulageneral}.
In this equation, {the} $\alpha$ variables denote the {weights} of sequences ({e.g., the weight} values reported in Table \ref{tbl:extractedweights}).
{The} $Markov$ function computes the miss ratio for {the} input values based on 
{a} Markov model.
Equation \ref{eq:missformula} shows the {HMA} miss ratio formula for the example trace file and its extracted weights (reported in Table \ref{tbl:extractedweights}).

\begin{figure}[h]
	\footnotesize
	\begin{align}
	\hspace{-.5cm}
	\label{eq:missformulageneral}
	HMA Miss Ratio =& \alpha_{\infty,\infty} * Markov(\infty,\infty) + \alpha_{0,0} * Markov(0,0)+ \nonumber \\ 
	&\alpha_{1,1} * Markov(1,1)+\alpha_{2,1} * Markov(2,1)+\ldots
	\end{align}
	\vspace{-.6cm}
\end{figure}

\begin{figure}[h]
	\footnotesize
	\vspace{-.6cm}
	\begin{align}
	\hspace{-.5cm}
	\label{eq:missformula}
	HMA &Miss Ratio =\nonumber \\ 
	&\! \frac{5}{11}\! *\! Markov(\infty,\infty)\! +\! \frac{1}{11}\! *\! Markov(0,0) +
	\frac{2}{11} * Markov(1,1) + \nonumber \\ 
	& \frac{2}{11} * Markov(2,2) + \frac{1}{11} * Markov(7,4)
	\end{align}
	\vspace{-.6cm}
\end{figure}

\begin{table}[t]
	\caption{Extracted sequences and their weights from {the} example trace file {in Fig.} \ref{fig:profiling}}
	\vspace{-.05cm}
	\label{tbl:extractedweights}
	\centering
	\bgroup
	\def\arraystretch{1.2}
	\begin{tabular}{|c||c|c|c|c|c|}
		\hline
		\scriptsize{Sequence} &   \scriptsize{\textless$\infty ,\infty$\textgreater} & \scriptsize{\textless$0,0$\textgreater} & \scriptsize{\textless$1,1$\textgreater} & \scriptsize{\textless$2,2$\textgreater} & \scriptsize{\textless$7,4$\textgreater}\\  \hhline{|=||=|=|=|=|=|}
		\scriptsize{Weight} $(\alpha_{r,u})$&   $\frac{5}{11}$   & $\frac{1}{11}$& $\frac{2}{11}$& $\frac{2}{11}$& $\frac{1}{11}$ \\ \hline
	\end{tabular}
	\egroup
\end{table}

\begin{figure}[t]
	\centering
	\includegraphics[scale=0.5]{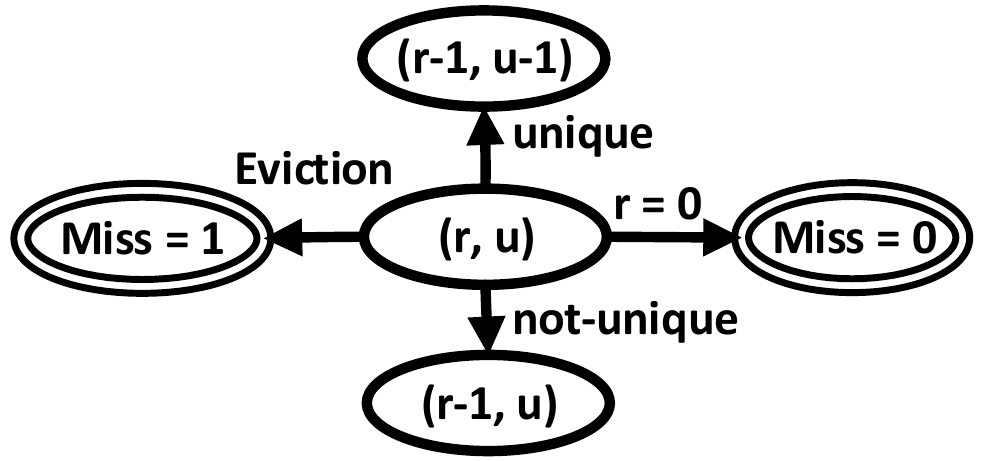}
	\caption{{Example} Markov model of a simple memory management mechanism. r 
	and u in {the} (r,u) pair denote the total and unique number of remaining 
	accesses in the examined access sequence, respectively.}
	\label{fig:simplemarkov}
		\vspace{-0.3cm}
\end{figure}

\begin{figure}[t]
	\centering
	\includegraphics[scale=0.47]{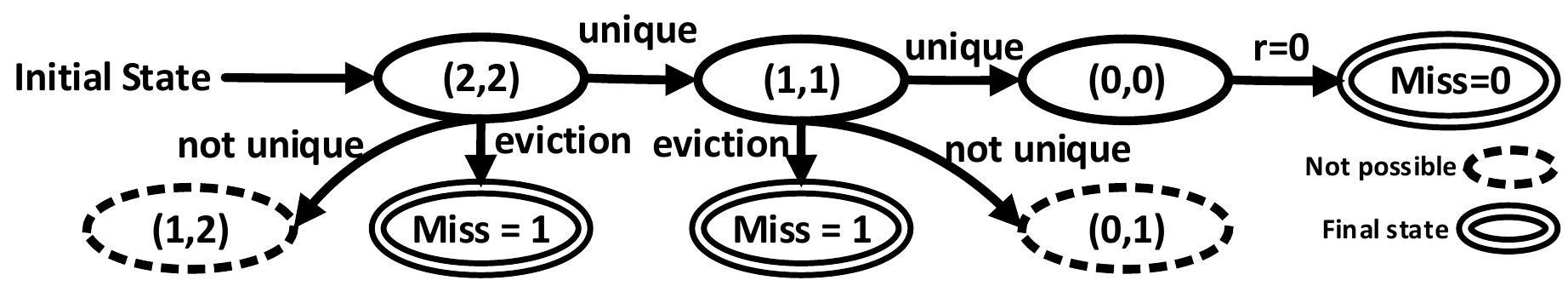}
	\hspace{-0.6cm}	
	\caption{Recursive calls to compute an arbitrary 
	sequence}
	\label{fig:simplemarkovsample}
		\vspace{-0.4cm}	
\end{figure}

\begin{figure}[!h]
	\footnotesize
	\vspace{-.3cm}
	\begin{align}
	\hspace{-.5cm}
	\label{eq:samplesolve}
	&Markov(2,2)  \\ \nonumber
	&= unique * eviction * 1 + (1-eviction) * unique * Markov(1, 1)\\ \nonumber
	&= unique\! * eviction  + (1-eviction)\! *\! unique * (unique * eviction * 1 + \\ \nonumber
	& \qquad (1-eviction) * unique * Markov (0,0)) \\ \nonumber
	&=  unique * eviction * (unique - unique * eviction + 1)
	\vspace{-.5cm}
	\end{align}
	\vspace{-.6cm}
\end{figure}

\begin{footnotesize}\circled{2}\end{footnotesize} \emph{Markov Solver} solves the Markov process for all sequences.
Since the equations in this step are dependent {only} on the initial states, 
the output can be reused for any trace file and/or HMA.
As an example, consider the simple {example} Markov process shown in 
Fig. \ref{fig:simplemarkov}, where $r$ and $u$ are two parameters of sequences.
For this example, Fig. \ref{fig:simplemarkovsample} depicts the recursive calls for calculating {the} miss ratio of a sequence.
Since the number of remaining requests ($r$) cannot be less than {the} number of remaining unique requests ($u$), many states, such as (1,2), are not possible and {thus,} are not considered in the computations.
Equation \ref{eq:samplesolve} shows the expanded formula for calculating 
{the} Markov process depicted in Fig. \ref{fig:simplemarkovsample}.
{The first access to a data page in a sequence is considered $unique$ while later accesses to the same data page are not $unique$.
The $eviction$ parameter denotes the probability that the target data page 
is evicted from the HMA on the next HMA miss access.
To understand the $eviction$ parameter, consider the LRU algorithm.
In LRU, the last data page in the LRU queue is evicted when a new data page is added and the memory is full.
Therefore, $eviction$ for all data pages will be equal to zero except the 
last data page in the queue, which will have $eviction$ equal to one.
The CLOCK algorithm, which employs a similar deterministic approach as LRU 
also has a single data page with \emph{eviction} equal to one, while all other 
data pages have \emph{eviction } equal to zero.}

\begin{footnotesize}\circled{3}\end{footnotesize} In the next step, the target 
HMA is analyzed to extract the transition probabilities {of} the Markov 
states.
The probabilities can be a function of {the} inner state of hybrid memory, 
i.e, position of {the} target data page in the queue.
To solve Equation \ref{eq:missformula}, \begin{footnotesize}\circled{4}\end{footnotesize} \emph{Seq. Solver} and \begin{footnotesize}\circled{5}\end{footnotesize} \emph{Miss/Migration Solver} receive the values for the variables of Markov equations.
Considering a very simple memory architecture {that} randomly evicts data 
pages ({i.e.,} $eviction = 1/size$), Equation \ref{eq:samplesolvewithsize} 
solves the sample miss ratio formula in Equation \ref{eq:samplesolve} for this 
memory architecture assuming $size$ is equal to 4.
Solution of the equation is divided into two steps to minimize the required 
computations by reusing the outputs of previous runs, i.e., a) running 
{an} HMA over various trace files and b) showing the effect of various 
migration probabilities on hit ratio.

\begin{figure}[!h]
	\footnotesize
	\vspace{-.5cm}
	\begin{align}
	\hspace{-.9cm}
	\label{eq:samplesolvewithsize}
	Markov(2,2) &= 1*\frac{1}{size} * (1 -\! 1*\frac{1}{size}\! + 1) \stackrel{size=4}{=} \frac{1}{4} * (1 - \! \frac{1}{4} +\! 1) =\! \frac{7}{16}
	\end{align}
	\vspace{-.6cm}
\end{figure}


%
\vspace{-0.4cm}
\section{Proposed Analytical Model}
\label{sec:proposed}
\vspace{-.15cm}
The proposed analytical model attempts to estimate the hit ratio of HMAs using 
Markov {decision} processes \cite{doi:10.1057/jors.1993.181}.
A {Markov} process is composed of a) initial states and their weights, b) 
transitions between states, and c) transition probabilities.
The initial states and their corresponding weights are determined by the 
data collected from the trace files {(as described in Section \ref{sec:inputs})}.
The transitions of the proposed model are the same across different HMAs while 
transition probabilities are dependent on the target HMA.
Section \ref{sec:states} describes the states of the proposed analytical 
model.
Section \ref{sec:transitions} and Section \ref{sec:probabilities} {present}
transitions and their probabilities, respectively.

\vspace{-.3cm}
\subsection{States}
\label{sec:states}
\vspace{-.1cm}
The states of the Markov model are defined recursively, where the depth of 
recursion is equal to the number of accesses in the 
currently-processed (r,u) pair.
Since in all transitions, the number of remaining requests ($r$) {decreases} by one, the proposed model will have a finite number of states under any HMA.
Each state in the Markov process of the proposed model {is identified 
with} four parameters, denoted as $(r,u,m,p)$:
\begin{itemize}[leftmargin=*]
	\item \textbf{Request (r):} Number of accesses in the remainder of the sequence
	\item \textbf{Unique (u):} Number of unique data pages in the remainder of the sequence
	\item \textbf{Memory (m):} current memory in which the target data page resides
	\item \textbf{Position (p):} current position of the target data page in 
	{the} memory queue
\end{itemize}

\begin{figure}[!h]
	\vspace{-.3cm}
			\centering
			\includegraphics[scale=0.9]{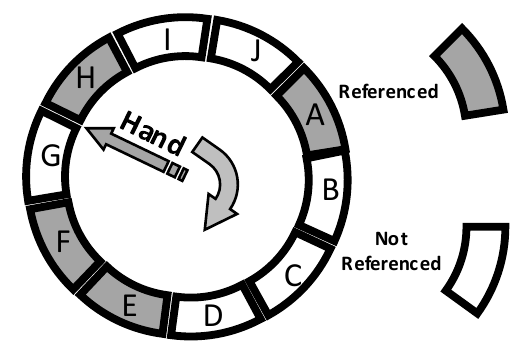}
			\caption{{An example} CLOCK data structure}
			\label{fig:clock}
\vspace{-.1cm}
\end{figure}

Note that Markov states are \emph{not} dependent on the data pages and their addresses.
For instance, the state (3, 2, DRAM, 0) can be used for both of the following access patterns: [1000, 1001, 1002, 1001, 1000] and [2500, 1200, 
3001, 3001, 2500].
The data page with position 0 is the last data page {that} will be evicted 
from the memory.
Similarly, the data page {whose} position value set to the number of data pages is the first data page to be evicted.
In other words, \emph{Position} {denotes} the number of data pages {that} remain in memory right after evicting the target data page.
For the LRU algorithm, this value is simply equal to the position of the target data page in the LRU queue.
For more complex algorithms such as CLOCK, the data pages can still {be} ordered based on the clock hand position and the referenced bit.
For instance, Fig. \ref{fig:clock} shows the state of a memory architecture managed by the CLOCK algorithm.
The value of \emph{Position} for target data page \emph{C} will be six since exactly after evicting the data page \emph{C}, 6 data pages remain in the memory (data pages A, D, E, F, G, H).
Thus, the CLOCK hand and any other parameters of the memory management algorithms can be encoded into \emph{Position}.

Each state in the proposed Markov model {needs} to store {the (r,u) 
pair} to calculate the transition probabilities and identify terminating 
states.
The states of the Markov model follow a target data page between two 
accesses to the target data page.
This is done to find the probability that the data page is 
evicted from the memory.
The remainder of accesses determines how many transitions we need to execute on 
the Markov state to reach the final state.
The number of unique remainder pages is employed to increase the accuracy of the proposed model.
The first access to a unique page will be a miss access, which requires evicting a data page from the memory (HMA).
The number of evicted data pages greatly affects the probability that the target data page is evicted from the memory.
Consider two sequences, both with 10 accesses, where the number of unique data pages 
in the first sequence is equal to three and in the second sequence is equal to 10.
The probability that the target data page is evicted from the memory is higher in the second sequence since more data pages should be evicted from the memory.

The relative position ($p$) of the target data page compared to the other data pages in HMA  is stored in states to calculate the transition probabilities.
Although the Markov model does not need additional information in 
the states, 
a {Boolean} variable ($m$) is also stored in each state to determine the 
residing memory of {the} target data page.
The current residing memory of the target data page can be calculated using the relative position of {the} data page in memory queue ($p$) and {the} size of {the} memory modules.
This, however, restricts the reusability of calculations to specific 
memory module sizes.
{Thus,} storing {the} residing memory of {the} target data page {reduces the computation.}

{In contrast} to modeling a single memory architecture, a hybrid memory model 
should have \emph{two} originating states since the corresponding data page 
might reside in either DRAM or NVM.
Therefore, our model calculates the miss probability of each 
access for both cases and uses their weighted average as the 
miss probability.
The weights are determined based on the equations presented in Section \ref{sec:inputs} and are calculated {\emph{only}} once per trace file, which enables {their reuse} for any HMA.

\begin{figure}[!h]
	\centering
	\includegraphics[scale=0.42]{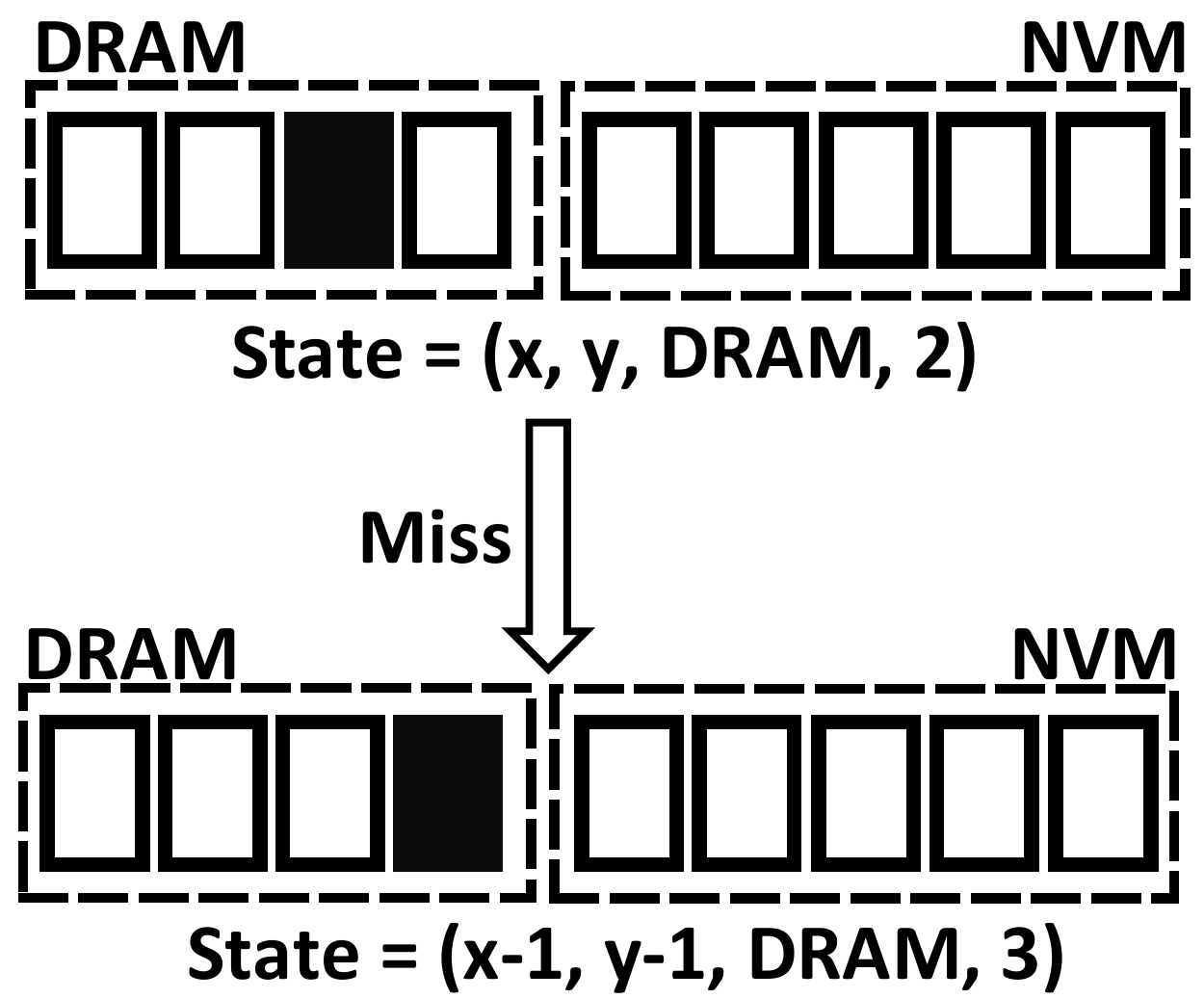}
	\hspace{-0.65cm}	
	\caption{Effect of a miss access on the Markov state}
	\label{fig:statessample}
	\vspace{-0.65cm}
\end{figure}

To clarify the states, let's consider an example HMA with four DRAM and five NVM data pages, shown in Fig. \ref{fig:statessample}.
Since the target data page is in the third position of {the} DRAM queue, it has
the state of {(x,y,DRAM,2)}.
Note that the position is zero-indexed.
A miss access will change the state to {(x-1,y-1,DRAM,3)} as depicted in Fig. \ref{fig:statessample}.
Each access reduces the number of remaining requests (x).
A miss request is to an address that has not been accessed before.
It is considered a \emph{unique}
request.
Therefore, the number of remaining unique data pages (y) is reduced by one.
Miss accesses result in moving the data page to DRAM, which results in shifting the target data page from position 2 to position 3.

\begin{table*}[!t]
		\vspace*{-0.3cm}
	\hspace*{-1cm}
	\caption{State transitions in the proposed model}
	\label{tbl:transitions}
	\centering
	\scriptsize
	\begin{tabular}{|c|c|c|c|c|c||c|c|c|c|c|c|}
		\hline \multicolumn{6}{|c||}{\bfseries DRAM} & \multicolumn{6}{|c|}{\bfseries NVM} \\
		\hline \bfseries No.  & \bfseries \specialcell{Unique\\$P(unique)$} & \bfseries \specialcell{Hit\\$P(hit)$} & \bfseries \specialcell{Before/After\\(Equation \ref{eq:phitbefore})} & \bfseries \specialcell{Mig.\\$P_{mig}$} & \bfseries Dest. & \bfseries No. & \bfseries \specialcell{Unique\\$P(unique)$} & \bfseries \specialcell{Hit\\$P(hit)$} & \bfseries \specialcell{Before/After\\(Equation \ref{eq:phitbefore})} & \bfseries \specialcell{Mig.\\$P_{mig}$} & \bfseries Dest. \\ 
		\hline \begin{tiny}\circled{1}\end{tiny}  & $\times$ & $\times$ & N/A & N/A & (r-1, u, D, p+1)  & \begin{tiny}\circled{11}\end{tiny} & $\times$ & $\times$ & N/A & N/A & (r-1, u, N, p+1)  \\ 
		\hline \begin{tiny}\circled{2}\end{tiny} & $\times$ & \checkmark & Before & N/A & (r-1, u, D, p)& \begin{tiny}\circled{12}\end{tiny} &  $\times$  & \checkmark& Before & N/A& (r-1, u, N, p) \\ 
		\hline \begin{tiny}\circled{3}\end{tiny}  &  $\times$ & \checkmark & After & $\times$ & \specialcell{\begin{tiny}hit in DRAM:\end{tiny} $0$\\\begin{tiny}hit in NVM: \space\space  \end{tiny}(r-1, u, D, p)} & \multirow{2}{*}{\begin{tiny}\circled{13}\end{tiny}} & \multirow{2}{*}{$\times$}  & \multirow{2}{*}{\checkmark} &  \multirow{2}{*}{After}  & \multirow{2}{*}{$\times$}  & \multirow{2}{*}{0}\\ 
		\cline{1-6} \begin{tiny}\circled{4}\end{tiny}  & $\times$ & \checkmark & After & \checkmark & \specialcell{\begin{tiny}hit in DRAM:\end{tiny} $0$\\\begin{tiny}hit in NVM: \space  \end{tiny}(r-1, u, D, p+1)} & \multirow{2}{*}{} & \multirow{2}{*}{}   &  \multirow{2}{*}{} &  \multirow{2}{*}{}   & \multirow{2}{*}{}   & \multirow{2}{*}{} \\ 
		\hline \begin{tiny}\circled{5}\end{tiny}  & \checkmark &$\times$ &  N/A &  N/A & (r-1, u-1, D, p+1) & \begin{tiny}\circled{14}\end{tiny} & \checkmark  & $\times$&  N/A&  N/A &  (r-1, u-1, N, p+1) \\ 
		\hline \begin{tiny}\circled{6}\end{tiny}  & \checkmark  & \checkmark  &Before & N/A & 0 & \begin{tiny}\circled{15}\end{tiny} &  \checkmark & \checkmark& Before& N/A& (r-1, u-1, N, p) \\ 
		\hline \begin{tiny}\circled{7}\end{tiny}  & \checkmark  & \checkmark  & After &$\times$ & \specialcell{\begin{tiny}hit in DRAM:\end{tiny} (r-1, u-1, D, p+1)\\\begin{tiny}hit in NVM: \space  \end{tiny}(r-1, u-1, D, p)} & \multirow{2}{*}{\begin{tiny}\circled{16}\end{tiny}} &  \multirow{2}{*}{\checkmark}  & \multirow{2}{*}{\checkmark}& \multirow{2}{*}{After}& \multirow{2}{*}{N/A} & \multirow{2}{*}{(r-1, u-1, N, p+1)} \\ 
		\cline{1-6} \begin{tiny}\circled{8}\end{tiny}  & \checkmark & \checkmark & After & \checkmark &(r-1, u-1, D, p+1) &\multirow{2}{*}{} & \multirow{2}{*}{}  & \multirow{2}{*}{}& \multirow{2}{*}{}& \multirow{2}{*}{} & \multirow{2}{*}{} \\
		\hline \begin{tiny}\circled{9}\end{tiny} & \multicolumn{5}{|c||}{$(\neg Hit \wedge evict) \Rightarrow state = (r,u,N\!V\!M,0) $} & \begin{tiny}\circled{17}\end{tiny} &\multicolumn{5}{|c|}{$(\neg Hit \wedge evict) \Rightarrow miss = 1 $} \\
		\hline \begin{tiny}\circled{10}\end{tiny} & \multicolumn{11}{|c|}{$(r  = 0 \quad or \quad u = 0) \Rightarrow miss = 0 $}\\
		\hline 
	\end{tabular} 
\vspace*{-0.4cm}
\end{table*}

\vspace{-.3cm}
\subsection{Transitions}
\label{sec:transitions}
\vspace{-.1cm}

Transitions between states are defined based on the possible conditions of the requests.
{Our model considers four} conditions of requests:
\emph{1) hit/miss}: {Requests} can be either hit or miss in the memory.
Miss accesses result in moving a new data page to the memory and therefore changing the position of the target data page.
\emph{2) before/after:} By considering the total order of pages in memory, each hit access can be either before or after the target data page, which might affect the target data page position.
\emph{3) uniqueness:} The first access to a data page in the sequence has a 
{distinct} outcome from the next access to the same data page {due to} 
the policies {employed in the HMA}.
\emph{4) migrations:} Requests can result in migrating the accessed data page between memories.
The discussed conditions are not independent of each other and many of their combinations may not be possible in the examined HMA.
For instance, the transition in Fig. \ref{fig:statessample} is a \emph{unique miss} access {(cases 3 and 1)}.
The other two conditions are not applicable since both are {defined only} 
when the access is a hit in the {HMA}.


Table \ref{tbl:transitions} shows the transitions of the proposed analytical model.
Since the transitions and their destination states differ based on the residing 
memory of {the} target data page, DRAM and NVM cases are shown separately.
``N/A''  values in Table \ref{tbl:transitions} depict \emph{Not Applicable} and/or {an} invalid combination of conditions.
We assign each transition a unique number, which {we will refer} to throughout the paper.
\emph{Unique} shows that the current request is a unique request in the current sequence.
\emph{Hit} shows that the current request is a hit in the {HMA} (DRAM or 
NVM).
If a request is a hit, {the} \emph{Before/After} column states whether it is hit 
in the data pages queue before or after the target data page.
{The} NVM queue is considered after {the} DRAM queue when comparing the 
positions of data pages.
{Section \ref{sec:beforeafter} details the \emph{Before/After} parameter.}
{Migration} \emph{Mig.} shows {if} the current request will result in {the} migration of 
its corresponding data page from NVM to DRAM.
\emph{Destination} {(Dest.)} depicts the destination state of the transition.
If the target data page is selected for eviction from DRAM, the destination state will be set to \textless r,u,NVM,0\textgreater .
Choosing the target data page for eviction from NVM will result in $miss = 1$ 
since the next access to an evicted data page will {always be} a miss.

\begin{figure}[!b]
			\centering
			\includegraphics[scale=0.29]{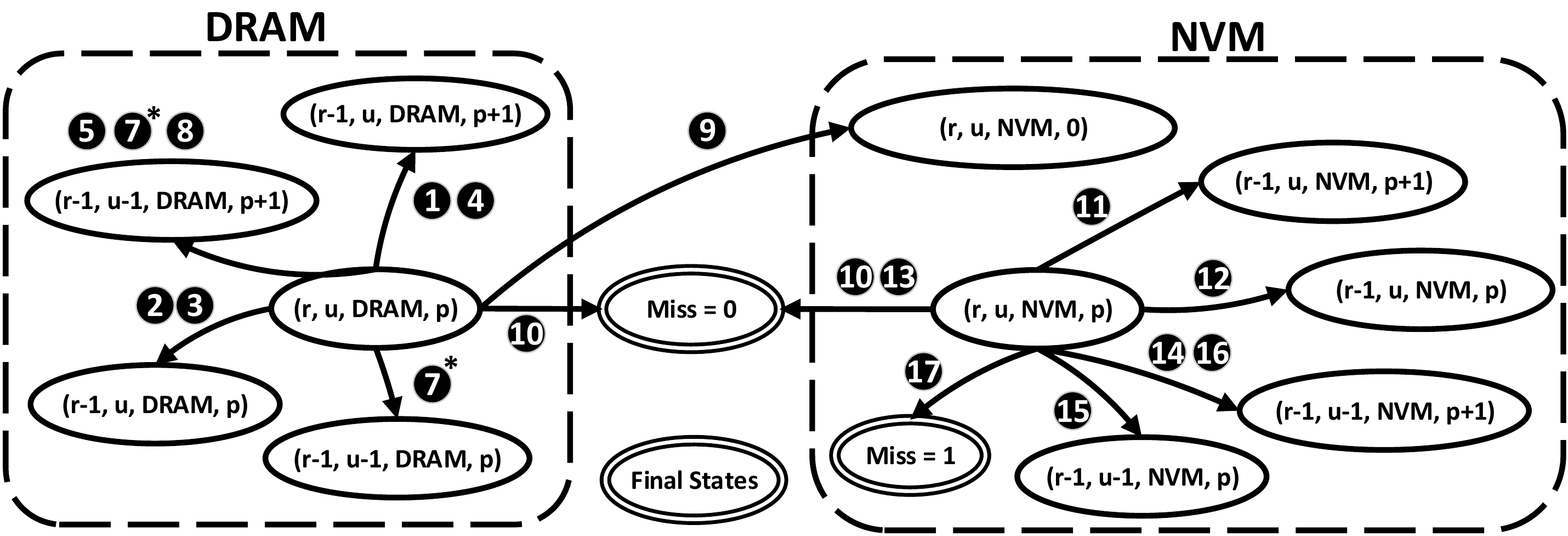}
			\\ \scriptsize *Transition has more than one destination state.
			\caption{Markov process of the proposed analytical model}
			\label{fig:markov}
			
\vspace{-0.55cm}
\end{figure}

The Markov process of the proposed analytical model is depicted in Fig. \ref{fig:markov}.
States are grouped based on the residing memory of the target data page 
(i.e., {the} third identifying parameter of a state).
The only interaction between DRAM and NVM for the target data page is through eviction from DRAM.
The Markov process has two final states.
{All} sequences will reach one of {the final states} in at most $r$ 
transitions.
This is {because we reduce} the $r$ parameter in all of the transitions 
(except transition 9) and {declare a state final} when $r$ reaches zero.
Transition 9 cannot create an infinite loop since there is no transition back from NVM to DRAM to complete such {a} loop.
The promotions (migrations from NVM to DRAM) are not present in the Markov 
process {because of} their intrinsic difference {from} demotions 
(migrations from DRAM to NVM).
A data page can be promoted {\emph{only}} when it is accessed.
{However}, sequences are constructed from accesses {to other data pages} 
between two consecutive accesses to the target data page and therefore, no 
access to the target data page exists {in the} sequence.
As such, it is not possible for the target data page to be promoted.
Promotions of the other data {pages, however,} are considered in the 
proposed model (transitions 4 and 8).

Transitions 9, 10, and 17 are exceptional transitions and their conditions will be examined before other transitions.
In miss accesses, \emph{Mig.} and \emph{Before/After} conditions are not applicable (transitions 1, 5, 11, 14).
Both conditions are defined {only} when the request is a hit in the HMA.
If a request is {a} hit, it can be {a} hit either before or after the 
target data page.
{A hit} before the target data page will \emph{not} have any effect {on} 
{the target data page's} position in {the} queue (transitions 2, 12, and 
15).
Transition 6 is not practical since data pages \emph{before} the target data 
page are accessed {at least once} and hence, cannot be unique.
A hit after the target data page in NVM (transition 16) will result in 
{increasing the position value in} {the} {state} regardless of 
{whether or 
not a} 
migration {happens}.
The output of {a hit after the target data page} in DRAM, however, depends 
on the probability of migration.
If an access {hits} in NVM and 1) the data page migrates to DRAM, the target 
data page will be shifted {into} {the} queue (transitions 4 and 8), {2) 
otherwise, if the page does not migrate to DRAM}, the target data page will 
stay in the same position as before (transitions 3 and 7).
A {non-unique} access cannot hit in DRAM if the target data page also 
resides in DRAM (transitions 3 and 4) since {non-unique} data pages {are 
either} before (as in {the \emph{Before/After}} parameter of Markov states) the target 
data page in DRAM or reside in NVM.
{Similar} to transition 6, transition 13 is also impractical.
{Non-unique pages} have been accessed before and therefore, either reside in 
DRAM or \emph{before} the target data page in NVM.

\vspace{-0.3cm}
\subsection{Transition probabilities}
\label{sec:probabilities}
\vspace{-.1cm}
In order to accurately estimate the miss probability of each state, the probabilities of transitions are also required.
The sum of probabilities of transitions going out of each state should be equal to one.
Since transitions are based on four parameters (uniqueness, hit, before/after, 
and migration), the probability of a transition is calculated based on the 
probability of {each of the four} parameters, which {we will present} in 
the remainder of this section.
\vspace{-0.2cm}
\subsubsection{Uniqueness}
In a given state with $r$ requests and $u$ unique data pages, $P(unique)$ denotes the probability that the next processed access is a unique access.
Our analysis shows that the number of unique accesses has {a} higher impact 
on {the HMA} hit ratio than the {order} of unique accesses inside the 
sequence.
This is also {demonstrated} in previous studies that employed sequence profiling \cite{Guo:2006:AMC:1140103.1140304}.
Therefore, {we consider each access to have} an equal probability to be $unique$.
Based on this assumption, the probability of uniqueness is calculated  
as: $P(unique) = \frac{u}{r}$.


\vspace{-0.2cm}
\subsubsection{Hit Probability}
Hit probability (HMA hit ratio) in {the} transition probabilities can {either}
be set to a constant value or remains as {a} variable in the equations.
Since predicting hit probability is the main goal of the proposed analytical model, using a constant value for this parameter will reduce the accuracy of the model.
Therefore, $P(hit)$ remains as a variable in the transition probabilities 
{of our model}.
Replacing $P(hit)$ with a constant value significantly reduces the required computational processing power to solve the final equation in \emph{Miss/Migration Solver}.
By using a constant value for $P(hit)$ and iteratively improving its accuracy, one can compute the final hit ratio.
This technique, however, requires an \emph{accurate} initial value {for} 
$P(hit)$ and is {beneficial only} when {the} \emph{miss/migration solver} 
cannot solve the formula {due to the high} degree of {the} polynomial 
equation.
In the proposed analytical model, we solve the {the Markov model} by leaving 
$P(hit)$ {as} a variable, to increase the accuracy of the model.

\vspace{-0.2cm}
\subsubsection{Before/After}
\label{sec:beforeafter}
In order to estimate the position of the target data page in {the} case of a hit in the HMA, {we} should {decide} whether the accessed data page is hit before or after the target data page.
The {\emph{Before/After}} parameter denotes the relative position of the accessed data page compared to the target data page if we sort the data pages based on their eviction time.
A data page is labeled \emph{before} if the target data page {is} evicted \emph{before} the accessed data page and vice versa.
For instance, considering the target data page C in Fig. \ref{fig:clock}, the data pages (I, J, and B) are labeled \emph{after} and other data pages will be labeled \emph{before}.
In the proposed analytical model, we use the stack distance data in $probArr$ 
to estimate {the} \emph{before/after} probability {of a data page}.
If the target data page resides in DRAM, $P(before)$ {is} calculated by 
summing all values of $probArr$ {in} positions before the target data page.
Since this probability is defined as $P(before|hit)$, we do not need to consider the miss probability in $P(before)$ or $P(after)$.
The migration probability, however, affects $P(before|hit)$ since the 
distribution of stack distance is dependent {on} the migration probability.
Therefore, $probArr$ is computed with respect to Equation \ref{eq:phitdram}.
In addition, the probabilities in $probArr$ are calculated as $P(i|hit)$ for an arbitrary position $i$, which enables us to sum their values without the need for additional calculations.
If the target data page resides in NVM, {a} hit data page might be hit in 
DRAM, before the target data page in NVM, or after the target data page.
The hit probability in DRAM is pre-calculated in Equation \ref{eq:phitdram} and the hit before the target data page in NVM is calculated using $probArr$.
Equation \ref{eq:phitbefore} shows the overall formula for estimating \emph{before/after} probability for an arbitrary data page residing in position $M$.

\begin{figure}[h]
	\footnotesize
	\vspace{-0.3cm}
	\begin{align}
	\label{eq:phitbefore}
	& P(before_{M}|hit) = \\ \nonumber
	& \begin{cases}
	\sum_{i = DRAM_{start}}^{M} ProbArr(i)       &  \text{if } M \in \text{DRAM}\\ \nonumber \\\nonumber
	P(hitD\!R\!A\!M|hit) \! +\!
	\sum_{i=N\!V\!M_{start}}^{M} ProbArr(i)  &  \text{if } M \in \text{ NVM} 
	\end{cases}
	\end{align}
	\vspace{-0.6cm}
\end{figure}

\vspace{-0.2cm}
\subsubsection{Migrations}
Migration probability is one of the design parameters of the examined HMAs, which depicts the probability that an access to NVM can result in migrating the corresponding data page to DRAM.
This probability can either {be} set to a) a fixed value to calculate the performance or b) a formula to calculate the effect of various migration policies on the performance.
The fixed value can be set 1) according to the empirical studies on the 
examined hybrid memory or 2) by substituting the migration variables with 
actual values from the HMA in the provided formula.
Section \ref{sec:setup} describes the process of extracting the 
migration probability from the examined hybrid memories.


\vspace{-0.3cm}
\subsection{AMAT \& NVM Lifetime Estimation}
\vspace{-0.1cm}
The proposed model estimates the {HMA} hit ratio {using} a trace file 
for a given HMA.
	By slightly modifying transitions and storing additional data, the proposed
	model can also estimate AMAT and NVM lifetime.

To estimate AMAT, \emph{a)} {the} number of accesses to each memory, \emph{b)} read/write ratio, \emph{c)} {the} number of migrations, and \emph{d)} {the} number of miss accesses are required.
The number of accesses to each memory (\emph{a}) can simply {be} computed by 
using $P_{HitDRAM|hit}$ and estimated hit ratio {by the proposed analytical 
model}.
The read/write ratio \emph{(b) is collected} {while parsing the} trace file.
The other two parameters (\emph{c} and \emph{d}) are already estimated by the proposed model.
This method, however, has {a} few limitations.
For example, it does not consider the concurrent processing of requests by the memory controller.
To address this shortcoming, the concurrency of the requests is included 
in the average response time of the memory when computing HMA 
average response time.
Thus, the error of the formula to calculate AMAT is low.

The NVM lifetime depends on a) the number of writes and 
b) the wear-leveling algorithm employed in the NVM.
{The wear-leveling algorithm is a design parameter of the NVM memory device.
It has almost a fixed effect on the NVM lifetime.
{Wear-leveling algorithms try to erase data blocks that are a) mostly read or b) have low erase count, to evenly distribute erase count across all blocks.
Both cases add additional writes to the NVM, compared to the normal garbage collection operations.
These algorithms limit the rate of erasing data blocks to reduce the impact of wear-leveling on NVM lifetime.
Therefore, we can consider a fixed overhead for wear-leveling algorithms in our evaluations.
The experimental results in previous studies also have such a fixed overhead in various workloads \cite{startgap,6853308,Chang:2015:MWP:2742143.2699831}.}
Thus, by estimating the NVM writes, we can estimate the NVM lifetime.}
NVM lifetime estimation requires a) {the} number of writes to NVM, b) {the} number of migrations to NVM, and c) {the} number of disk-to-NVM page copy operations.
The first parameter can be estimated by the number of accesses to NVM and {the} write ratio of {the} trace file.
The second parameter is estimated during {the} hit ratio estimation process 
by counting the number of times transition 9 is triggered.
To calculate the third parameter, {the} page fault destination of the examined HMA is considered, which can
have one of the three scenarios: \emph{a)} {the} data page is copied from disk 
to DRAM in {the} case of a page fault, \emph{b)} {the} data page is copied from disk 
to NVM, and \emph{c)} {the} data page is copied to either DRAM or NVM based on 
request type (read/write).
In the first scenario, {the} number of disk to NVM page copy operations will 
be zero.
{In} the second scenario, it will be
equal to the number of miss accesses.
{In} the last scenario, it will be computed by the number of miss accesses
and {the} read/write ratio of the trace file.

\vspace{-.3cm}
\section{Experimental Setup \& Results}
\label{sec:experiments}
\vspace{-.1cm}
In this section, {we describe} {our} evaluation {methodology} {and} experimental 
results {of our} model, {compared to} {the traditional} simulation method.
Section \ref{sec:setup} {presents our} experimental setup.
{Section \ref{sec:accuracy} reports accuracy and Section 
\ref{sec:computation} reports the computation time.}
{Section \ref{sec:application} presents an application of the proposed 
analytical model.
Finally, Section \ref{sec:overheads} discusses the overhead of using the proposed analytical model.}

\vspace{-.2cm}
\subsection{{Experimental Setup \& HMA Modeling}}
\label{sec:setup}
\vspace{-.05cm}
We evaluate our analytical model using two recent HMAs (TwoLRU 
\cite{date16} and CLOCK-DWF \cite{clockdwf}).
These HMAs {enable us to measure the} accuracy of the proposed 
model for HMAs based on LRU and CLOCK.
All experiments are conducted on a server with 3.3 GHz CPU and 192 GB memory running {a} Linux kernel.
For fair comparison, we use a single CPU core to perform 
computations of {both} our model {and} for the {baseline} simulation method.
The input trace files are captured by running {the} PARSEC benchmark suite 
\cite{parsec} in {the} COTSon \cite{cotson} full system simulator.
Table \ref{tbl:workloads} reports the characteristics of the trace files.

\begin{table}[!h]
			\caption{Workload characteristics}
			\scriptsize
			\label{tbl:workloads}
			\centering
			\begin{tabular}{|c|c|c|c|}
				\hline
				\textbf{Workload} & \textbf{Unique Pages} & \textbf{Read Requests} & \textbf{Write Requests}\\ \hline\hline
				Blackscholes & 5,188& 26,242 (100\%) & 0 (0\%) \\ \hline
				Bodytrack & 25,304& 658,606 (62\%) & 403,835 (38\%) \\ \hline
				Canneal &164,768 & 24,432,900 (98\%) & 653,623 (2\%)\\ \hline
				Dedup & 512,460& 17,187,130 (71\%) & 6,998,314 (29\%)\\ \hline
				Facesim &210,368 & 11,730,278 (66\%) & 6,137,519 (34\%) \\ \hline
				Ferret &68,904 & 54,538,546  (89\%) &  7,033,936 (11\%) \\ \hline
				Fluidanimate &266,120 & 9,951,202 (69\%)& 4,492,775 (31\%) \\ \hline
				Freqmine & 156,108& 8,427,181 (69\%)& 3,947,122 (31\%)\\ \hline
				Raytrace & 57,116& 1,807,142 (83\%)&  370,573 (17\%)\\ \hline
				Streamcluster &15,452 & 168,666,464 (99.8\%)& 448,612 (0.2\%)\\ \hline
				Vips & 115,380& 5,802,657  (59\%) &4,117,660  (41\%)\\ \hline
				X264 & 80,232& 14,669,353 (74\%)& 5,220,400 (26\%)\\ \hline
			\end{tabular}
\end{table}

\begin{figure*}[!b]
			\centering
			\subfloat[blackscholes]{\includegraphics[width=0.242\textwidth]{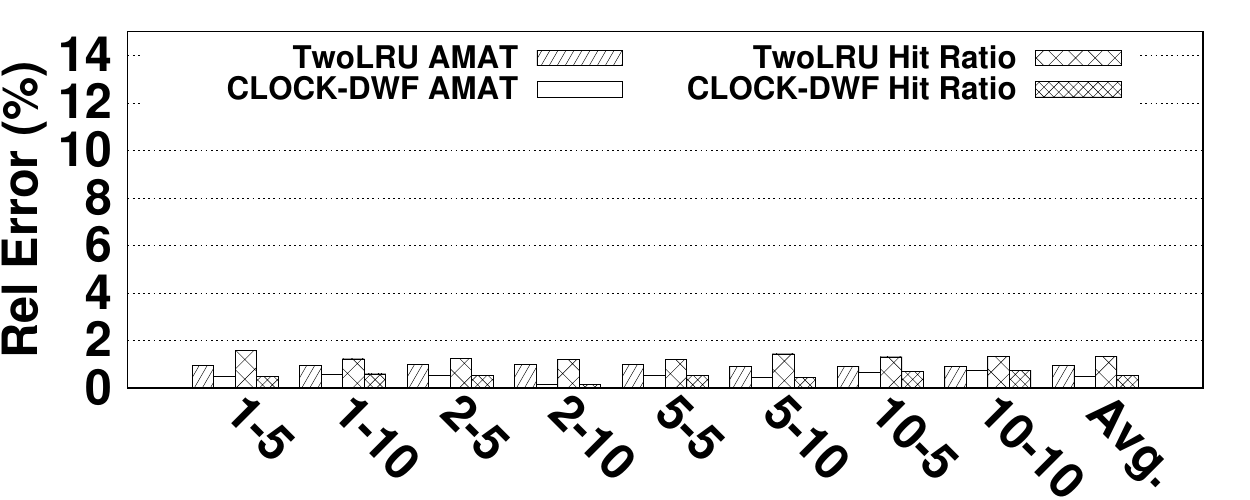}
				\label{fig:blackscholes}}
			\hfill
			\subfloat[bodytrack]{\includegraphics[width=0.242\textwidth]{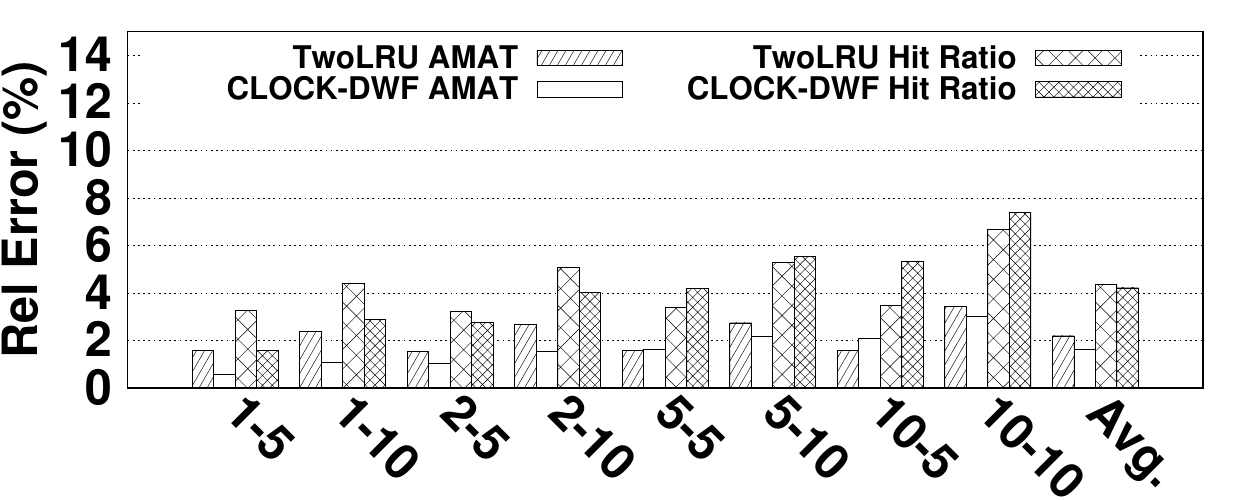}
				\label{fig:bodytrackerror}}
			\hfill
			\subfloat[canneal]{\includegraphics[width=0.242\textwidth]{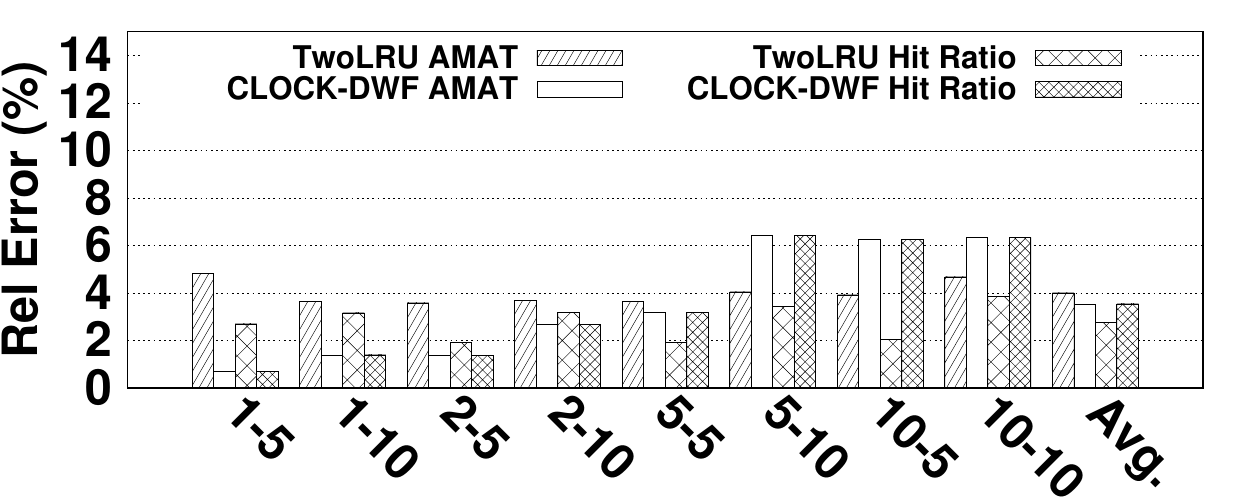}
				\label{fig:cannealerror}}
			\hfill
			\subfloat[dedup]{\includegraphics[width=0.242\textwidth]{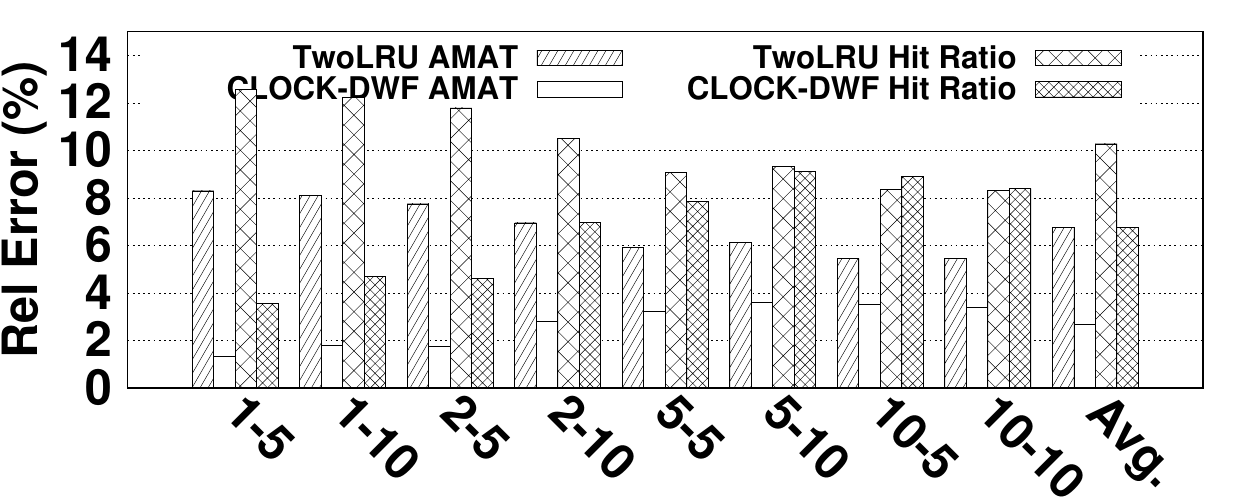}
				\label{fig:deduperror}}
			\hfill
			\vspace{-.5cm}
			\subfloat[facesim]{\includegraphics[width=0.242\textwidth]{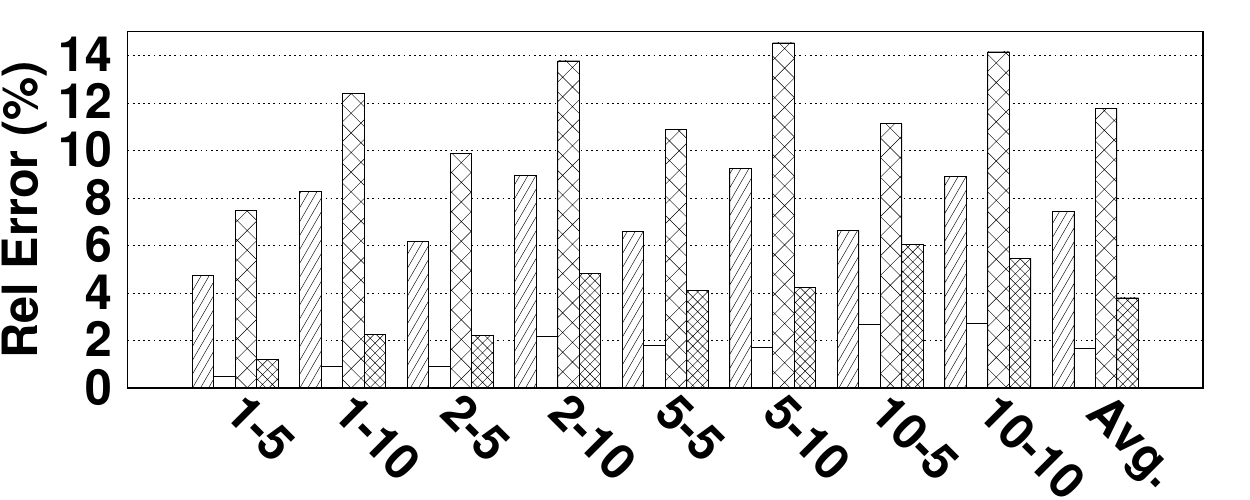}
				\label{fig:facesimerror}}
			\hfill
			\subfloat[ferret]{\includegraphics[width=0.242\textwidth]{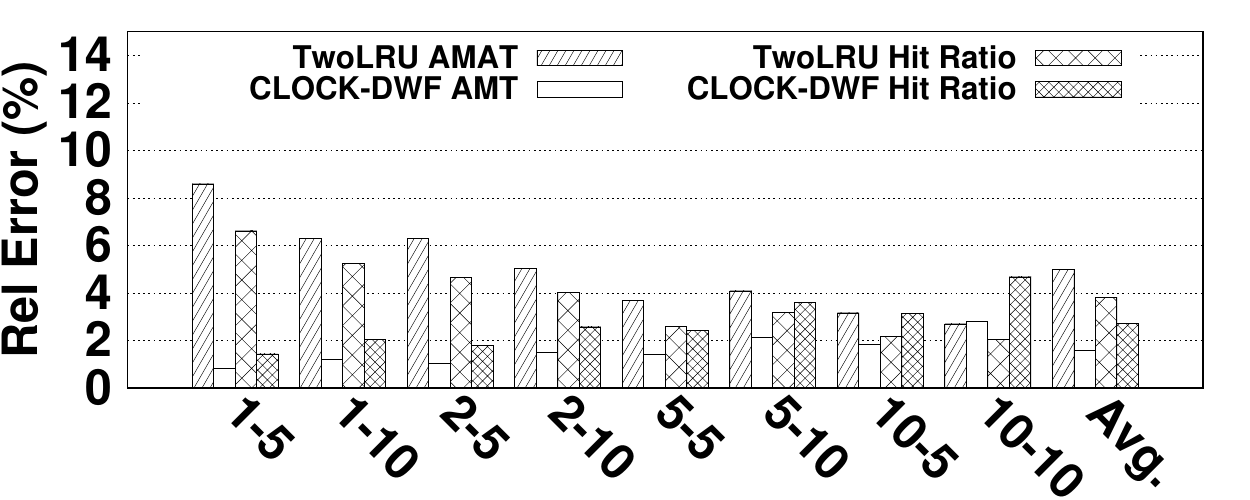}
				\label{fig:ferreterror}}
			\hfill
			\subfloat[fluidanimate]{\includegraphics[width=0.242\textwidth]{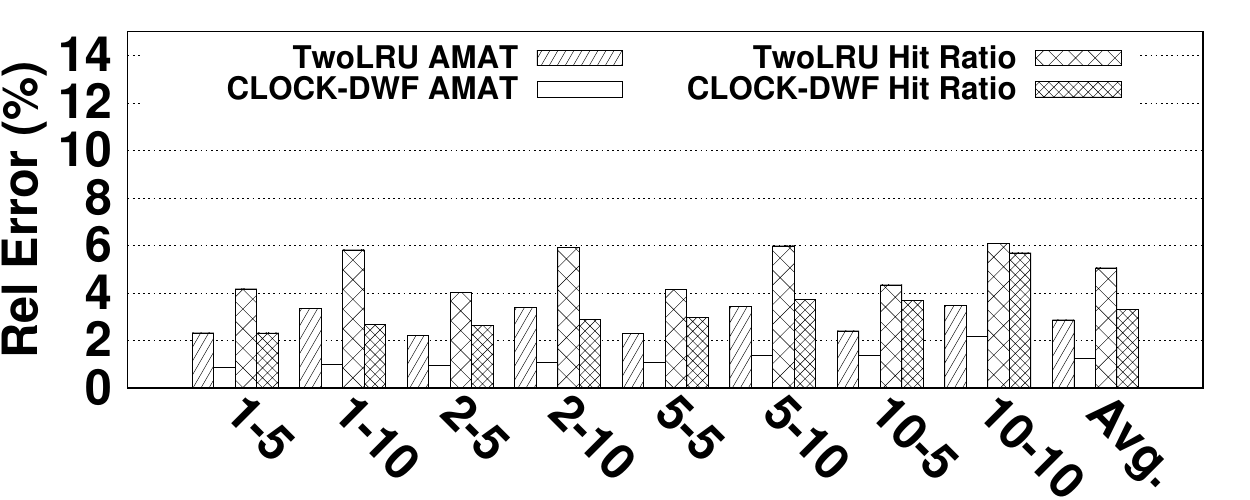}
				\label{fig:fluidanimateerror}}
			\hfill
			\subfloat[freqmine]{\includegraphics[width=0.242\textwidth]{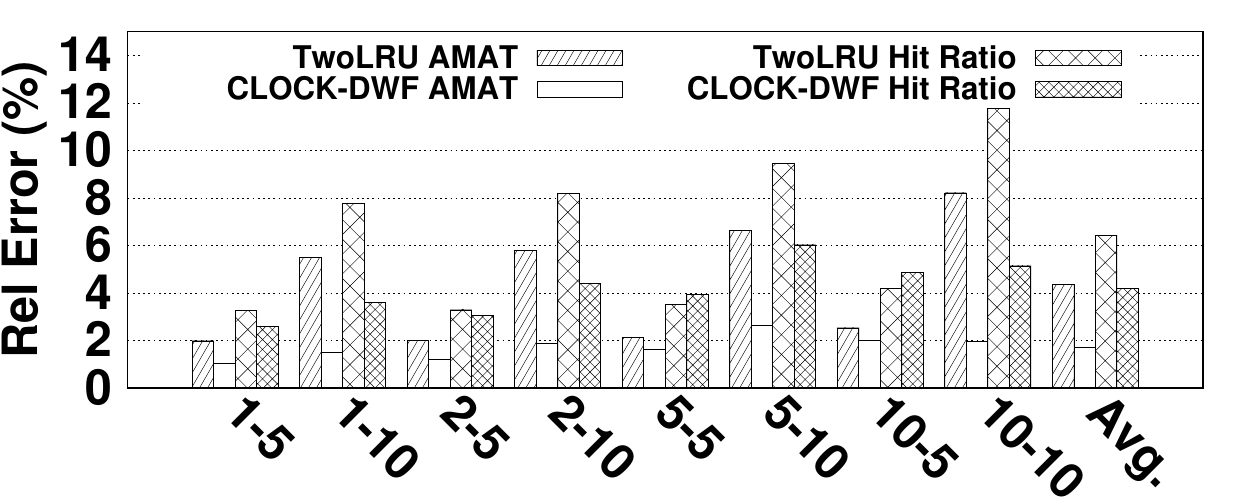}
				\label{fig:freqmineerror}}
			\hfill
			\vspace{-.5cm}
			\subfloat[raytrace]{\includegraphics[width=0.242\textwidth]{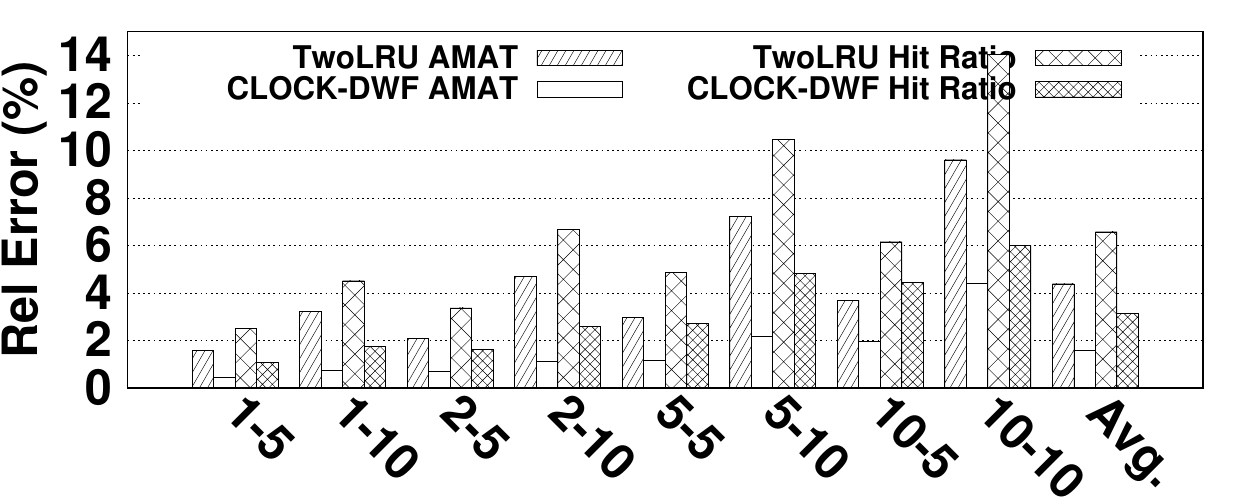}
				\label{fig:raytraceerror}}
			\hfill
			\subfloat[streamcluster]{\includegraphics[width=0.242\textwidth]{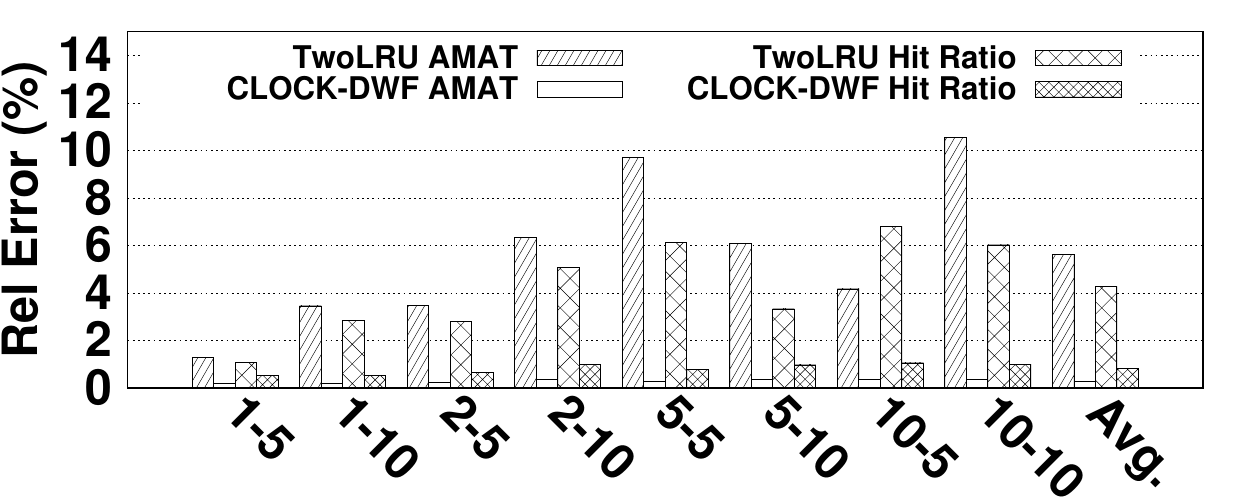}
				\label{fig:streamclustererror}}
			\hfill
			\subfloat[vips]{\includegraphics[width=0.242\textwidth]{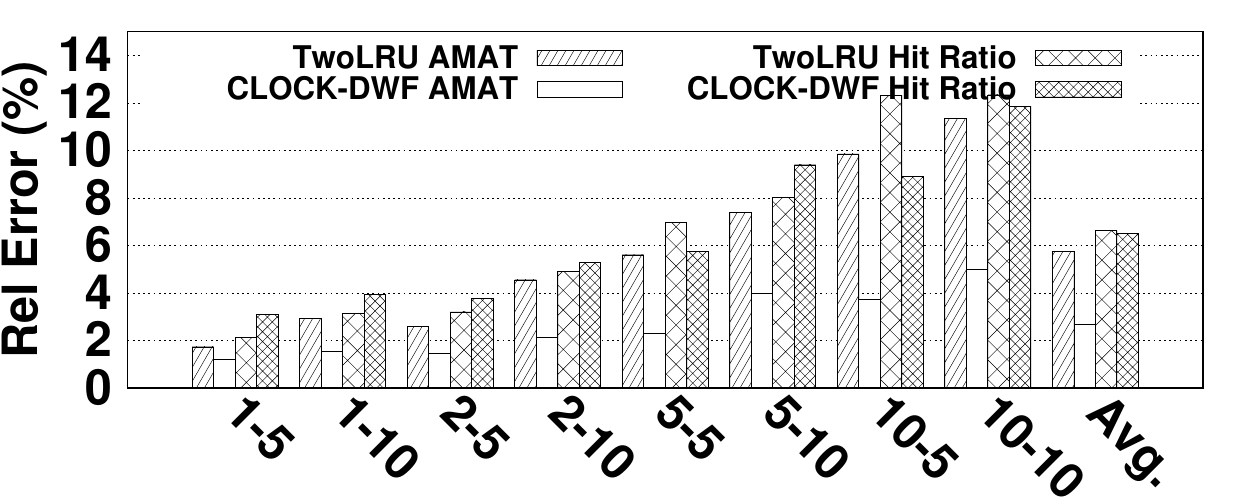}
				\label{fig:vipserror}}
			\hfill
			\subfloat[x264]{\includegraphics[width=0.242\textwidth]{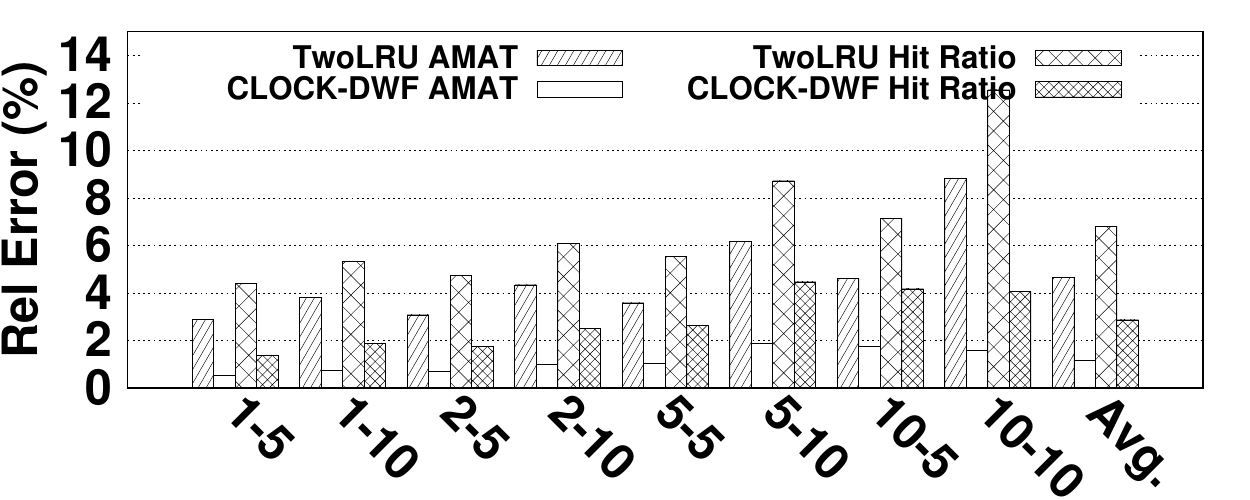}
				\label{fig:x264error}}
			\vspace{-.2cm}
			\caption{Error of proposed model for AMAT and hit ratio compared to 
			{simulation, for} various memory sizes. {A-B} values {on the x-axis} denote 
			DRAM and NVM sizes of {A\%} and {B\%} of the {workload's} 
			working set size, respectively.}
			\label{fig:totalerrorperf}
	\vspace{-0.5cm}
\end{figure*}

Most hybrid memories (including TwoLRU and CLOCK-DWF) implement a variation 
of either LRU or CLOCK as their DRAM and NVM eviction policies. 
Both policies have a deterministic algorithm for finding the 
victim data page. In LRU, the last data page in the queue and in CLOCK, 
the first data page with the reference bit set to 0 after 
the clock handle is evicted.
Thus, only one data page is eligible for eviction, which has an 
eviction probability of 1.
The eviction probability of all other data pages is 0.
TwoLRU employs LRU as the eviction policy of both DRAM and NVM.
Hence, the eviction probability for this architecture is the same as LRU.
The NVM clock in CLOCK-DWF is a simple CLOCK algorithm without any 
modifications and thus we use the simple eviction probability based 
on the clock algorithm in the experiments.
Although the CLOCK algorithm in DRAM {determines} the victim data page 
based on the values of counters stored for data pages, only 
one page is eligible for eviction.
The proposed analytical model, however, does \emph{not} store the actual values 
of counters and cannot determine the exact victim data page.
Instead of the actual values, an approximate order of data pages based on the counter values is stored by the proposed analytical model.
Based on this information, we can estimate the victim data page using two options: 
a) select the data page with the smallest counter value, and b) set a 
probability for data pages based on their order of counter values.
In the experiments, we choose the first option since it simplifies the 
computations and its output is closer to the actual output of CLOCK-DWF.

{The migration policy determines the probability that an access in 
NVM results in migrating the accessed data page to DRAM.
The migration probability is extracted from the examined hybrid memory by 
analyzing the migration policy and conditions resulting in a migration.
CLOCK-DWF migrates a data page from NVM to DRAM on a write 
access to the page.}
{Thus, the probability of migration is equal to the probability that an 
access is a write access}.
{In the proposed analytical model, we consider the write ratio of accesses in the trace file as the write probability and hence, 
the migration probability of CLOCK-DWF.
TwoLRU employs a counter-based method to decide whether or not the accessed data page should be migrated from NVM to DRAM.
Accurately formulating this method requires storing additional data, which significantly increases the memory and computation overhead of the proposed model.}
{To avoid such cost, we approximate the complex migration policy into a 
simple probability function.}
{To construct this function, we analyze the migration probability under 
various threshold counters, and construct a table {(Table \ref{tbl:migprobs})} that}
{translates the threshold counters to migration probabilities.}
{We use this table in our proposed analytical model.}
{Since we employ this simple function in our model for TwoLRU, we expect our 
model would have slightly higher accuracy in modeling of CLOCK-DWF than 
modeling TwoLRU.}
\\
\begin{table}[!h]
			\centering
			\caption{Migration probabilities for various thresholds in TwoLRU}
			\label{tbl:migprobs}
			\begin{tabular}{|c|c|}
				\hline
				\textbf{Threshold} & \textbf{Migration Probability} \\ \hline\hline
				1 & 0.16\\ \hline
				4 & 0.13\\ \hline
				8 & 0.08\\ \hline
				16 & 0.05\\ \hline
			\end{tabular}
\end{table}

\vspace{-.3cm}
\subsection{Accuracy}
\label{sec:accuracy}
\vspace{-.1cm}
The main goal of the proposed analytical model is to accurately estimate the 
hit ratio and performance of main memory under various {HMA} configurations.
{We evaluate the} accuracy of the proposed model {using} the error in 
the estimated hit ratio
and {estimated} performance, compared to the {baseline} simulation method.
Relative error is defined as the relative difference between {the} hit ratio 
{estimated} by the proposed model {($Hit_{estimated}$)} and the hit ratio 
measured {using} simulation {($Hit_{simulation}$)}, as denoted in 
Equation \ref{eq:error}.
{Relative error in estimated performance is measured similarly.}
In addition, {we measure the error of our model in} estimating NVM lifetime.

\begin{figure}[h]
	\vspace{-.3cm}
	\footnotesize
	\begin{align}
	\label{eq:error}
	& RelErrorHitRatio = \left| \frac{Hit_{estimated} -Hit_{simulation}}{Hit_{simulation}}  \right|
	\end{align}
\vspace{-.3cm}
\end{figure}

Fig. \ref{fig:totalerrorperf} depicts the relative error of the proposed model 
for estimating the hit ratio of {various} HMAs in {a variety of} 
workloads.
DRAM and NVM sizes are set relative to the working set {sizes} of benchmarks.
In most experiments, the proposed analytical model estimates the main memory hit ratio by less than 5\% error.
The average error is 4.61\% while the maximum error is 13.6\%.
{Due to} the simplifications {we employ} in {modeling its} migration 
algorithm, {TwoLRU} has {a} higher error rate compared to CLOCK-DWF, 
which also has a more steady error rate across various memory sizes.
\emph{Dedup} has a high average error {rate} in both HMAs, which is 
{because } it does not {have} {an} {easy-to-model} memory 
access 
distribution, unlike other programs in the PARSEC benchmark {suite}.
The abnormal behavior of {the} \emph{dedup} benchmark also {results} in high error 
in estimating $P_{HitDRAM|hit}$ as depicted in Fig. \ref{fig:pdramerror}.
Comparing Fig. \ref{fig:pdramerror} and Fig. \ref{fig:totalerrorperf} shows that most of the error in estimating $P_{HitDRAM|hit}$ is masked throughout {later} calculations {in our analytical model}.
Therefore, a more accurate and computation-intensive method for estimating $P_{HitDRAM|hit}$ will not significantly increase the accuracy of
{our} proposed model.

The error in estimating {the} hit ratio under different memory sizes does not follow a fixed pattern across workloads.
{We observe increasing}, decreasing, and stable error ratio {trends in} 
Fig. \ref{fig:totalerrorperf} {as we increase the} memory size.
Workloads with a {relatively} steady access distribution, such as 
\emph{canneal}, maintain {a stable} error ratio
across different memory sizes in TwoLRU.
Our analysis shows that workloads such as \emph{raytrace} and \emph{vips} have 
anomalies in their access distribution, which results in {higher} error 
{with larger} memory {sizes}.
Anomalies are less visible in CLOCK-DWF since it uses a more {deterministic} 
algorithm compared to TwoLRU.
Another interesting pattern in the experiments is the effect of {the} NVM/DRAM ratio on the hit ratio error.
In \emph{streamcluster}, experiments with {a} 1:1 DRAM/NVM ratio have higher error compared to {a} 1:2 ratio.
This is {due to} the {higher number of migrations in 1:2 ratio 
experiments.}


The main memory hit ratio cannot be used as the main source for analyzing the 
performance of HMAs since HMAs {tend} to have different performance behavior 
on NVMs with {different} performance characteristics.
To address this issue, the proposed model reports the number of accesses to each memory, which can be used to estimate the AMAT.
The formula for AMAT estimation is depicted in Equation \ref{eq:amat}.
{$Rlat_{x}$ and $Wlat_{x}$ denote the read and write latencies for device x, respectively.
Table \ref{tbl:notation}} {describes the notation} {employed in Equation 
\ref{eq:amat}.}
{We extract the} actual values for {average} DRAM and NVM response times 
from \cite{date16} and {show them} in Table \ref{tbl:memory}.
Fig. \ref{fig:totalerrorperf} presents the relative error of the proposed model 
in AMAT compared to {the baseline} simulation {method}, over various benchmarks and memory sizes.
The relative error in AMAT, which is more meaningful for HMA designers than {the relative error in} hit 
ratio, is lower {than the average error in hit ratio estimation} in 
all experiments.
This is {due to} the effect of very long latency of disk subsystem accesses for missed data pages on AMAT.
The proposed model has only 2.99\% average error in estimating performance 
of HMAs while the highest error is 11.3\%, in the \emph{vips} program.\footnote{{Appendix B presents the absolute} error values to estimate AMAT.}

\begin{figure}[!h]
	\scriptsize
	\vspace{-0.6cm}
	\begin{align}
	\label{eq:amat}
	& AMAT = \\ \nonumber
	& \frac{R\!l\!a\!t_{D}\! *\! R_{D}\! +\! W\!l\!a\!t_{D}\! *\! W_{D}\! +\! R\!l\!a\!t_{N}\! *\! R_{N}\! +\! W\!l\!a\!t_{N}\!\! *\!\! W_{N}\! \! +\!\! M\!i\!s\!s\! *\! R\!l\!a\!t_{Disk}}{\#\_of\_requests}
	\end{align}
	\vspace{-0.4cm}
\end{figure}

\begin{table}[!h]
	\caption{Latency of memory devices used in experiments 
		\cite{date16}}
	\vspace{-0.2cm}
	\label{tbl:memory}
	\centering
	\begin{tabular}{|c|c|}
		\hline
		\textbf{Device} &   \textbf{Latency r/w$(\eta s)$} \\  \hline\hline
		DRAM &   50/50  \\ \hline
		NVM & 100/350  \\     \hline
		Disk & 5,000,000  \\     \hline
	\end{tabular}
\end{table}

\begin{figure*}[!b]
	\centering
	\subfloat[blackscholes]{\includegraphics[width=0.242\textwidth]{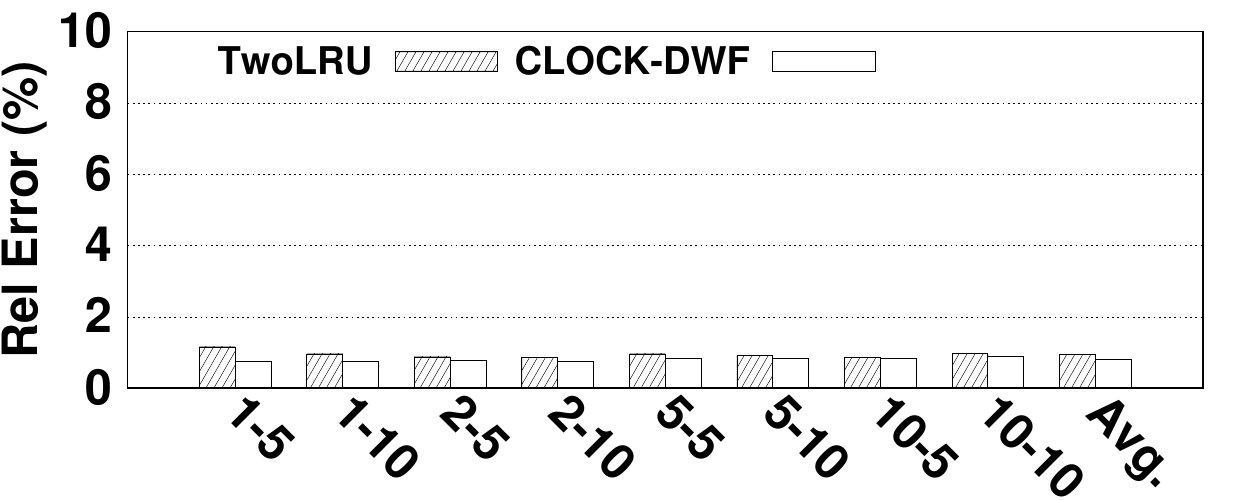}
		\label{fig:blackscholes-lifetime}}
	\hfill
	\subfloat[bodytrack]{\includegraphics[width=0.242\textwidth]{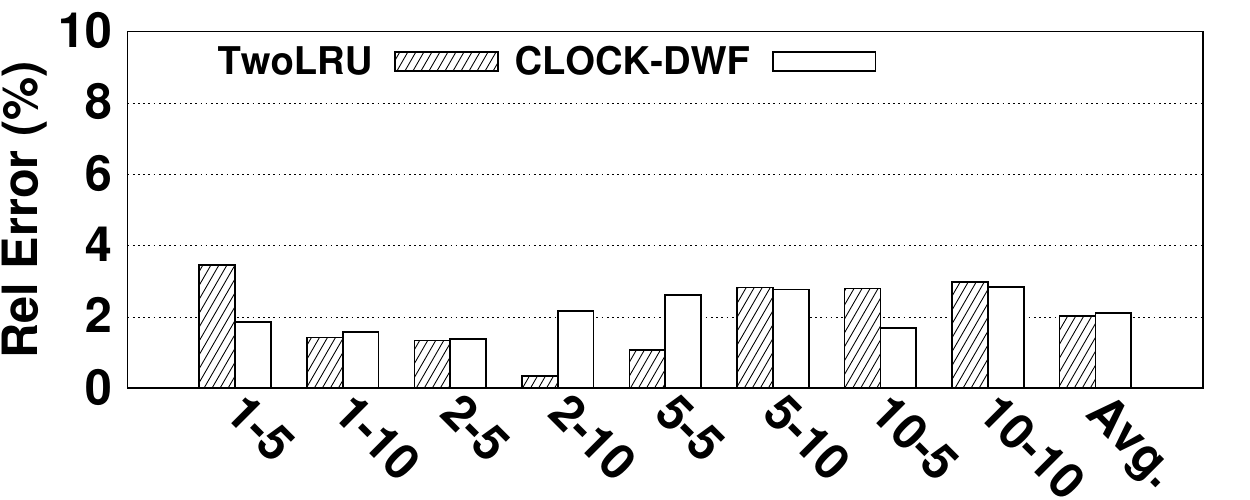}
		\label{fig:bodytrackerror-lifetime}}
	\hfill
	\subfloat[canneal]{\includegraphics[width=0.242\textwidth]{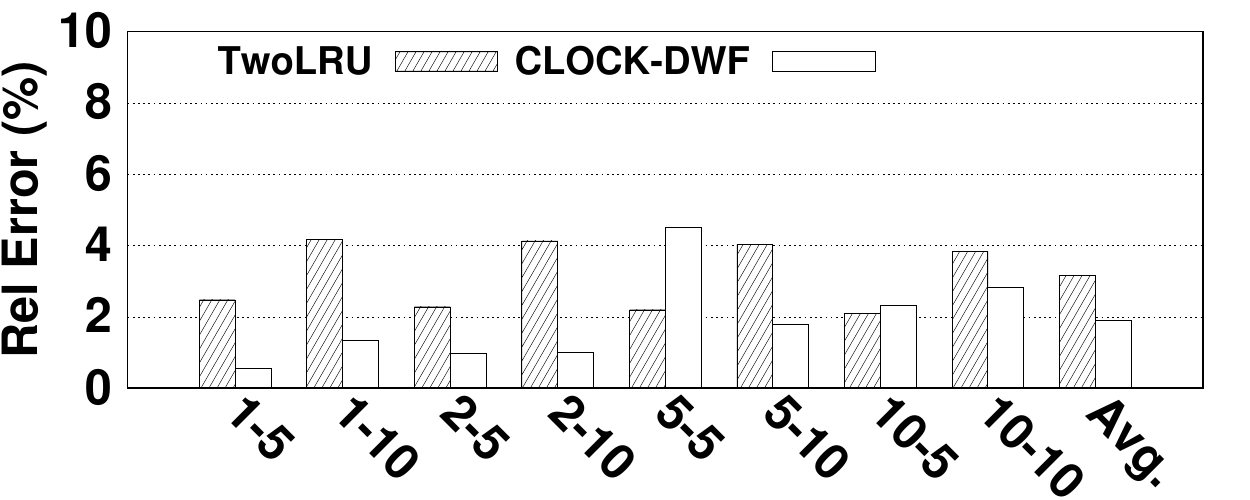}
		\label{fig:cannealerror-lifetime}}
	\hfill
	\subfloat[dedup]{\includegraphics[width=0.242\textwidth]{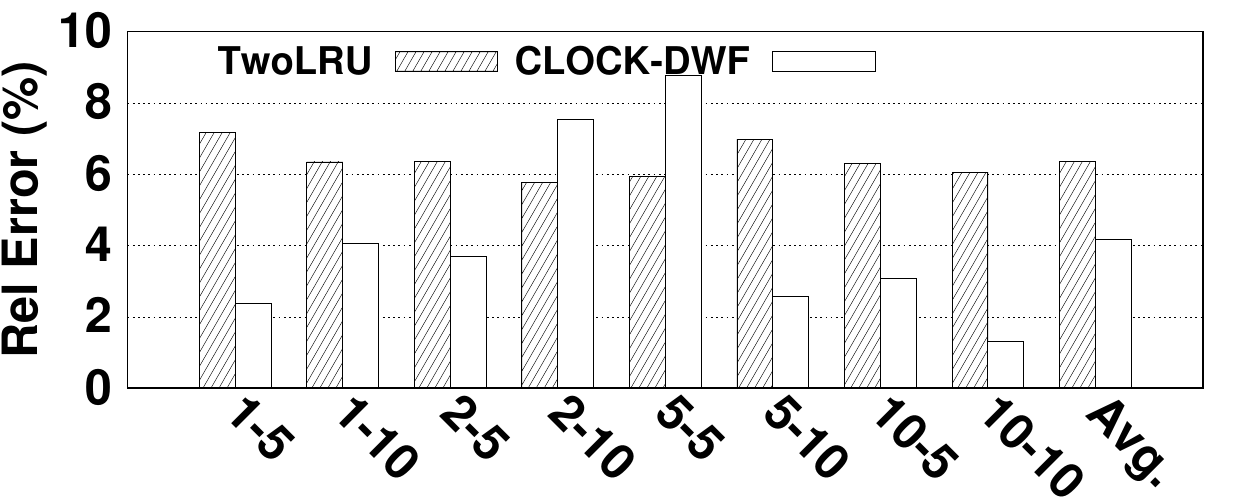}
		\label{fig:deduperror-lifetime}}
	\hfill
	\vspace{-.5cm}
	\subfloat[facesim]{\includegraphics[width=0.242\textwidth]{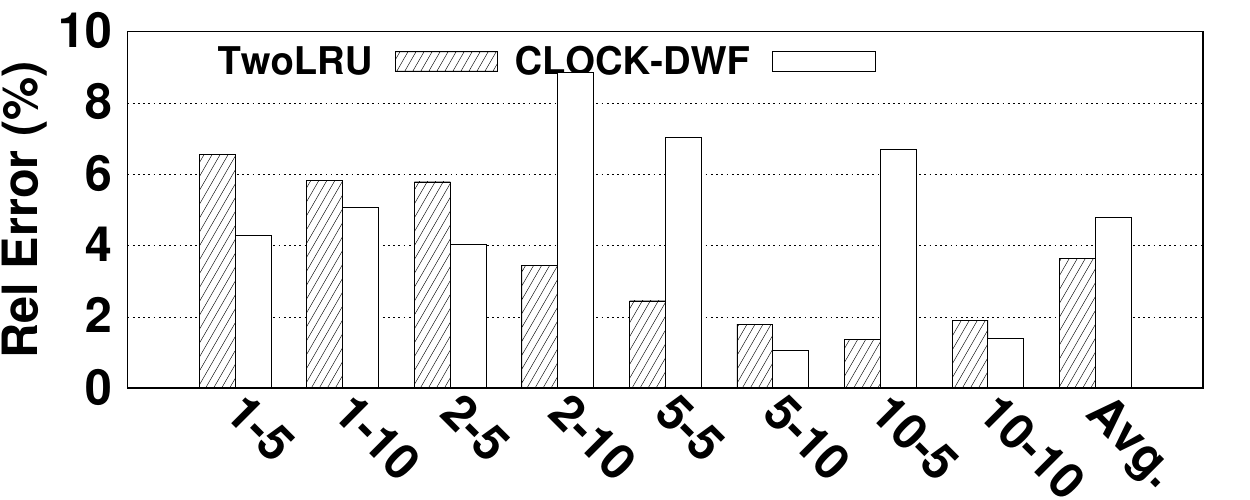}
		\label{fig:facesimerror-lifetime}}
	\hfill
	\subfloat[ferret]{\includegraphics[width=0.242\textwidth]{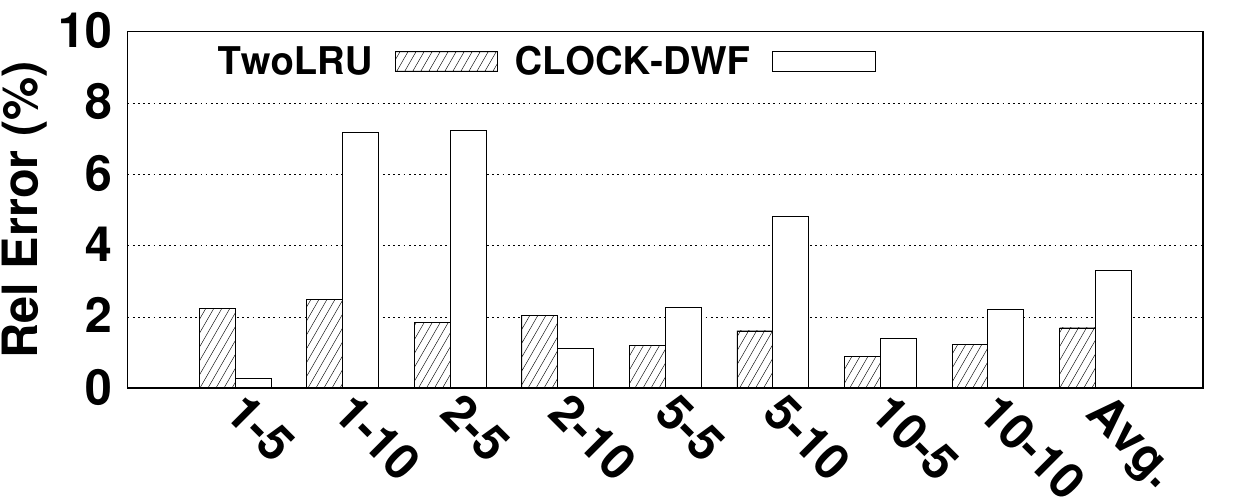}
		\label{fig:ferreterror-lifetime}}
	\hfill
	\subfloat[fluidanimate]{\includegraphics[width=0.242\textwidth]{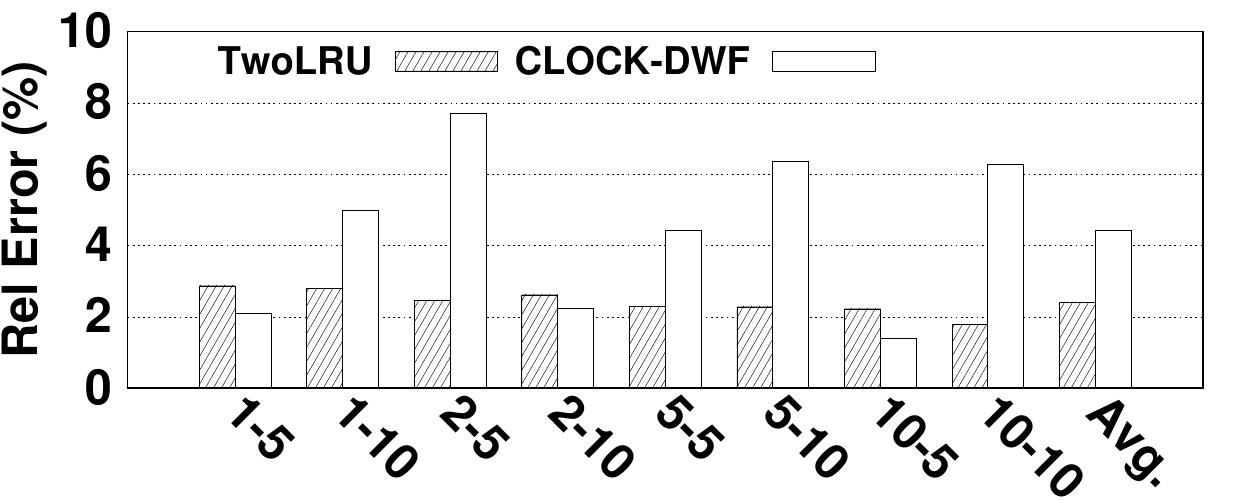}
		\label{fig:fluidanimateerror-lifetime}}
	\hfill
	\subfloat[freqmine]{\includegraphics[width=0.242\textwidth]{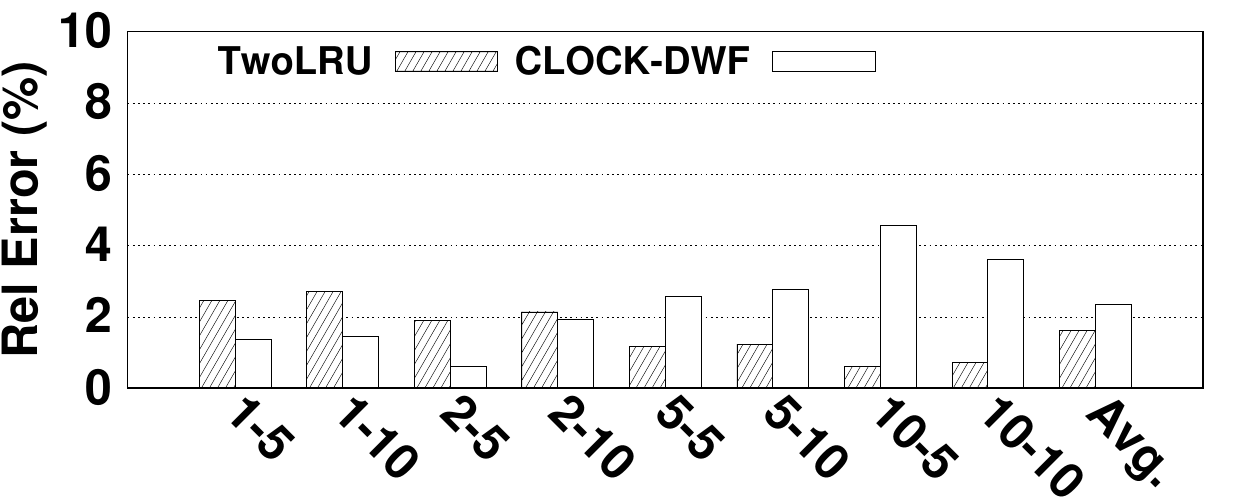}
		\label{fig:freqmineerror-lifetime}}
	\hfill
	\vspace{-.5cm}
	\subfloat[raytrace]{\includegraphics[width=0.242\textwidth]{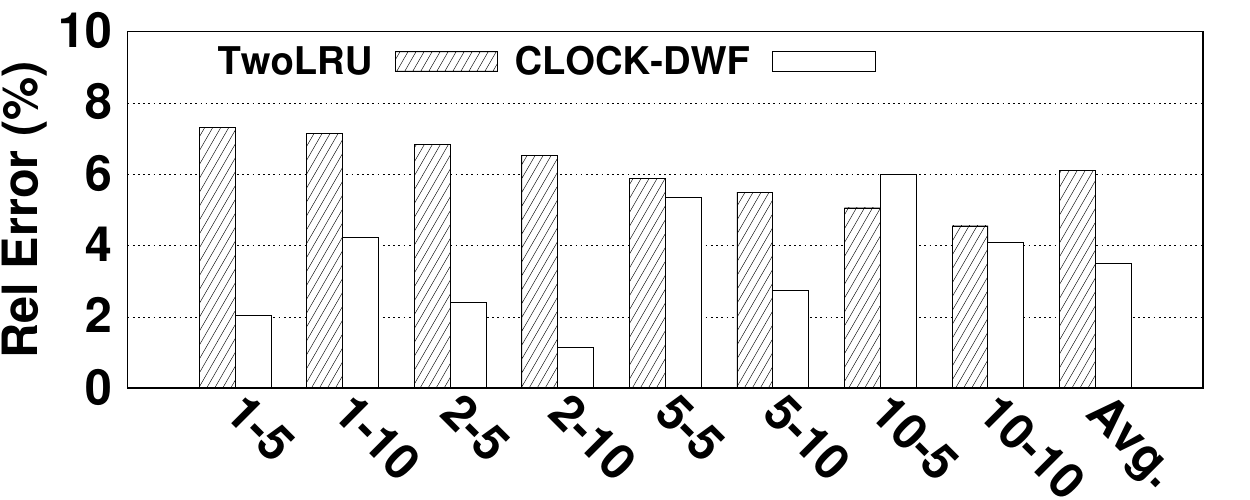}
		\label{fig:raytraceerror-lifetime}}
	\hfill
	\subfloat[streamcluster]{\includegraphics[width=0.242\textwidth]{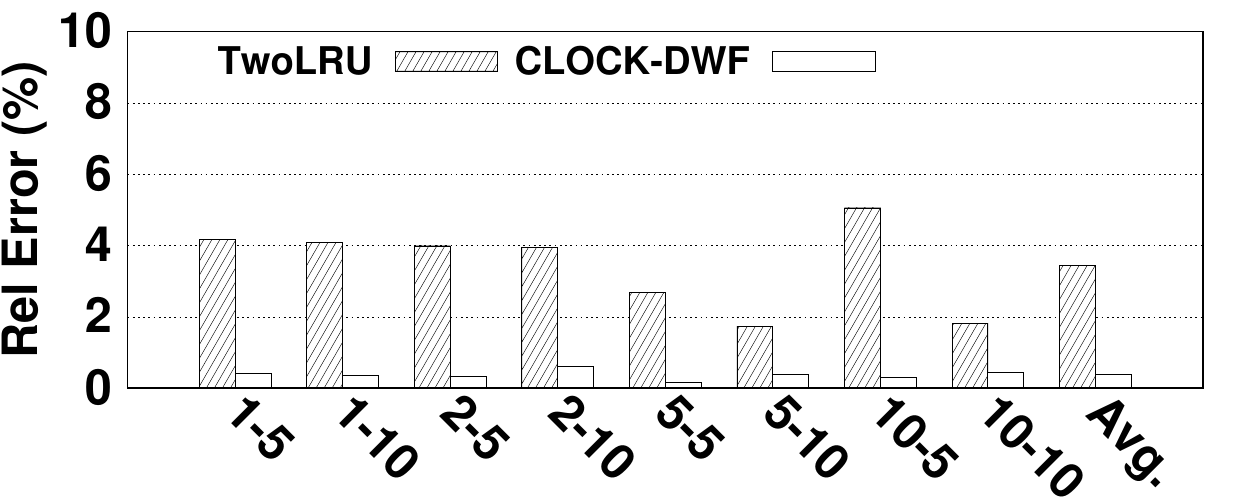}
		\label{fig:streamclustererror-lifetime}}
	\hfill
	\subfloat[vips]{\includegraphics[width=0.242\textwidth]{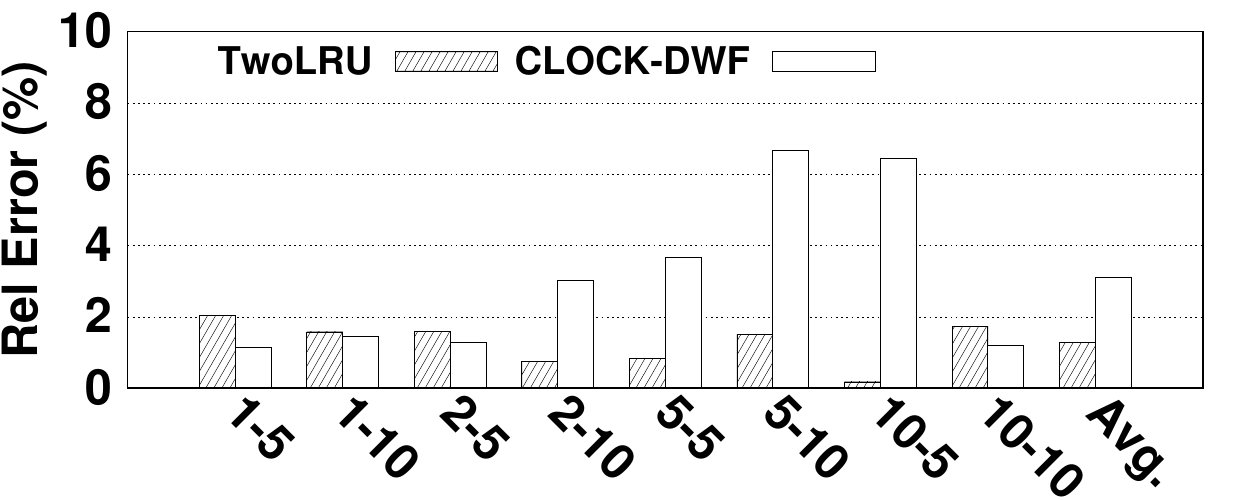}
		\label{fig:vipserror-lifetime}}
	\hfill
	\subfloat[x264]{\includegraphics[width=0.242\textwidth]{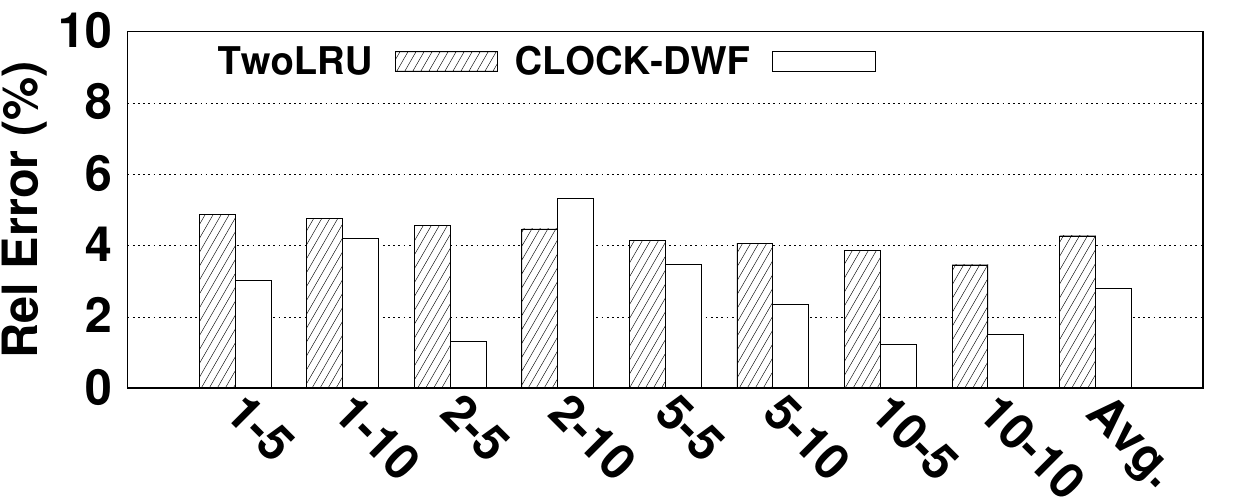}
		\label{fig:x264error-lifetime}}
	\vspace{-.2cm}
	\caption{Error of proposed model for NVM lifetime compared to simulation 
		method for various memory sizes. {A-B} values {on the x-axis} denote DRAM and NVM sizes 
		of {A\%} and {B\%} of the {workload's} working set size, 
		respectively.}
	\label{fig:totalerrorlifetime}
	\vspace{-0.5cm}
\end{figure*}

To evaluate the accuracy of the proposed model {in} estimating NVM lifetime, 
{the} number of write requests
issued to NVM is calculated using Equation \ref{eq:lifetime}.
$Pagefactor$ is the number of writes required for migrating a page, which is set to 64.
{$Mig_{toNVM}$ denotes the number of migrations from DRAM to NVM.}
CLOCK-DWF does not issue write requests directly to NVM and all writes in NVM are {due to a)} migrations from DRAM to NVM
and {b)} moving data pages from disk to NVM.
Therefore, evaluating the NVM lifetime in CLOCK-DWF directly evaluates the accuracy of estimating {the} number of migrations to NVM.
Unlike CLOCK-DWF, TwoLRU issues write requests to NVM.
Since estimating {the} number of write requests is relatively
easier than estimating migrations, {our model} is expected to have slightly 
higher accuracy {for} TwoLRU compared to {for} CLOCK-DWF.
Fig. \ref{fig:totalerrorlifetime} depicts the accuracy results for both examined HMAs.
CLOCK-DWF does not follow a fixed pattern across workloads and memory sizes.
The spikes in error rates
are generally cases where the number of migrations is low.
The error {in} estimating {the} NVM lifetime in TwoLRU follows a more stable pattern across different memory sizes
compared to estimating hit ratio and AMAT.
Both HMAs have \emph{high} relative error {rates} in {the} \emph{Dedup} 
benchmark, similar to the relative error in hit ratio {due to} the unique 
characteristics of this benchmark.
{The} average error of estimating {the} NVM lifetime is 2.93\%, while {the} maximum 
error is 8.8\%, which is lower than {the} error
in estimating hit ratio and AMAT.

\begin{figure}[h]
	\scriptsize
	\vspace{-0.6cm}
	\begin{align}
	\label{eq:lifetime}
	& Writes = W_{N} + Mig_{toNVM} * Pagefactor
	\end{align}
	\vspace{-0.6cm}
\end{figure}

\vspace{-.2cm}
\subsection{{Model Execution Time}}
\label{sec:computation}
{The proposed model consists of offline and online phases.
During the offline phase, all heavy computations for (r, u) 
pairs, which we described in Section \ref{sec:extract}, are conducted.}
{In the online (runtime) phase, only frequency of occurrence of pairs is 
computed to obtain the results.}
{Thus, the online phase does not have high complexity compared to 
simulation}, {which requires simulating all requests.}
In simulation methods, any modification in the trace file and/or the configuration of the examined architecture requires a re-run of the simulation, which is very time-consuming.
The proposed analytical model can reuse most of the required computations 
when such changes happen, {thereby} significantly reducing the 
runtime per experiment.
{As a practical example to show the required number of simulations, consider running TwoLRU on 12 traces as conducted in the previous work proposing this technique \cite{date16}.
For each workload, at least six combinations of thresholds need to be evaluated to find the most suitable threshold value.
Therefore, at least 72 simulations need to be run.
If different NVM sizes (e.g., 10 different values) are also required, this number will increase to more than 720 simulations.}
To calculate the reduction {in experimental evaluation} time due to using our proposed 
analytical 
model, {we run} each workload 50, 100, and 1000 times with various DRAM/NVM 
sizes 
and HMA configurations.

\begin{figure}[t]
	\centering
	\includegraphics[scale=0.65]{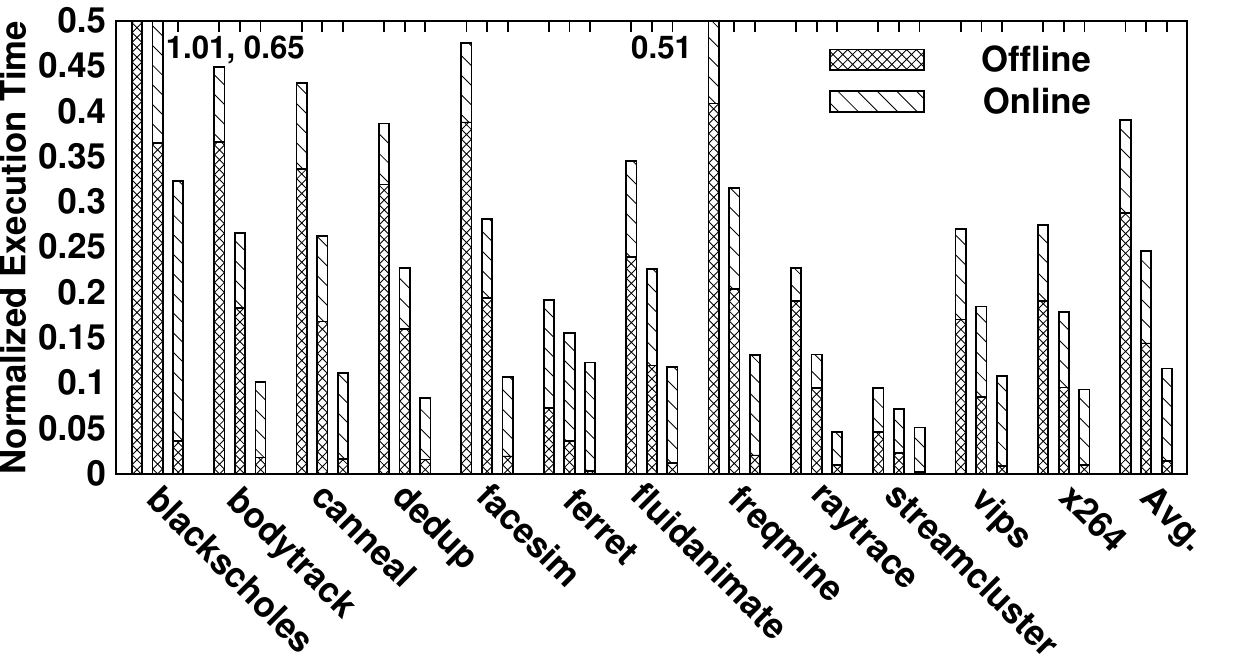}
	\caption{Normalized execution time of {the} proposed model compared to 
	simulation for three {numbers of} \emph{experiments per workload} (from 
	left to right 50, 100, and 1000)}
	\label{fig:time}
	\vspace{-0.5cm}
\end{figure}

Fig. \ref{fig:time} depicts the normalized execution time of the proposed 
analytical model compared to {simulation.}
The overhead of offline calculations {in} the proposed model is 
{amortized} over all experiments and is included in the execution time of 
the proposed model.
Since efficient data structures are employed throughout simulations, and 
both HMAs have almost equal simulation time, we normalize the simulation time 
to the average execution time of both HMAs	in Fig. \ref{fig:time}.
According to this figure, the online phase execution time is only 10\% of the simulation {time}, on average.
This shows that the runtime complexity of our proposed analytical 
model is much lower compared to simulation.
{The computations of the online phase can also be easily parallelized, 
which would further reduce the execution time of the online phase.}
{We leave this for future works.}
In workloads with larger number of requests, such as \emph{streamcluster}, the 
proposed model reduces the execution time by more than 10x {over simulation 
when we run} more than 1000 experiments.
Smaller workloads, such as \emph{blackscholes} {see smaller reduction in} 
execution time since their simulations are relatively fast.
{The highest} execution time {reductions} {are seen in} programs with long 
simulation {times}, which are more challenging programs to simulate {to 
begin with}.

\begin{figure}[!h]
		\centering
		\includegraphics[scale=0.68]{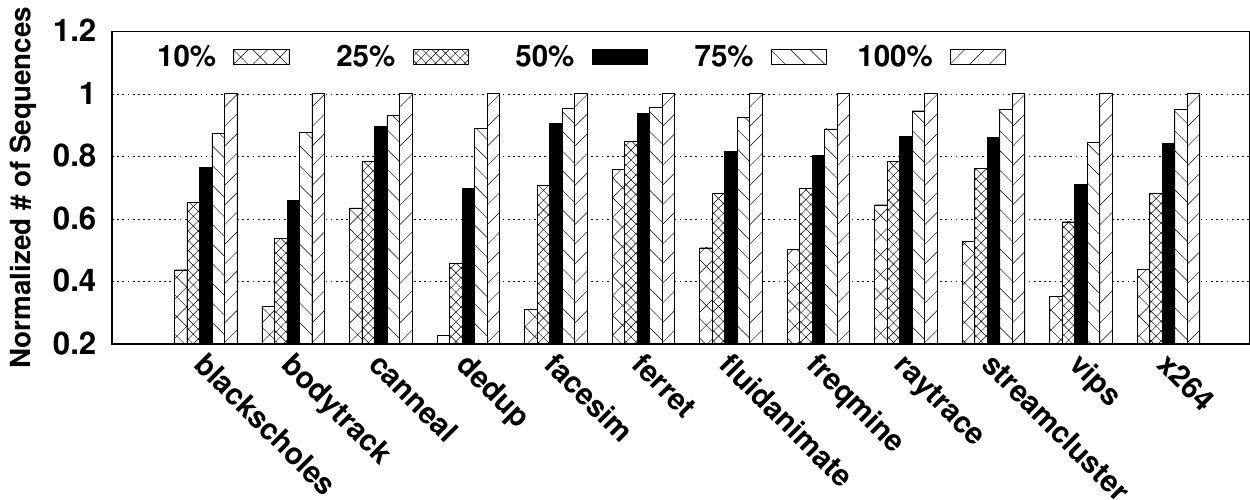}
		\caption{Normalized number of sequences observed for various percentages of 
			{the trace file length}}
		\label{fig:growth}
\vspace{-.3cm}
\end{figure}

{Another advantage of our proposed model over simulation is 
that its computational complexity grows more slowly with trace size.}
{The runtime complexity, i.e., the required computations to calculate the HMA hit ratio is not proportional to the number of distinct data pages accessed throughout the trace file.
Accessing thousands of different data pages \emph{only} increases the values of i and u in the extracted (i,u) pairs from the trace files.
For instance, consider a trace file accessing 1,000 different data pages.
The profiler extracts X (i,u) pairs from the trace file, where the runtime algorithm needs to solve each of them.
By adding Y accesses to the Z new distinct data pages to the trace file, the number of extracted (i,u) pairs by the profiler does not necessarily increase.
Since adding an access to a new data page in a sequence currently belonging to the (i,u) pair only reduces the number of occurrences of the (i,u) pair and also increases the number of occurrences of (i+1, u+1) pair.
There is, however, a small probability that the (i+1, u+1) pair does not exist in the extracted pairs from the original trace file.
In this case, the runtime complexity increases to compute the output of {the} (i+1, u+1) pair (from computing X pairs to X+1 pairs).
Note that the (i,u) pairs where u is greater than the size of the main memory are always a \emph{miss} and hence, the number of possible (i,u) pairs is limited.
In practice, due to the locality of the accesses of an application, the values of both i and u are rather small in most of the extracted pairs.}
To show the effect of increasing the trace size on the number of sequences and 
thus, on the execution time of the proposed model, we split the trace 
files and compute the number of required sequences for 10\%, 25\%, 50\%, 75\% 
and 100\% of the trace files.
As Fig. \ref{fig:growth} shows, the number of sequences does not 
significantly increase when trace size increases {once the trace file length is larger}.
Thus, {the} proposed analytical model has significant advantage over simulation 
{when using large trace files.
We conclude that the computational complexity of our proposed 
analytical model is much lower than that of simulation.}

\vspace{-.3cm}
\subsection{{Model Application}}
\label{sec:application}
\begin{figure}[!h]
	\hspace{-.5cm}
		\centering
		\includegraphics[scale=0.7]{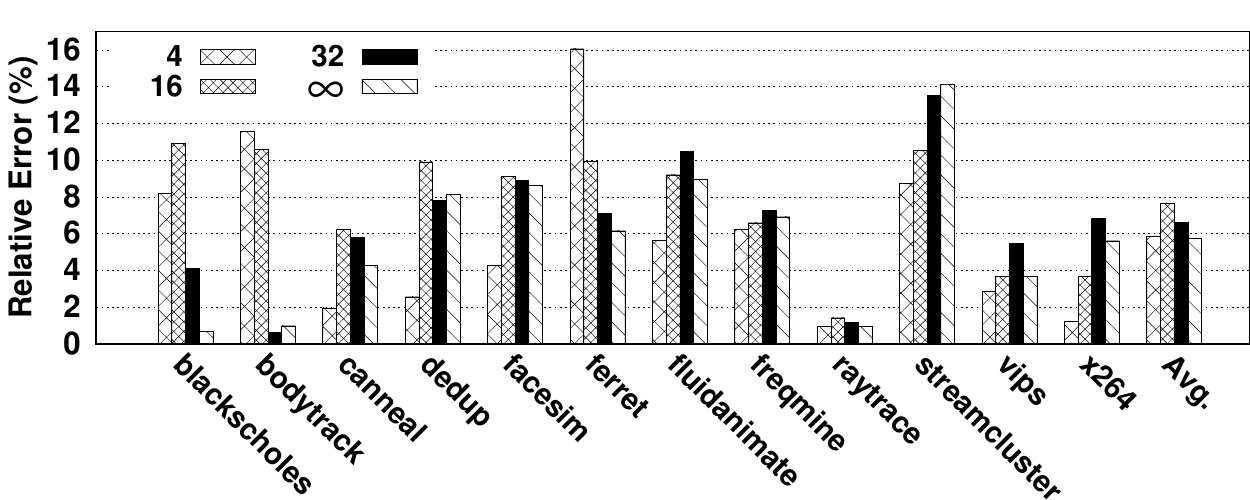}
		\caption{Relative error of performance estimation using the  
			formula provided by the proposed model compared to executing the proposed model 
			for each threshold value.}
		\label{fig:migerror}
\vspace{-.3cm}
\end{figure}

{As mentioned before, the proposed analytical model enables system} 
{architects to more} {easily identify the trade-offs and weaknesses of 
different hybrid memory architectures.
For instance, the analytical model can provide a formula that shows} 
the correlation between the migration threshold used in 
TwoLRU and {the} overall performance of the HMA.
{Using this data, we can predict the performance of TwoLRU without running the proposed analytical model for various migration thresholds, which further reduces the execution time of the proposed model.
Fig. \ref{fig:migerror} shows the accuracy of our model's performance 
estimates when we use} {the formula provided by our model.}
{We generate the formula using a migration threshold of \emph{one}.}
{We calculate performance for other threshold values based on the 
formula, without running the HMA through the proposed analytical model.
In most workloads, the relative error is less than 10\%, which shows the 
accuracy of the provided formula in estimating the performance of 
TwoLRU.
System architects can employ such a formula to easily find a} 
{suitable migration threshold for a vast range of workloads without} 
{the need for running simulations or even running the proposed 
analytical model for each possible migration threshold value.}

\vspace{-.3cm}
\subsection{Model Overheads}
\label{sec:overheads}
Most of the required computations of the proposed model can be {performed} 
offline and are reusable for any examined hybrid memory architecture and/or 
trace file.
The output of such computations should be stored {in} persistent storage for 
{repeated use}.
To reduce the required storage space, the {generated formulas for Markov 
states are} stored in gzip format on a hard disk, which can be extracted when 
needed.
{A} current dataset of {these formulas} alongside their outputs require less 
than 50GB of storage space, which can be either loaded into the main memory 
before starting
experiments or loaded on-demand.
{The storage overhead is independent} {of} {the memory size and} 
{trace size}.
{We employ a} \emph{dynamic programming} {approach 
\cite{bertsekas1995dynamic} in the proposed analytical model to reduce the 
required computations.}
{We use most of the storage space for storing the intermediate computation 
results.}
{The proposed model does not require all these data to be fully loaded in 
the memory. 
The proposed model can still operate without losing any accuracy with the cost of slightly higher computation time, if we do not have a large enough memory to hold all data.}

\section{Conclusion}
\label{sec:conclusion}
Hybrid DRAM-NVM main memory architectures have been proposed by system designers to exploit NVM benefits while reducing the effect of their negative characteristics.
Designing an efficient hybrid memory architecture requires {an exploration of} the design space of its architectural parameters, which is very time-consuming using traditional simulation methods.
This paper, for the first time, presents an analytical model for hybrid 
memory architectures, which significantly reduces the computation time for 
estimating the performance and lifetime of hybrid memories.
The proposed model is based on Markov decision processes and employs recursive state definitions.
Our model is capable of modeling various eviction and memory management policies and considers the interaction of memory modules when migrating data pages.
The proposed model is designed in such a way that most of the calculations can be reused across different trace files and/or HMAs.
Our model can also provide a formula to {evaluate} the effect of {different} migration thresholds on the overall HMA performance, which further reduces the required computations for analyzing HMAs.
Our experimental results demonstrate that the proposed model can accurately estimate the main memory hit ratio, and NVM lifetime with an average error of only 4.61\% and 2.93\%, respectively, while reducing
the computation time {by} up to 90\% over simulation based methods.

\ifCLASSOPTIONcompsoc
  \section*{Acknowledgments}
\else
  \section*{Acknowledgment}
\fi
This work has been partially supported by Iran National Science Foundation (INSF) under grant number 96006071 and by ETH Zurich and donations from various industrial partners, including Alibaba, Google, Huawei, HPDS, Intel, Microsoft, and VMware.
%
%
%
%



\bibliographystyle{IEEEtran}
\bibliography{IEEEabrv,ref}
\vskip -1.5\baselineskip plus -1fil
\begin{IEEEbiography}[{\includegraphics[width=1in,height=1.25in,clip,keepaspectratio]{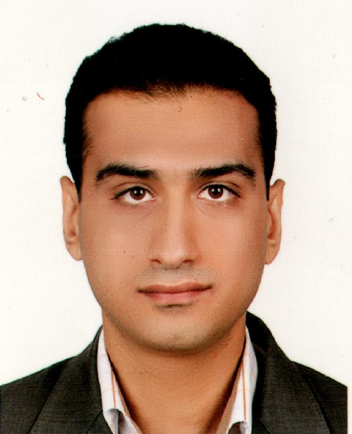}}]{Reza Salkhordeh}
	received the B.Sc. degree in computer engineering from Ferdowsi University of Mashhad in 2011, and M.Sc. degree in computer engineering from Sharif University of Technology (SUT) in 2013. He has been a member of \emph{Data Storage, Networks, and Processing} (DSN) lab since 2011. He was also a member of Iran National Elites Foundation from 2012 to 2015. He has been the director of Software division in HPDS corporation since 2015. He is currently a Ph.D. candidate at SUT. His research interests include operating systems, solid-state drives, memory systems, and data storage systems.
\end{IEEEbiography}
\vskip -3.5\baselineskip plus -1fil
\vfill
\begin{IEEEbiography}[{\includegraphics[width=1in,height=1.25in,clip,keepaspectratio]{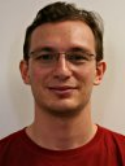}}]{Onur Mutlu}
is a Professor of Computer Science at ETH Zurich. He is
also a faculty member at Carnegie Mellon University, where he
previously held the Strecker Early Career Professorship.  His current
broader research interests are in computer architecture, systems,
hardware security, and bioinformatics. A variety of techniques he,
along with his group and collaborators, has invented over the years
have influenced industry and have been employed in commercial
microprocessors and memory/storage systems. He obtained his PhD and MS
in ECE from the University of Texas at Austin and BS degrees in
Computer Engineering and Psychology from the University of Michigan,
Ann Arbor. He started the Computer Architecture Group at Microsoft
Research (2006-2009), and held various product and research positions
at Intel Corporation, Advanced Micro Devices, VMware, and Google.  He
received the inaugural IEEE Computer Society Young Computer Architect
Award, the inaugural Intel Early Career Faculty Award, US National
Science Foundation CAREER Award, Carnegie Mellon University Ladd
Research Award, faculty partnership awards from various companies, and
a healthy number of best paper or "Top Pick" paper recognitions at
various computer systems, architecture, and hardware security
venues. He is an ACM Fellow "for contributions to computer
architecture research, especially in memory systems", IEEE Fellow for
"contributions to computer architecture research and practice", and an
elected member of the Academy of Europe (Academia Europaea). His
computer architecture and digital circuit design course lectures and
materials are freely available on YouTube, and his research group
makes a wide variety of software and hardware artifacts freely
available online. For more information, please see his webpage at
https://people.inf.ethz.ch/omutlu/.
\end{IEEEbiography}

\vskip -3.5\baselineskip plus -1fil
\vfill
\vspace{.6cm}

\begin{IEEEbiography}[{\includegraphics[width=1in,height=1.25in,clip,keepaspectratio]{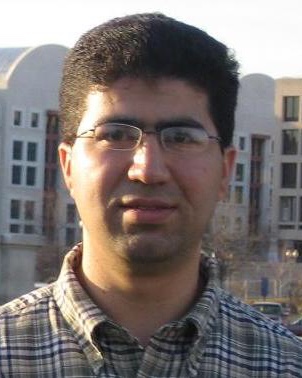}}]{Hossein Asadi}
	(M'08, SM'14) received his B.Sc. and M.Sc. degrees in computer engineering from the SUT, Tehran, Iran, in 2000 and 2002, respectively, and his Ph.D. degree in electrical and computer engineering from Northeastern University, Boston, MA, USA, in 2007. 
	
	He was with EMC Corporation, Hopkinton, MA, USA, as a Research Scientist and Senior Hardware Engineer, from 2006 to 2009. From 2002 to 2003, he was a member of the Dependable Systems Laboratory, SUT, where he researched hardware verification techniques. From 2001 to 2002, he was a member of the Sharif Rescue Robots Group. He has been with the Department of Computer Engineering, SUT, since 2009, where he is currently a tenured Associate Professor. He is the Founder and Director of the \emph{Data Storage, Networks, and Processing} (DSN) Laboratory, Director of Sharif \emph{High-Performance Computing} (HPC) Center, the Director of Sharif \emph{Information and Communications Technology Center} (ICTC), and the President of Sharif ICT Innovation Center. He spent three months in the summer 2015 as a Visiting Professor at the School of Computer and Communication Sciences at the Ecole Poly-technique Federele de Lausanne (EPFL). He is also the co-founder of HPDS corp., designing and fabricating midrange and high-end data storage systems. He has authored and co-authored more than eighty technical papers in reputed journals and conference proceedings. His current research interests include data storage systems and networks, solid-state drives, operating system support for I/O and memory management, and reconfigurable and dependable computing.
	
	Dr. Asadi was a recipient of the Technical Award for the Best Robot Design from the International RoboCup Rescue Competition, organized by AAAI and RoboCup, a recipient of Best Paper Award at the 15th CSI International Symposium on \emph{Computer Architecture \& Digital Systems} (CADS), the Distinguished Lecturer Award from SUT in 2010, the Distinguished Researcher Award and the Distinguished Research Institute Award from SUT in 2016, and the Distinguished Technology Award from SUT in 2017. He is also recipient of Extraordinary Ability in Science visa from US Citizenship and Immigration Services in 2008. He has also served as the publication chair of several national and international conferences including CNDS2013, AISP2013, and CSSE2013 during the past four years. Most recently, he has served as a Guest Editor of IEEE Transactions on Computers, an Associate Editor of Microelectronics Reliability, a Program Co-Chair of CADS2015, and the Program Chair of CSI National Computer Conference (CSICC2017). 
\end{IEEEbiography}

\appendices

\vspace{-.3cm}
\section{HMA Assumptions}
\label{sec:hmaassumption}
{We present} the assumptions {we make about} HMAs that can be evaluated by {our} proposed analytical model.
{We first present our definition of an HMA.}
Then, based on this definition, {we present} the assumptions {we made about} HMAs {that} enable us to accurately predict {the HMA performance and lifetime.}

\subsection{HMA Definition}
{An HMA} can be defined by a tuple \textless mapping, Upd, Mig, EvicP\textgreater:
\textbf{\emph{(a)}} \emph{mapping} stores {the} required information for managing memory (per data page), such as position in {the} LRU queue or reference and dirty bits alongside {the} clock handle position in the {CLOCK} algorithm.
Two subsets of \emph{mapping} can also be {present} to depict data pages residing in DRAM and NVM which are denoted as \emph{mapping.DRAM} and \emph{mapping.NVM}, respectively.
\textbf{\emph{(b)}} For each access to the memory subsystem, \emph{Upd(mapping, Page, \textoverline{NewMapping})}\footnote{Parameters with overline are outputs.} function called, which reconfigures the internal placement of data pages (setting ref bit in {CLOCK} and moving the data page to the \emph{Most Recently Used} (MRU) position in {the} LRU {queue}).
\emph{Upd} function calls other functions to decide {which pages to replace} in case of migration or free space shortage in memory.
\textbf{\emph{(c)}} Any required movement of data pages between memories {is} governed by {the} \emph{Mig(mapping, \textoverline{MigPage})} function, which decides whether or not a data page should be moved to another memory module.
\textbf{\emph{(d)}} In case of a miss access, \emph{EvicP(mapping, Page, \textoverline{Prob})} is called to calculate the eviction probability for data pages and deciding {which} data page should be evicted from the memory.

\begin{figure*}[t]
	\centering
	\subfloat[blackscholes]{\includegraphics[width=0.242\textwidth]{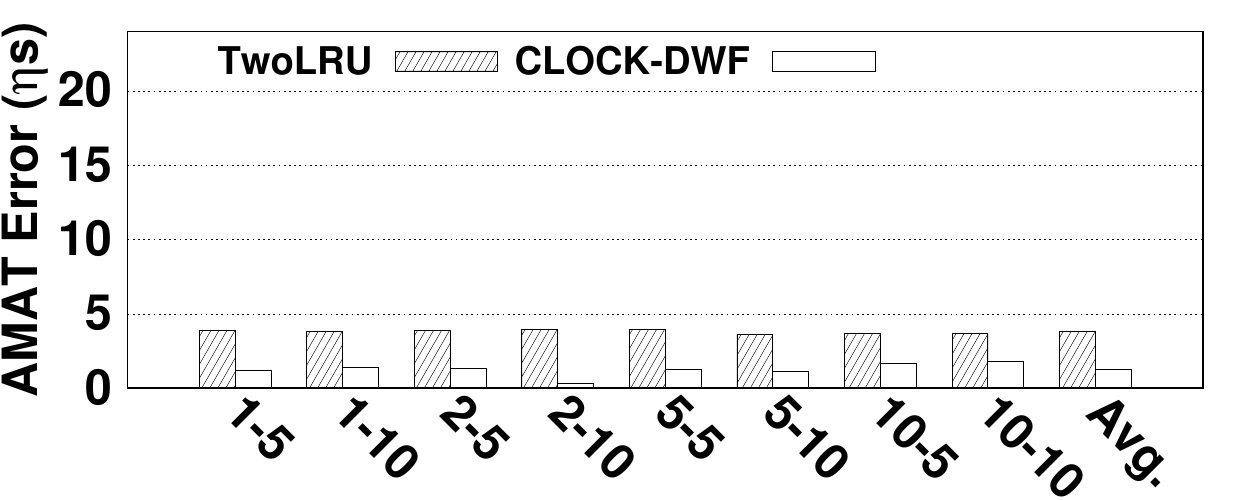}
		\label{fig:blackscholes-lifetime}}
	\hfill
	\subfloat[bodytrack]{\includegraphics[width=0.242\textwidth]{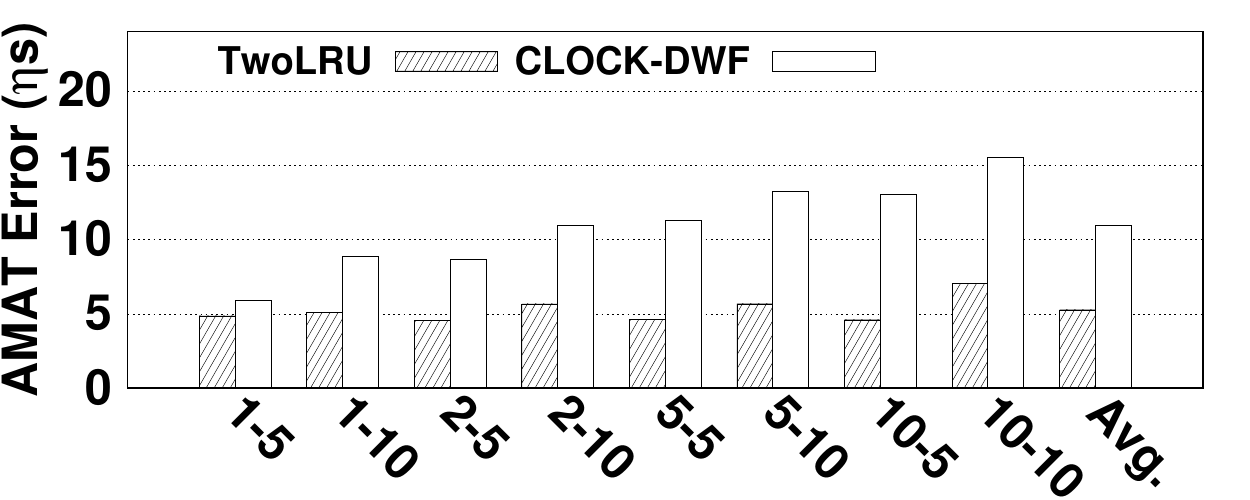}
		\label{fig:bodytrackerror-lifetime}}
	\hfill
	\subfloat[canneal]{\includegraphics[width=0.242\textwidth]{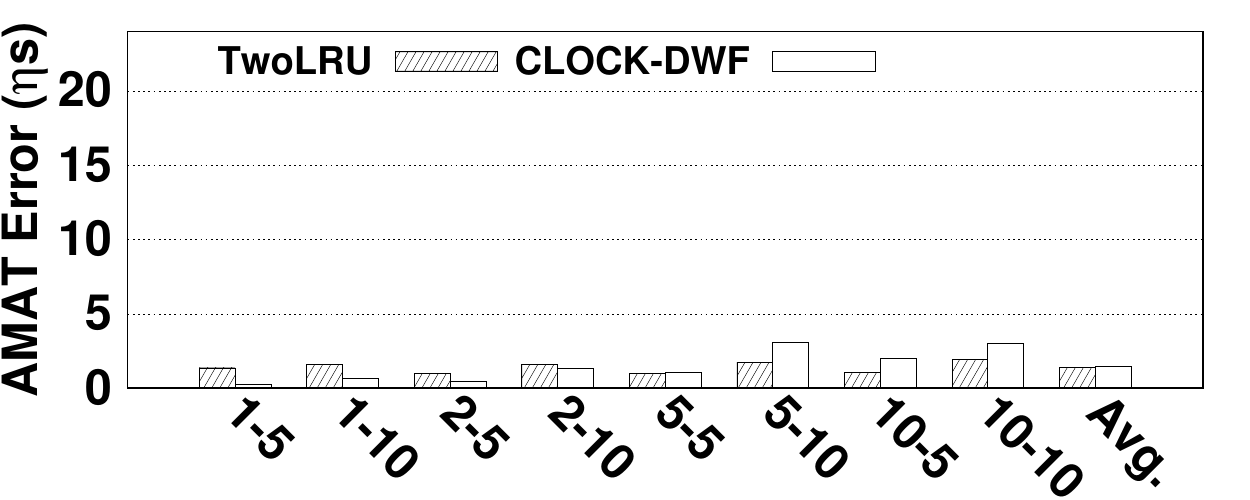}
		\label{fig:cannealerror-lifetime}}
	\hfill
	\subfloat[dedup]{\includegraphics[width=0.242\textwidth]{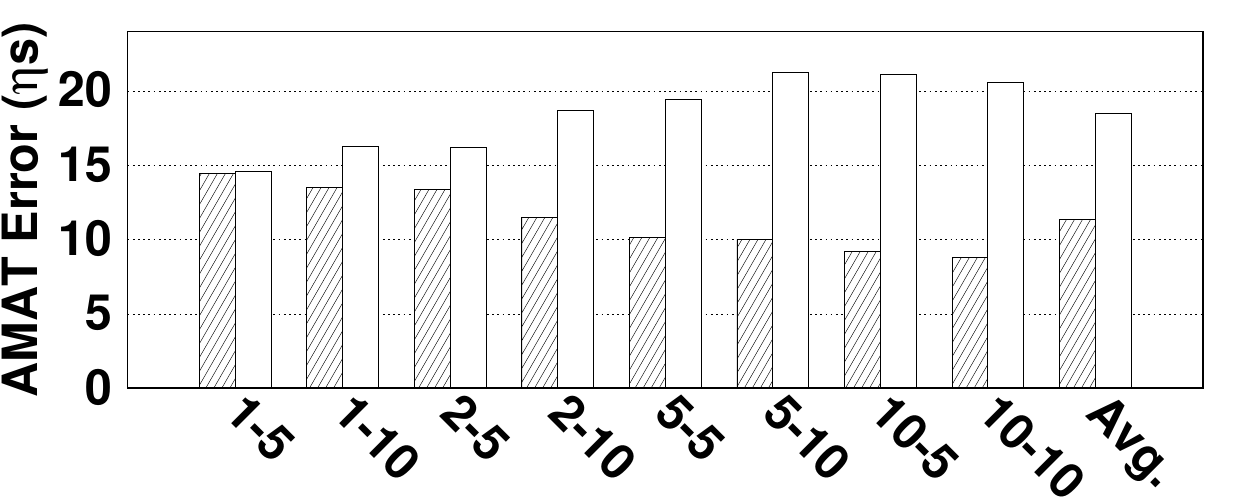}
		\label{fig:deduperror-lifetime}}
	\hfill
	\vspace{-.5cm}
	\subfloat[facesim]{\includegraphics[width=0.242\textwidth]{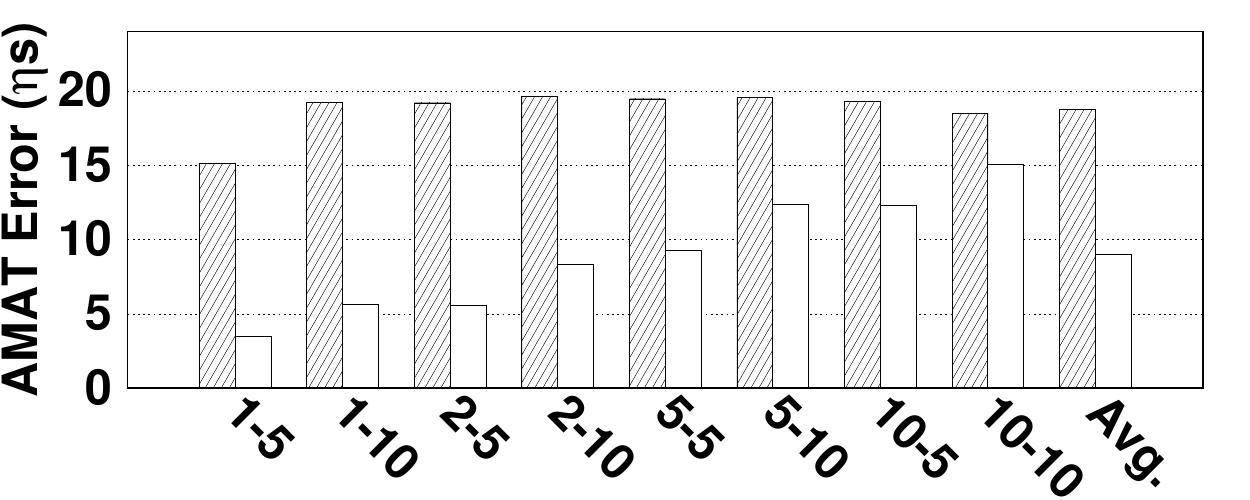}
		\label{fig:facesimerror-lifetime}}
	\hfill
	\subfloat[ferret]{\includegraphics[width=0.242\textwidth]{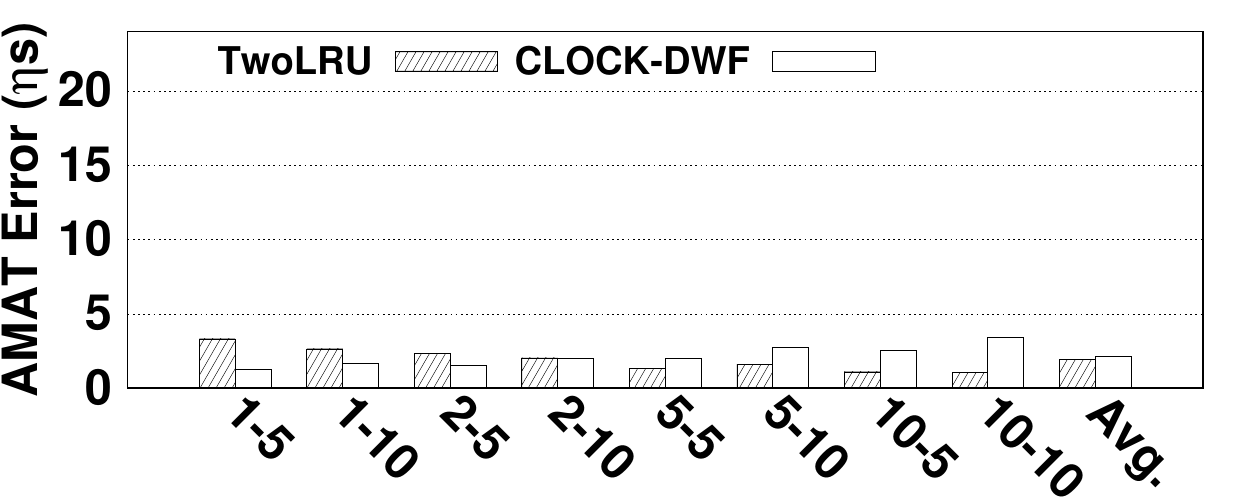}
		\label{fig:ferreterror-lifetime}}
	\hfill
	\subfloat[fluidanimate]{\includegraphics[width=0.242\textwidth]{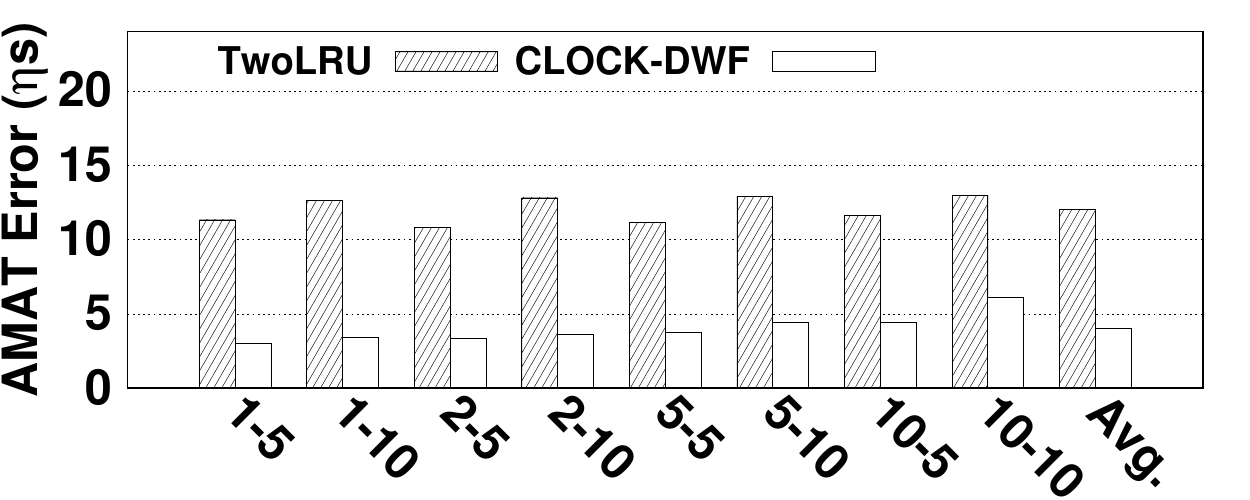}
		\label{fig:fluidanimateerror-lifetime}}
	\hfill
	\subfloat[freqmine]{\includegraphics[width=0.242\textwidth]{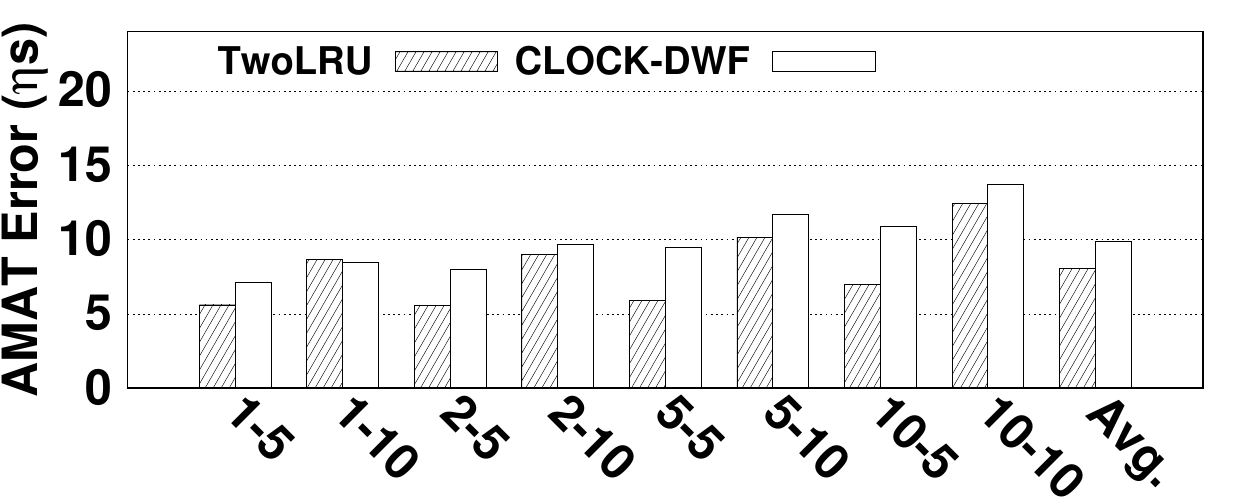}
		\label{fig:freqmineerror-lifetime}}
	\hfill
	\vspace{-.5cm}
	\subfloat[raytrace]{\includegraphics[width=0.242\textwidth]{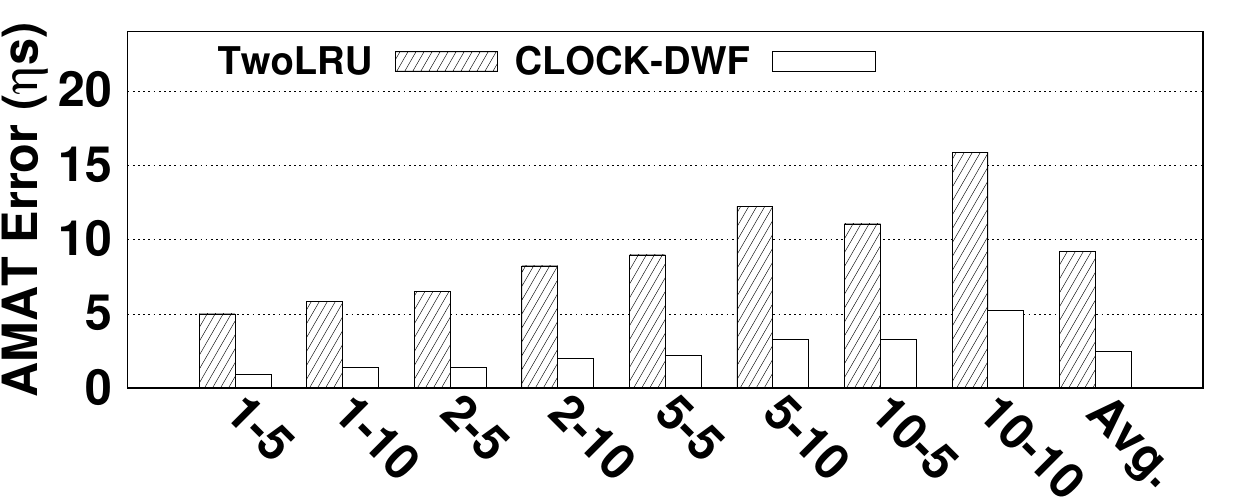}
		\label{fig:raytraceerror-lifetime}}
	\hfill
	\subfloat[streamcluster]{\includegraphics[width=0.242\textwidth]{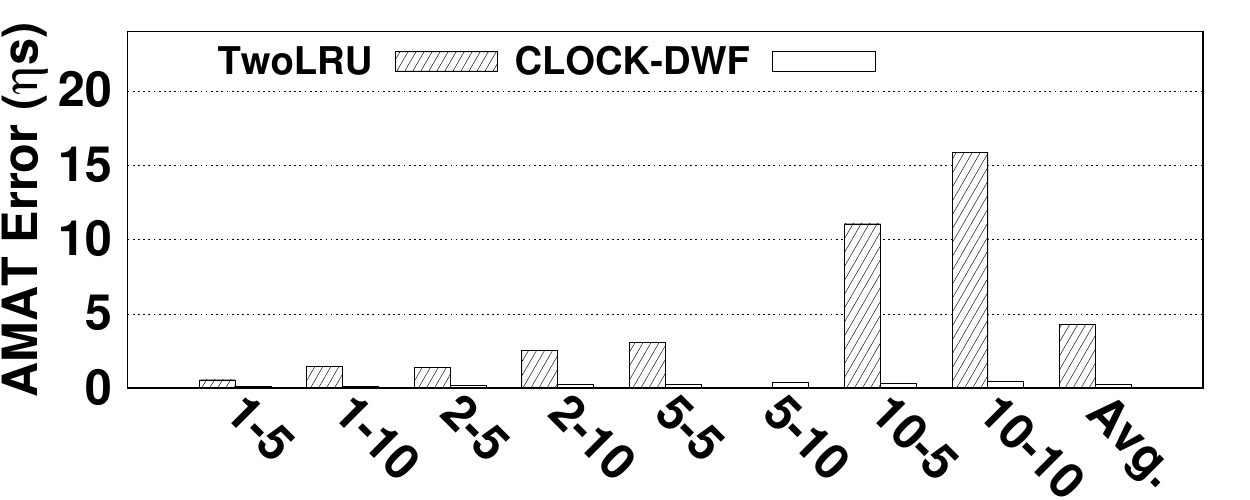}
		\label{fig:streamclustererror-lifetime}}
	\hfill
	\subfloat[vips]{\includegraphics[width=0.242\textwidth]{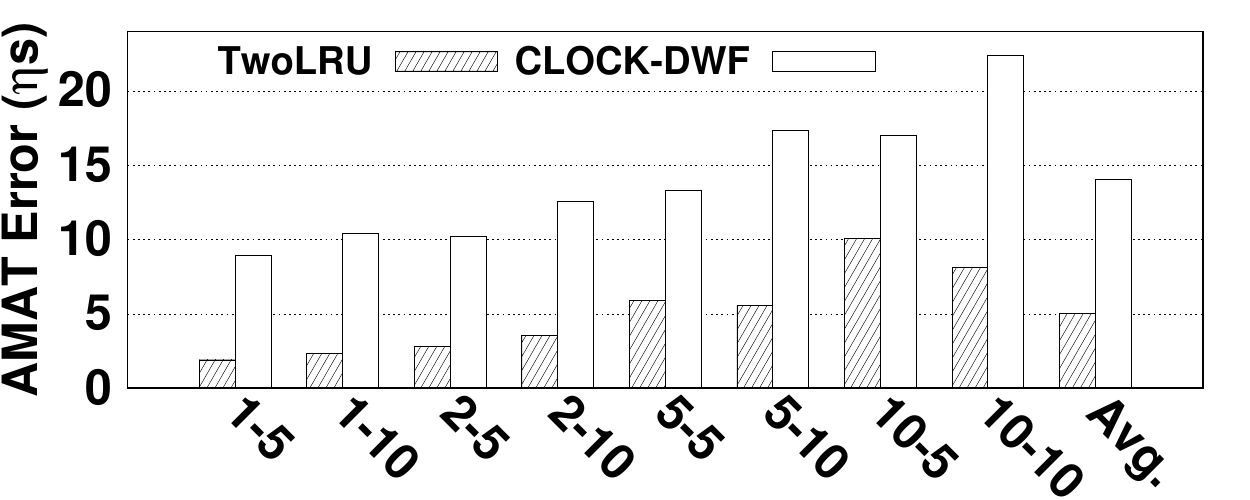}
		\label{fig:vipserror-lifetime}}
	\hfill
	\subfloat[x264]{\includegraphics[width=0.242\textwidth]{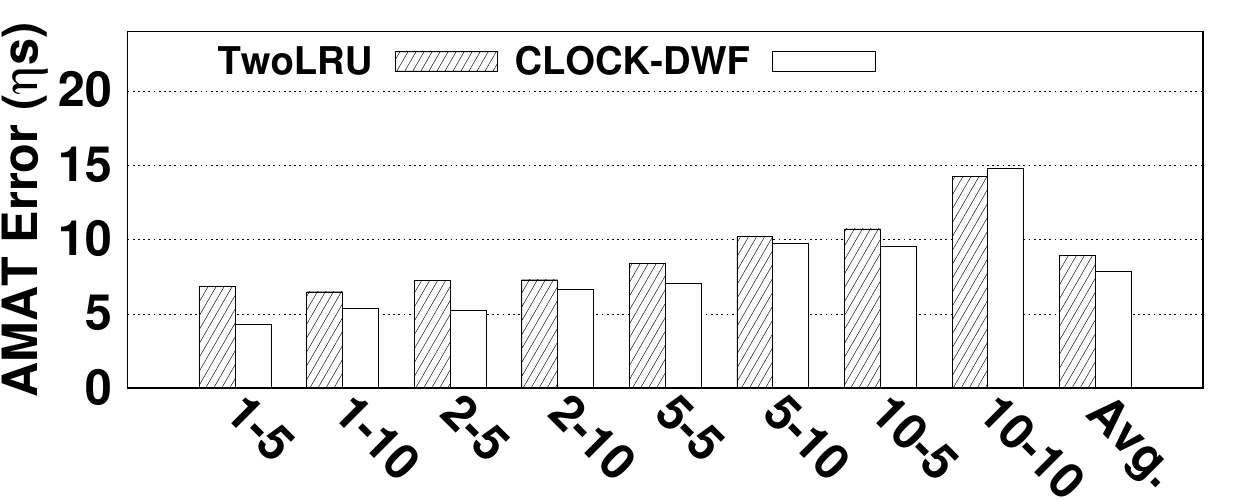}
		\label{fig:x264error-lifetime}}
	\vspace{-.2cm}
	\caption{Absolute Response Time}
	\label{fig:absresponse}
	\vspace{-0.5cm}
\end{figure*}

\begin{figure*}[t]
	\centering
	\subfloat[blackscholes]{\includegraphics[width=0.242\textwidth]{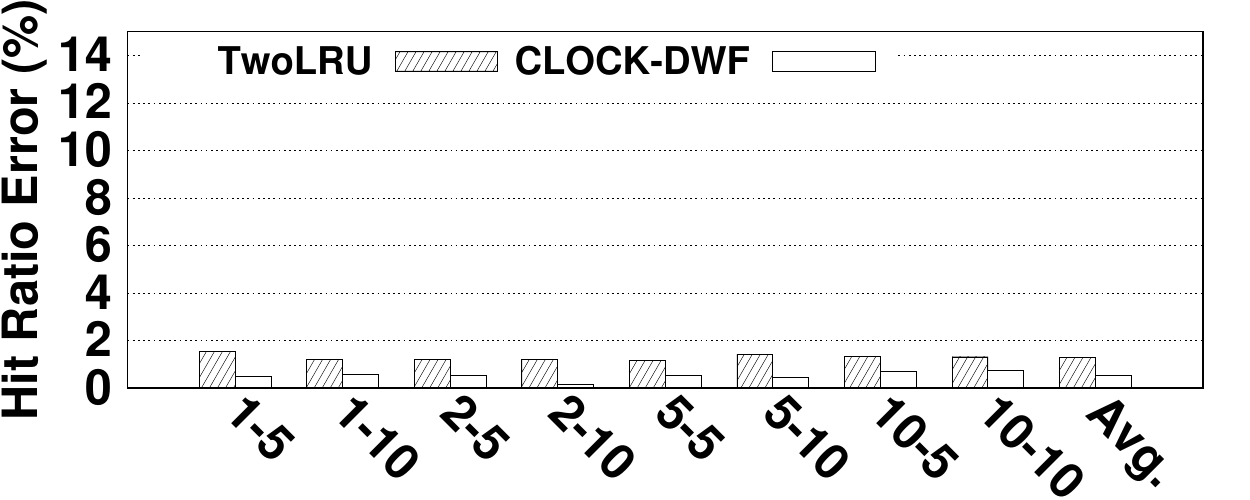}
		\label{fig:blackscholes-lifetime}}
	\hfill
	\subfloat[bodytrack]{\includegraphics[width=0.242\textwidth]{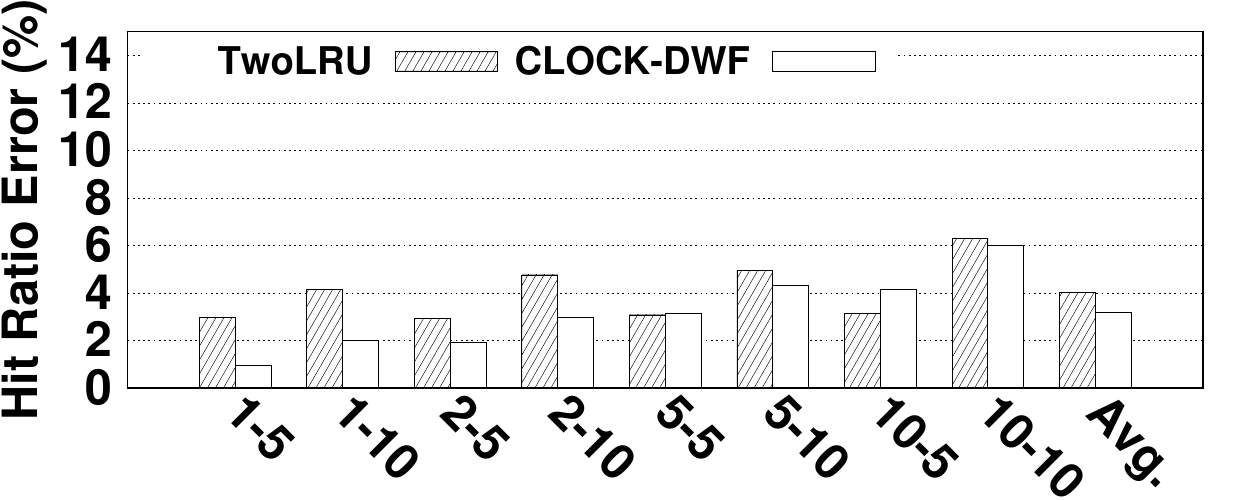}
		\label{fig:bodytrackerror-lifetime}}
	\hfill
	\subfloat[canneal]{\includegraphics[width=0.242\textwidth]{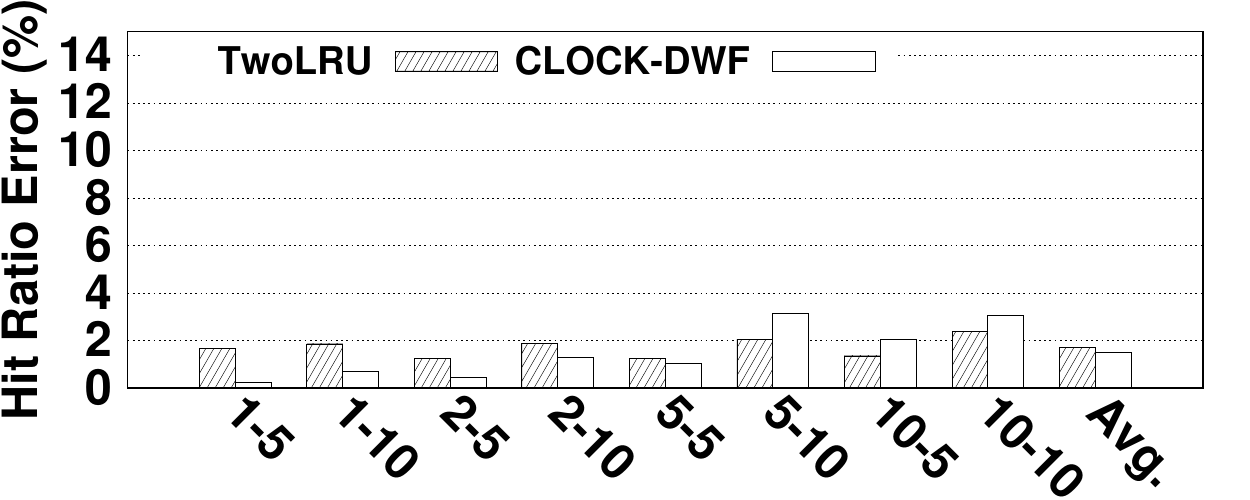}
		\label{fig:cannealerror-lifetime}}
	\hfill
	\subfloat[dedup]{\includegraphics[width=0.242\textwidth]{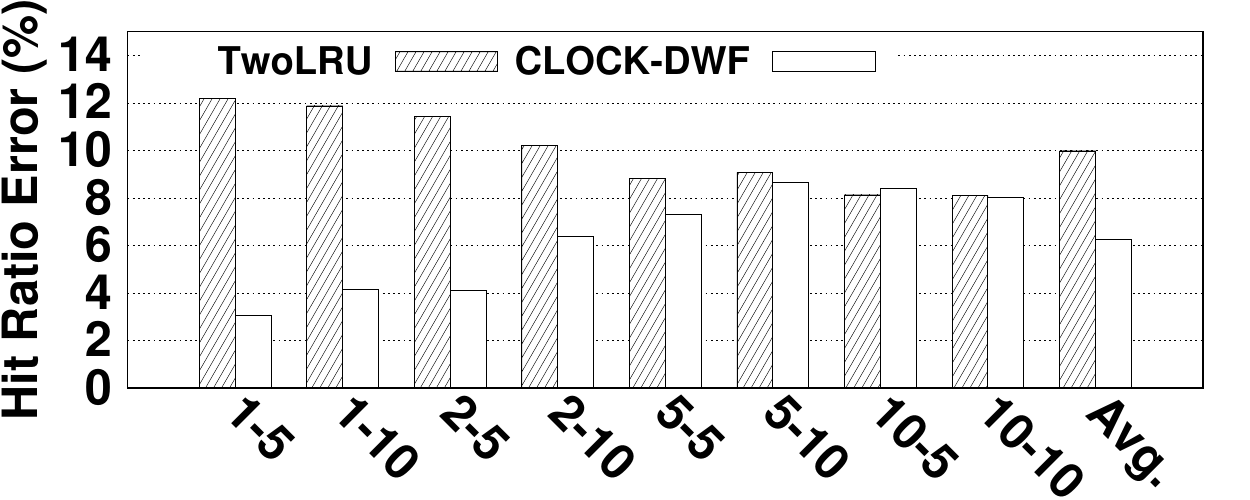}
		\label{fig:deduperror-lifetime}}
	\hfill
	\vspace{-.5cm}
	\subfloat[facesim]{\includegraphics[width=0.242\textwidth]{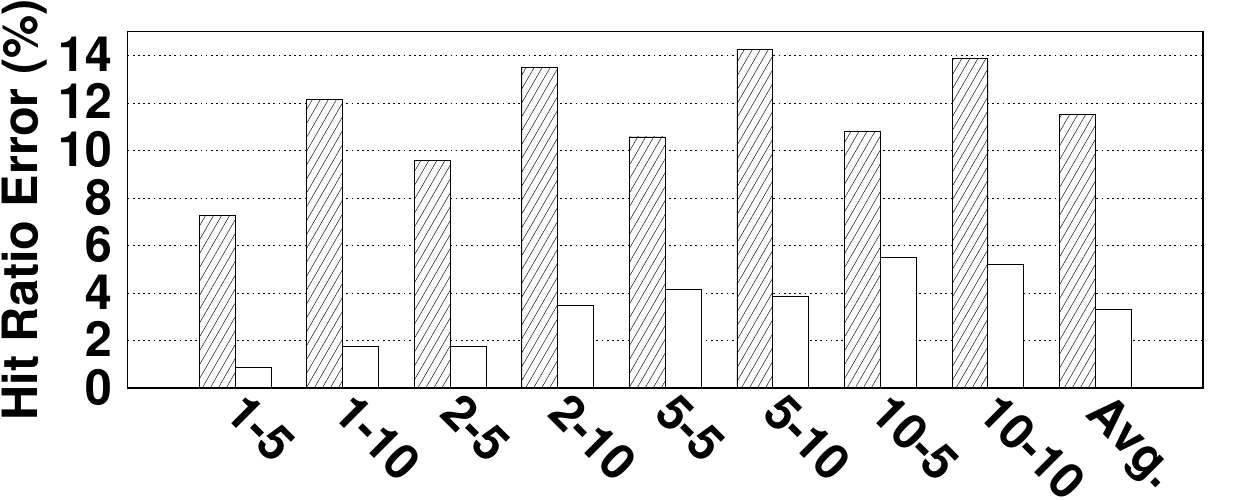}
		\label{fig:facesimerror-lifetime}}
	\hfill
	\subfloat[ferret]{\includegraphics[width=0.242\textwidth]{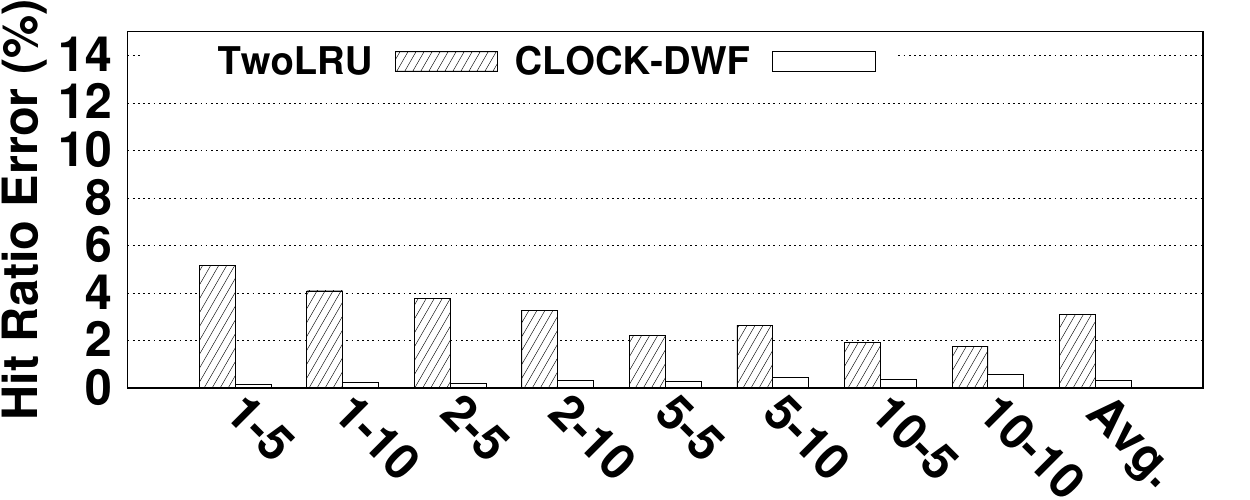}
		\label{fig:ferreterror-lifetime}}
	\hfill
	\subfloat[fluidanimate]{\includegraphics[width=0.242\textwidth]{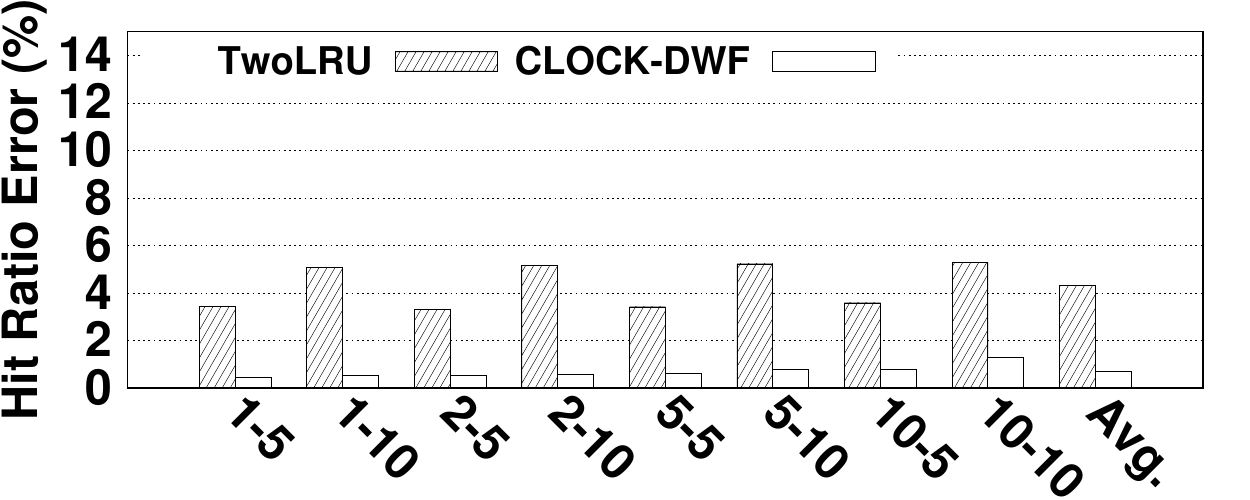}
		\label{fig:fluidanimateerror-lifetime}}
	\hfill
	\subfloat[freqmine]{\includegraphics[width=0.242\textwidth]{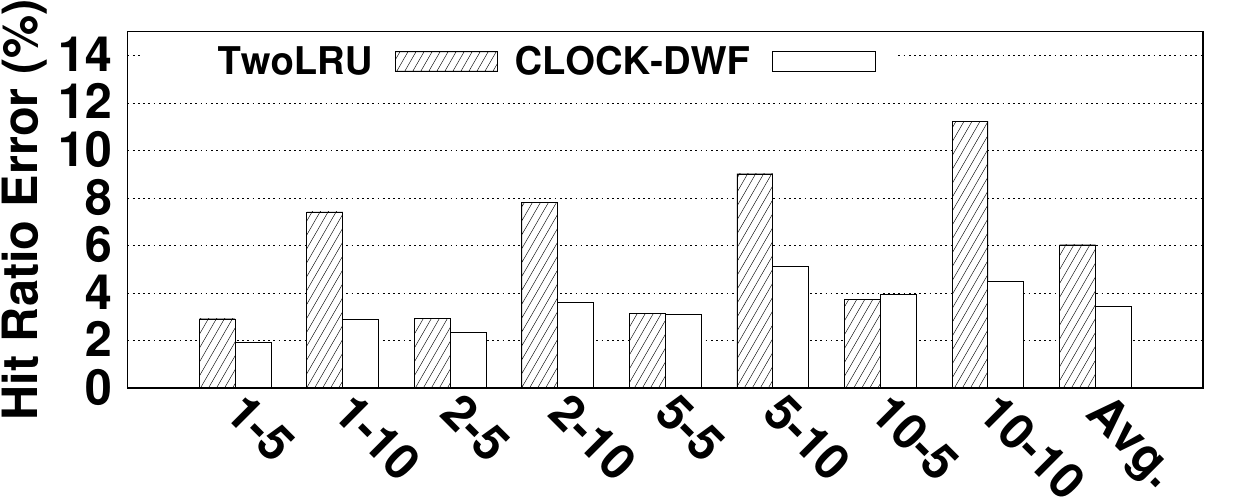}
		\label{fig:freqmineerror-lifetime}}
	\hfill
	\vspace{-.5cm}
	\subfloat[raytrace]{\includegraphics[width=0.242\textwidth]{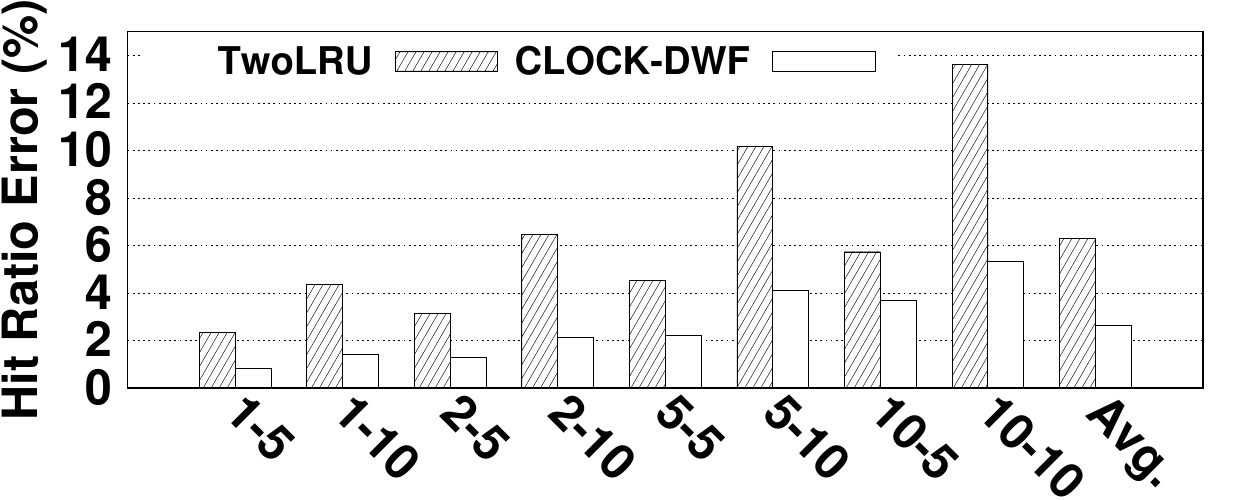}
		\label{fig:raytraceerror-lifetime}}
	\hfill
	\subfloat[streamcluster]{\includegraphics[width=0.242\textwidth]{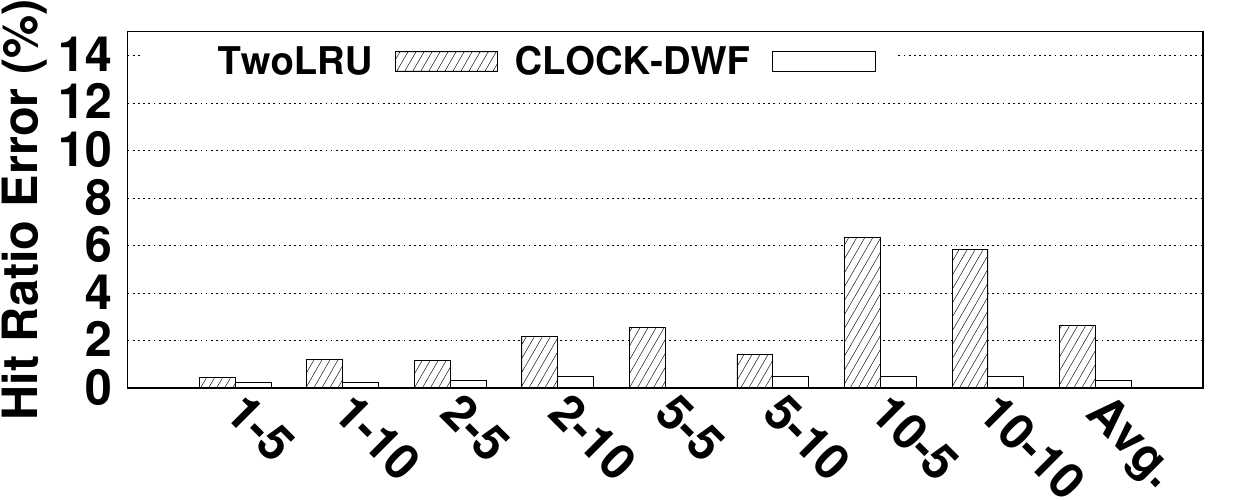}
		\label{fig:streamclustererror-lifetime}}
	\hfill
	\subfloat[vips]{\includegraphics[width=0.242\textwidth]{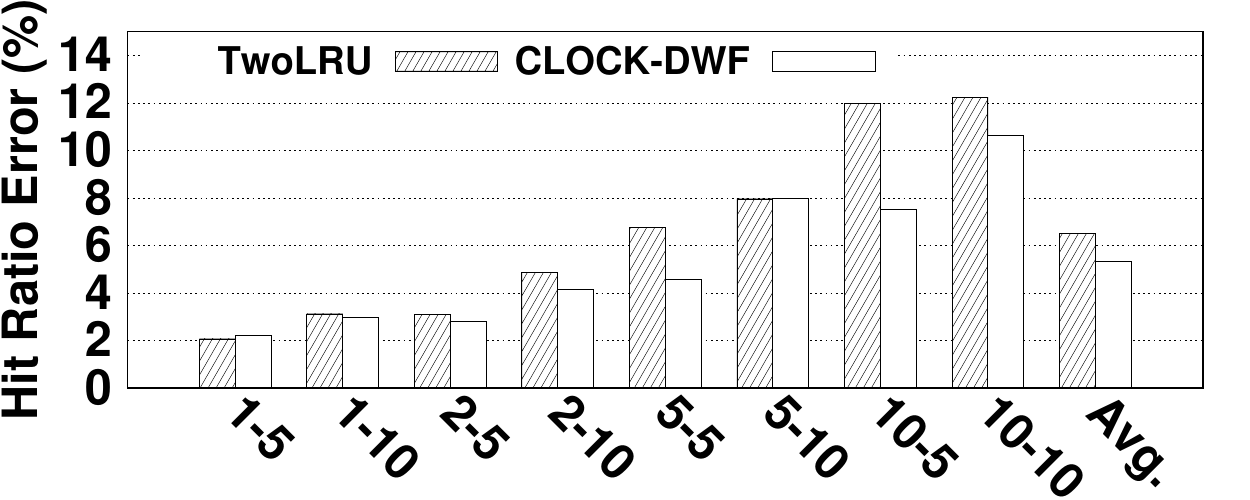}
		\label{fig:vipserror-lifetime}}
	\hfill
	\subfloat[x264]{\includegraphics[width=0.242\textwidth]{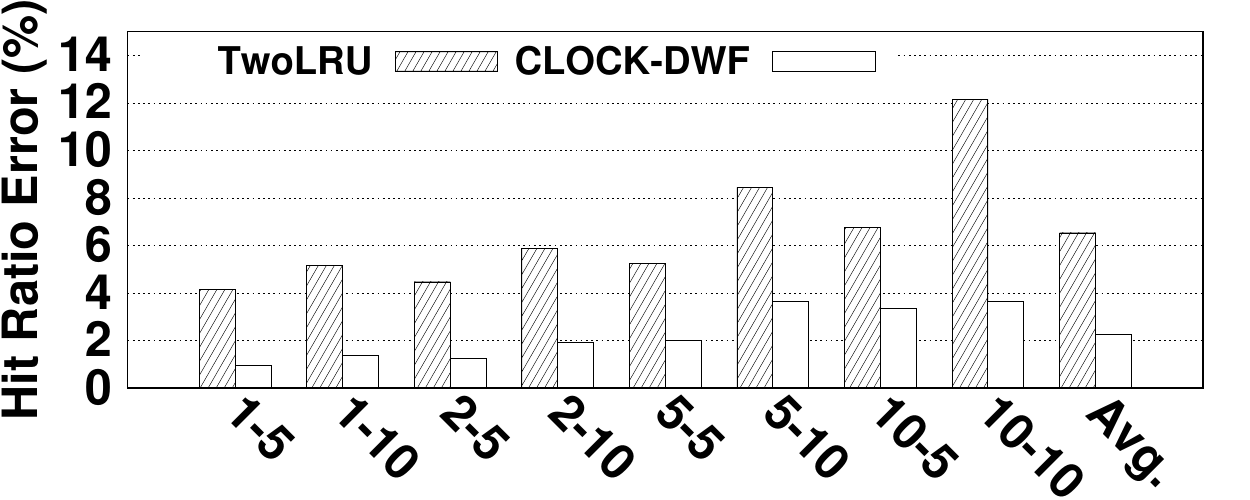}
		\label{fig:x264error-lifetime}}
	\vspace{-.2cm}
	\caption{Absolute Hit Ratio Values}
	\label{fig:abshitratio}
	\vspace{-0.5cm}
\end{figure*}

\subsection{Assumptions}
The HMAs that can be evaluated by the proposed model should have the following criteria:
\begin{itemize}[leftmargin=*]
	\item Data page eviction probabilities should be only dependent on the corresponding information in \emph{mapping} (Equation \ref{eq:eviction} {below}).
	\item Accessed data pages should have a static predefined mapping information , i.e., the first position of {the} queue in LRU algorithm (denoted as \emph{Hit\_Mapping} in Equation \ref{eq:hitaccess} {below}).
	\item \emph{mapping.NVM} and \emph{mapping.DRAM} should be totally ordered under \emph{less than or equal} ($\leq$) as denoted by Equation \ref{eq:ordered} {below}.
	\item A page can be migrated from NVM to DRAM ({i.e.,} promotion) only when it is accessed (Equation \ref{eq:migration} {below}).
	\item Evicted data pages from DRAM should be moved to NVM and as such, in the updated mapping (denoted as NM in Equation \ref{eq:demotion} {below}), they should be treated similar to a hit access in NVM (Equation \ref{eq:demotion} {below}).
	\item Evicted data pages from NVM are removed from memory (Equation \ref{eq:nvmeviction} {below}).
	\item Accessed data page will be assigned the highest position in $mapping$ (Equation \ref{eq:static} {below}).
\end{itemize}

\begin{figure}[h]
	\footnotesize
	\vspace{-.2cm}
	\begin{align}
	\label{eq:eviction}
	EvicP(page) &\propto mapping[page] \nonumber \\
	\forall p \in mapping&: p \neq page \Rightarrow evicP(page) \perp mapping[p]
	\end{align}
	\vspace{-.2cm}
	\begin{align}
	\label{eq:hitaccess}
	\forall mapping,& \forall page : U\!pd(mapping, page, new\_mapping) \nonumber \\
	\Rightarrow &new\_mapping[page] = \text{Hit\_Mapping}
	\end{align}
	\vspace{-.2cm}
	\begin{align}
	\label{eq:ordered}
	&\forall M \in \{D\!R\!A\!M, N\!V\!M\}, \forall P1, P2, P3 \in mapping.M: \nonumber \\
	&(P1 \leq P2 \wedge P2 \leq P1 \Rightarrow P1 = P2)  \wedge 
	(P1 \leq P2 \wedge P2 \leq P3 \Rightarrow  \nonumber \\ 
	& P1 \leq P3) \wedge (P1 \leq P2 \vee P2 \leq P1)
	\end{align}
	\vspace{-.2cm}
	\begin{align}
	\label{eq:migration}
	U\!pd(mapping,& page, new\_mapping) \wedge mig(mapping, mig\_page) \nonumber \\
	\Rightarrow &mig\_page = N\!U\!L\!L \vee mig\_page = page
	\end{align}
	\vspace{-.2cm}
	\begin{align}
	\label{eq:demotion}
	&P \in D\!R\!A\!M \wedge U\!pd(M, P2, N\!M) \wedge eviction\_victim = P2 \nonumber \\
	&\Rightarrow N\!M[P] = \text{Hit\_Mapping} \wedge P \in N\!M.N\!V\!M
	\end{align}
	\vspace{-.2cm}
	\begin{align}
	\label{eq:nvmeviction}
	&P \in N\!V\!M \wedge Upd(M, P2, N\!M) \wedge eviction\_victim = P\nonumber \\
	&\Rightarrow P \not\in N\!M
	\end{align}
	\vspace{-.2cm}
	\begin{align}
	\label{eq:static}
	& \forall M \in \{D\!R\!A\!M, N\!V\!M\}, \forall p \in {mapping}.M : p \leq \text{Hit\_Mapping}
	\end{align}
	\vspace{-.5cm}
\end{figure}

\vspace{-.3cm}
\section{Absolute Error Values}
\label{sec:asberror}
{We report} absolute error values for hit ratio and AMAT.
Relative errors for both hit ratio and AMAT are presented in Section \ref{sec:accuracy}.
The absolute error of AMAT for various benchmarks is 7.14\ $\eta s$ on average, reported in Fig. \ref{fig:absresponse}.
The maximum absolute error for estimating AMAT {is for the} \emph{vips} benchmark in CLOCK-DWF, which is 22.35 $\eta s$.
In \emph{bodytrack}, TwoLRU and CLOCK-DWF have almost the same relative AMAT error as depicted in Fig. 14
while the absolute AMAT error is almost twofold in CLOCK-DWF.
This is due to the difference in number of {disk} and memory accesses in {the} two examined HMAs.

The average {absolute} hit ratio error is 5.31\% and 2.48\% for TwoLRU and CLOCK-DWF, respectively as demonstrated in Fig. \ref{fig:abshitratio}.
The \emph{dedup} benchmark has highest absolute hit ratio error (8.11\% on average) in both examined HMAs, which is due to anomalies in the access distribution of this benchmark.
Such {anomalies} also exist in estimating AMAT and NVM lifetime of \emph{dedup} benchmark.
The highest {absolute} hit ratio error {(14.2\%)}, however, belongs to \emph{facesim} {for} TwoLRU.
CLOCK-DWF has a lower maximum hit ratio error of 10.6\%, which belongs to \emph{vips}.
\end{document}